\documentclass[12pt,epsf]{article}

\usepackage{times}
\usepackage{graphicx}
\usepackage{amsfonts}
\usepackage{amsmath}
\usepackage{amssymb}
\usepackage{amstext}
\usepackage{amsthm}

\usepackage{epsfig}
\psfigdriver{dvips}
\usepackage{latexsym}
\usepackage[matrix,curve]{xypic}
\usepackage{rotating}
\rotdriver{dvips}

\begin{document}

\baselineskip 6mm
\renewcommand{\thefootnote}{\fnsymbol{footnote}}


\newcommand{\nc}{\newcommand}
\newcommand{\rnc}{\renewcommand}


\rnc{\baselinestretch}{1.24}    
\setlength{\jot}{6pt}       
\rnc{\arraystretch}{1.24}   

\makeatletter
\rnc{\theequation}{\thesection.\arabic{equation}}
\@addtoreset{equation}{section}
\makeatother



\nc{\be}{\begin{equation}}

\nc{\ee}{\end{equation}}

\nc{\bea}{\begin{eqnarray}}

\nc{\eea}{\end{eqnarray}}

\nc{\xx}{\nonumber\\}

\nc{\ct}{\cite}

\nc{\la}{\label}

\nc{\eq}[1]{(\ref{#1})}

\nc{\newcaption}[1]{\centerline{\parbox{6in}{\caption{#1}}}}

\nc{\fig}[3]{

\begin{figure}
\centerline{\epsfxsize=#1\epsfbox{#2.eps}}
\newcaption{#3. \label{#2}}
\end{figure}
}


\def\CA{{\cal A}}
\def\CC{{\cal C}}
\def\CD{{\cal D}}
\def\CE{{\cal E}}
\def\CF{{\cal F}}
\def\CG{{\cal G}}
\def\CH{{\cal H}}
\def\CK{{\cal K}}
\def\CL{{\cal L}}
\def\CM{{\cal M}}
\def\CN{{\cal N}}
\def\CO{{\cal O}}
\def\CP{{\cal P}}
\def\CR{{\cal R}}
\def\CS{{\cal S}}
\def\CU{{\cal U}}
\def\CV{{\cal V}}
\def\CW{{\cal W}}
\def\CY{{\cal Y}}
\def\CZ{{\cal Z}}


\def\IB{{\hbox{{\rm I}\kern-.2em\hbox{\rm B}}}}
\def\IC{\,\,{\hbox{{\rm I}\kern-.50em\hbox{\bf C}}}}
\def\ID{{\hbox{{\rm I}\kern-.2em\hbox{\rm D}}}}
\def\IF{{\hbox{{\rm I}\kern-.2em\hbox{\rm F}}}}
\def\IH{{\hbox{{\rm I}\kern-.2em\hbox{\rm H}}}}
\def\IN{{\hbox{{\rm I}\kern-.2em\hbox{\rm N}}}}
\def\IP{{\hbox{{\rm I}\kern-.2em\hbox{\rm P}}}}
\def\IR{{\hbox{{\rm I}\kern-.2em\hbox{\rm R}}}}
\def\IZ{{\hbox{{\rm Z}\kern-.4em\hbox{\rm Z}}}}


\def\a{\alpha}
\def\b{\beta}
\def\d{\delta}
\def\ep{\epsilon}
\def\ga{\gamma}
\def\k{\kappa}
\def\l{\lambda}
\def\s{\sigma}
\def\t{\theta}
\def\w{\omega}
\def\G{\Gamma}


\def\half{\frac{1}{2}}
\def\dint#1#2{\int\limits_{#1}^{#2}}
\def\goto{\rightarrow}
\def\para{\parallel}
\def\brac#1{\langle #1 \rangle}
\def\curl{\nabla\times}
\def\div{\nabla\cdot}
\def\p{\partial}


\def\Tr{{\rm Tr}\,}
\def\det{{\rm det}}


\def\vare{\varepsilon}
\def\zbar{\bar{z}}
\def\wbar{\bar{w}}
\def\what#1{\widehat{#1}}


\def\ad{\dot{a}}
\def\bd{\dot{b}}
\def\cd{\dot{c}}
\def\dd{\dot{d}}
\def\so{SO(4)}
\def\bfr{{\bf R}}
\def\bfc{{\bf C}}
\def\bfz{{\bf Z}}

\begin{titlepage}


\hfill\parbox{3.7cm} {KIAS-P08059 \\
{\tt arXiv:0809.4728}}

\vspace{15mm}

\begin{center}
{\Large \bf  Emergent Spacetime and The Origin of Gravity}

\vspace{10mm}
Hyun Seok Yang \footnote{hsyang@kias.re.kr}
\\[10mm]

{\sl School of Physics, Korea Institute for Advanced Study,
Seoul 130-012, Korea}

\end{center}

\thispagestyle{empty}

\vskip1cm


\centerline{\bf ABSTRACT}
\vskip 4mm
\noindent

We present an exposition on the geometrization of the
electromagnetic force. We show that, in noncommutative (NC)
spacetime, there always exists a coordinate transformation to
locally eliminate the electromagnetic force, which is precisely the
Darboux theorem in symplectic geometry. As a consequence, the
electromagnetism can be realized as a geometrical property of
spacetime like gravity. We show that the geometrization of the
electromagnetic force in NC spacetime is the origin of gravity,
dubbed as the emergent gravity. We discuss how the emergent gravity
reveals a novel, radically different picture about the origin of
spacetime. In particular, the emergent gravity naturally explains
the dynamical origin of flat spacetime, which is absent in Einstein
gravity. This spacetime picture turns out to be crucial for a
tenable solution of the cosmological constant problem.
\\



\vspace{1cm}

\today

\end{titlepage}

\renewcommand{\thefootnote}{\arabic{footnote}}
\setcounter{footnote}{0}

\section{Deformation Theory}

One of the main trends in modern physics and mathematics is to study
a theory of deformations. Deformations are performed first to
specify a particular structure (e.g., complex, symplectic, or
algebraic structures) which one wants to deform, and then to
introduce a deformation parameter $[\hbar]$ such that the limit
$[\hbar] \to 0$ recovers its parent theory. The most salient
examples of the deformation theories are Kodaira-Spencer theory,
deformation quantization, quantum group, etc. in mathematics and
quantum mechanics, string theory, noncommutative (NC) field theory,
etc. in physics. Interestingly, consequences after the deformation
are often radical: A theory with $[\hbar] \neq 0$ is often
qualitatively different from its parent theory and reveals a
unification of physical or mathematical structures (e.g.,
wave-particle duality, mirror symmetry, etc.).

Let us focus on the deformation theories appearing in physics. Our
mission is to deform some structures of a point-particle theory in
classical mechanics. There could be several in general, but the most
salient ones among them are quantum mechanics, string theory and NC
field theory, which we call $\hbar$-deformation,
$\alpha^\prime$-deformation and $\theta$-deformation, respectively.
The deformation parameter $[\hbar]$ (which denotes a generic one) is
mostly a dimensionful constant and plays a role of a conversion
factor bridging two different quantities, e.g., $p = 2 \pi \hbar/
\lambda$ for the famous wave-particle duality in quantum mechanics.
The introduction of the new constant $[\hbar]$ into the theory is
not a simple addition but often a radical change of the parent
theory triggering a new physics. Let us reflect the new physics
sprouted up from the $[\hbar]$-deformation, which never exists in
the $[\hbar]=0$ theory.

Quantum mechanics is the formulation of mechanics in NC phase space
\be \la{nc-phase}
[x^i, p_k] = i \hbar \delta^i_k.
\ee
The deformation parameter $\hbar$ is to deform a commutative Poisson
algebra of observables in phase space into NC one. This
$\hbar$-deformation (quantum mechanics) has activated revolutionary
changes of classical physics. One of the most prominent physics is
the wave-particle duality whose striking physics could be embodied
in the two-slit experiment.

String theory can be regarded as a deformation of point-particle
theory in the sense that zero-dimensional point particles are
replaced by one-dimensional extended objects, strings, whose size is
characterized by the parameter $\alpha^\prime$. This
$\alpha^\prime$-deformation also results in a fundamental change of
physics, which has never been observed in a particle theory. It is
rather a theory of gravity (or grandiloquently a theory of
everything). One of the striking consequences due to the
$\alpha^\prime$-deformation is `T-duality', which is a symmetry
between small and large distances, symbolically represented by
\be \la{t-duality}
R \leftrightarrow \frac{\alpha^\prime}{R}.
\ee
The T-duality is a crucial ingredient for various string dualities
and mirror symmetry.

NC field theory is the formulation of field theory in NC spacetime
\be \la{nc-spacetime}
[y^a, y^b]_\star = i \theta^{ab}.
\ee
See \ct{nc-review,szabo} for a review of this subject. We will
consider only space-noncommutativity throughout the paper in spite
of the abuse of the term `NC spacetime' and argue in Section 4.1
that ``Time" emerges in a different way. This NC spacetime arises
from introducing a symplectic structure $B = \half B_{ab} dy^a
\wedge dy^b$ and then quantizing the spacetime with its Poisson
structure $\theta^{ab} \equiv (B^{-1})^{ab}$, treating it as a
quantum phase space. In other words, the spacetime
\eq{nc-spacetime} becomes a NC phase space. Therefore the NC field
theory, which we call $\theta$-deformation, is mathematically very
similar to quantum mechanics. They are all involved with a NC
$\star$-algebra generated by Eq.\eq{nc-phase} or
Eq.\eq{nc-spacetime}. Indeed we will find many parallels. Another
naive observation is that the $\theta$-deformation (NC field theory)
would be much similar to the $\alpha^\prime$-deformation from the
viewpoint of deformation theory since the deformation parameters
$\alpha^\prime$ and $\theta$ equally carry the dimension of
$(\rm{length})^2$. A difference is that the $\theta$-deformation is
done in the field theory framework. We will further elaborate the
similarity in this paper.

What is a new physics due to the $\theta$-deformation ? A remarkable
fact is that translations in NC directions are an inner automorphism
of NC $\star$-algebra $\CA_\theta$, i.e., $e^{ik \cdot y} \star
\widehat{f}(y) \star e^{-ik \cdot y} = \widehat{f}(y +  \theta \cdot k)$
for any $\widehat{f}(y) \in \CA_\theta$ or, in its infinitesimal
form,
\be \la{inner-der}
[y^a, \widehat{f}(y)]_\star = i \theta^{ab} \p_b \widehat{f}(y).
\ee
In this paper we will denote NC fields (or variables) with the hat
as in Eq.\eq{inner-der} but we will omit the hat for NC coordinates
$y^a$ in Eq.\eq{nc-spacetime} for notational convenience. We will
show that the $\theta$-deformation is seeding in it the physics of
the $\alpha^\prime$-deformation as well as the $\hbar$-deformation,
so to answer the question in the Table 1.

\begin{center}
\begin{tabular}{|c|c|c|}
  \hline
  Theory & Deformation & New physics \\
  \hline
  Quantum mechanics & $\hbar$ & wave-particle duality \\
  \hline
  String theory & $\alpha^\prime$ & T-duality \\
  \hline
  NC field theory & $\theta^{ab}$ &  {\Large ?} \\
  \hline
\end{tabular} \\
\vspace{0.5cm}
Table 1. $[\hbar]$-deformations and their new physics
\end{center}

This paper is organized as follows. In Section 2 we review the
picture of emergent gravity presented in \ct{hsy2} with a few
refinements. First we consolidate some results well-known from
string theory to explain why there always exists a coordinate
transformation to locally eliminate the electromagnetic force as
long as D-brane worldvolume $M$ supports a symplectic structure $B$,
i.e., $M$ becomes a NC space. That is, the NC spacetime admits a
novel form of the equivalence principle, known as the Darboux
theorem, for the geometrization of the electromagnetism. It turns
out \ct{hsy2} that the Darboux theorem as the equivalence principle
in symplectic geometry is the essence of emergent gravity. See the
Table 2. In addition we add a new observation that the
geometrization of the electromagnetism in the $B$-field background
can be nicely understood in terms of the generalized geometry
\ct{gcg-hitchin,gcg-gualtieri}. Recently there have been
considerable efforts
\ct{madore-mourad,vacaru,langman-szabo,rivelles,eg1-mine,sty-ys,hsy1,hsy2,
mpla,hsy3,hsy-cc,eg-others,madore-poisson,steinacker,harald-wien,steinacker06}
to construct gravity from NC field theories. The emergent gravity
has also been suggested to resolve the cosmological constant problem
and dark energy \ct{pad,hsy-cc}.

In Section 3, we put the arguments in Section 2 on a firm foundation
using the background independent formulation of NC gauge theory
\ct{sw,seiberg}. In Sec. 3.1, we first clarify based on the argument in \ct{hsy3}
that the emergent gravity from NC gauge theory is essentially a
large $N$ duality consistent with the AdS/CFT duality \ct{ads-cft}.
And then we move onto the geometric representation of NC field
theory using the inner automorphism \eq{inner-der} of the NC
spacetime \eq{nc-spacetime}. In Sec. 3.2, we show how to explicitly
determine a gravitational metric emerging from NC gauge fields and
show that the equations of motion for NC gauge fields are mapped to
the Einstein equations for the emergent metric. This part consists
of our main new results generalizing the emergent gravity in
\ct{hsy1,hsy2} for self-dual gauge fields. In the course of the
derivation, we find that NC gauge fields induce an exotic form of
energy, dubbed as the Liouville energy-momentum tensor. A simple
analysis shows that this Liouville energy mimics the several aspects
of dark energy, so we suggest the energy as a plausible candidate of
dark energy. In Sec. 3.3, the emergent gravity is further
generalized to the nontrivial background of nonconstant $\theta$
induced by an inhomogenous condensation of gauge fields. In Sec.
3.4, we discuss the spacetime picture revealed from NC gauge fields.
We also confirm the observation in \ct{hsy-cc} that the emergent
gravity reveals a remarkably beautiful and consistent picture about
the dynamical origin of flat spacetime.

In Section 4 we speculate how to understand ``Time" and matter
fields in the context of emergent geometry. As a first step, we
elucidate in Sec. 4.1 how the well-known `minimal coupling' of
matters with gauge fields can be understood as a symplectic geometry
in phase space. There are two important works \ct{sternberg,dyson}
for this understanding. Based on the symplectic geometry of
particles, in Sec. 4.2, we suggest a K-theory picture for matter
fields such as quarks and leptons adopting the Fermi-surface
scenario in \ct{horava,volovik} where non-Abelian gauge fields are
understood as collective modes acting on the matter fields.

In Section 5, we address the problem on the existence of spin-2
bound states which presupposes the basis of emergent gravity.
Although we don't know any rigorous proof, we outline some positive
evidences for the bound states using the relation to the AdS/CFT
duality. We further notice an interesting similarity between the BCS
superconductivity \ct{bcs} and the emergent gravity about some
dynamical mechanism for the spin-0 and spin-2 bound states,
respectively. See the Table 3. We also discuss the issues on the
Lorentz symmetry breaking and the nonlocality in NC field theories
from the viewpoint of emergent spacetime.

In Section 6, we summarize the message uncovered by the emergent
gravity picture with some closing remarks.

The calculational details in Section 3 are deferred to two
Appendices. In Appendix A we give a self-contained proof of the
equivalence between self-dual NC electromagnetism and self-dual
Einstein gravity, first shown in \ct{hsy1}, for completeness. The
self-dual case will provide a clear picture to appreciate what the
emergent gravity is, which will also be useful to consider a general
situation of emergent gravity. In Appendix B the equivalence is
generalized to arbitrary NC gauge fields.

\section{Geometrization of Forces}

One of the guiding principles in modern physics is the
geometrization of forces, i.e., to view physical forces as a
reflection of the curvature of the geometry of spacetime or internal
space. In this line of thought, gravity is quite different from the
other three forces - the electromagnetic, the weak, and the strong
interactions. It is a manifestation of the curvature of spacetime
while the other three are a manifestation of the curvature of
internal spaces. If it makes sense to pursue a unification of
forces, in which the four forces are different manifestations of a
single force, it would be desirable to reconcile gravity with the
others and to find a general categorical structure of physical
forces: Either to find a rationale that gravity is not a fundamental
force or to find a framework that the other three forces are also
geometrical properties of spacetime. We will show these two features
are simultaneously realized in NC spacetime, at least, for the
electromagnetism.

\subsection{Einstein's happiest thought}

The geometrization of forces is largely originated with Albert
Einstein, whose general theory of relativity is to view the gravity
as a metric field of spacetime which is determined by the
distribution of matter and energy. The remarkable vision of gravity
in terms of the geometry of spacetime has been based on the local
equivalence of gravitation and inertia, or the local cancellation of
the gravitational field by local inertial frames - the equivalence
principle. Einstein once recalled that the equivalence principle was
the happiest thought of his life.

The equivalence principle guarantees that it is ``always" possible
at any spacetime point of interest to find a coordinate system, say
$\xi^\alpha$, such that the effects of gravity will disappear over a
differential region in the neighborhood of that point. (Precisely
speaking, the neighborhood should be taken small enough so that the
variation of gravity within the region may be neglected.) For a
particle moving freely under the influence of purely gravitational
force, the equation of motion in terms of the freely falling
coordinate system $\xi^\alpha$ is thus
\be \la{eq-free-spacetime}
\frac{d^2 \xi^\alpha}{d\tau^2} = 0
\ee
with $d\tau$ the proper time
\be \la{metric-inertial}
d\tau^2 = \eta_{\alpha\beta}d \xi^\alpha d\xi^\beta.
\ee
We will use the metric $\eta_{\alpha\beta}$ with signature $(-++
\cdots)$ throughout the paper.

Suppose that we perform a coordinate transformation to find the
corresponding equations in a laboratory at rest, which may be
described by a Cartesian coordinate system $x^\mu$. The freely
falling coordinates $\xi^\alpha$ are then functions of the $x^\mu$,
that is, $\xi^\alpha = \xi^\alpha(x)$. The freely falling particle
in the laboratory coordinate system now obeys the equation of motion
\be \la{eom-lab}
\frac{d^2 x^\mu}{d \tau^2} + \Gamma^\mu_{\nu \lambda} \frac{d x^\nu}{d \tau}
\frac{d x^\lambda}{d \tau} = 0
\ee
where
\be \la{curved-distance}
 d\tau^2 = g_{\mu\nu} (x) dx^\mu dx^\nu
\ee
and
\be \la{curved-metric}
g_{\mu\nu} (x) = \eta_{\alpha\beta} \frac{\p \xi^\alpha}{\p x^\mu}
\frac{\p \xi^\beta}{\p x^\nu}.
\ee
It turns out that Eq.\eq{eom-lab} is the geodesic equation moving on
the shortest possible path between two points through the curved
spacetime described by the metric \eq{curved-metric}. In the end the
gravitational force manifests itself only as the geometry of
spacetime.

In accordance with the principle of general covariance the laws of
physics must be independent of the choice of spacetime coordinates.
That is, Eq.\eq{eom-lab} is true in all coordinate systems. For
example, under a coordinate transformation $x^\mu
\to {x^\prime}^\mu$, the metric transforms into
\be \la{metric-tr}
 g^\prime_{\mu\nu} (x^\prime) =
 \frac{\p x^\lambda}{\p {x^\prime}^\mu} \frac{\p x^\sigma}{\p {x^\prime}^\nu}
 g_{\lambda\sigma}(x)
\ee
and Eq.\eq{eom-lab} transforms into the geodesic equation in the
spacetime described by the metric \eq{metric-tr}. The significance
of the equivalence principle in conjunction with the principle of
covariance lies in its statement that there ``always" exists a
locally inertial frame at an arbitrary point $P$ in spacetime where
$g^\prime_{\alpha\beta} (P) = \eta_{\alpha\beta}$ and
${\Gamma^\prime}^\mu_{\alpha\beta}(P)=0$. But the second derivatives
of $g^\prime_{\alpha\beta}$ at $P$ cannot all be set to zero unless
the spacetime is flat. This coordinate system is precisely the
freely falling coordinates $\xi^\alpha$ in
Eq.\eq{eq-free-spacetime}, i.e., $\xi^\alpha = {x^\prime}^\alpha
(x)$, so the metric at $P$ in the original system can consistently
be written as the form \eq{curved-metric}.

But a routine calculation using the metric \eq{curved-metric} leads
to identically vanishing curvature tensors. Thus one may claim that
the geometry described by the metric \eq{curved-metric} is always
flat. Of course it should not be the case. Remember that the metric
\eq{curved-metric} in the $x$-coordinate system should be understood at a
point $P$ since it has been obtained from the local inertial frame
$\xi^\alpha$ where $g^\prime_{\alpha\beta} (P) = \eta_{\alpha\beta}$
and ${\Gamma^\prime}^\mu_{\alpha\beta}(P)=0$ are satisfied only at
that point. In order to calculate the curvature tensors correctly,
one needs to extend the local inertial frame at $P$ to an
infinitesimal neighborhood. A special and useful realization of such
a local inertial frame is a Riemann normal coordinate system
\ct{big-gravity} (where we choose the point $P$ as a coordinate origin, i.e.,
$\xi^\alpha|_P = x^\mu|_P = 0$)
\be \la{normal-coordinate}
\xi^\alpha (x) = x^\alpha + \half \Gamma^\alpha_{\mu\nu}(P) x^\mu
x^\nu + \frac{1}{6}\bigl(\Gamma^\alpha_{\mu\beta}
\Gamma^\beta_{\nu\lambda} + \p_\lambda \Gamma^\alpha_{\mu\nu} \bigr)(P)
x^\mu x^\nu x^\lambda + \cdots,
\ee
which can be checked using Eq.\eq{metric-tr} with the identification
${x^\prime}^\alpha = \xi^\alpha$.  One can then arrive at a metric
\be \la{normal-metric}
g^\prime_{\alpha\beta} (x) = \eta_{\alpha\beta} -
\frac{1}{3} R_{\alpha\mu\beta\nu}(P) x^\mu x^\nu  - \frac{1}{6}
D_\lambda R_{\alpha\mu\beta\nu}(P) x^\lambda x^\mu x^\nu + \cdots.
\ee

\subsection{Darboux theorem as the equivalence principle in symplectic geometry}

What about other forces ? Is it possible to realize, for example,
the electromagnetism as a geometrical property of spacetime like
gravity ? To be specific, we are wondering whether or not there
``always" exists any coordinate transformation to eliminate the
electromagnetic force at least locally. The usual wisdom says no
since there is no analogue of the equivalence principle for the
geometrization of the electromagnetic force. But one has to recall
that this wisdom has been based on the usual concept of geometry,
i.e., Riemannian geometry in commutative spacetime. Surprisingly,
the conventional wisdom turns out to be no longer true in NC
spacetime, which is based on symplectic geometry in sharp contrast
to the Riemannian geometry.

We will show that it is ``always" possible to find a coordinate
transformation to eliminate locally the electromagnetic force if and
only if spacetime supports a symplectic structure, viz., NC
spacetime. To be definite, we will proceed with string theory
although an elegant and rigorous approach can be done using the
formalism of deformation quantization \ct{kontsevich}. See \ct{hsy2}
for some arguments based on the latter approach.

A scheme to introduce gauge fields in string theory is by means of
boundary interactions or via boundary conditions of open strings,
aside from through the Kaluza-Klein compactifications in type II or
heterotic string theories. With a compact notation, the open or
closed string action reads as \footnote{Although we will focus on
the open string theory, our arguments in this section also hold for
a closed string theory where the string worldsheet $\Sigma$ is a
compact Riemann surface without boundary, so the last term in
Eq.\eq{open-action} is absent.}
\be \la{open-action}
S = \frac{1}{4 \pi \alpha^\prime} \int_{\Sigma} |d X|^2  -
\int_{\Sigma} B -
\int_{\p \Sigma} A,
\ee
where $X: \Sigma \to M$ is a map from an open or closed string
worldsheet $\Sigma$ to a target spacetime $M$ and $B(\Sigma) =
X^*B(M)$ and $A(\p \Sigma) = X^* A(M)$ are pull-backs of spacetime
fields to the worldsheet $\Sigma$ and the worldsheet boundary $\p
\Sigma$, respectively.

The string action \eq{open-action} respects the following local
symmetries.

(I) Diff(M)-symmetry:
\be \la{diff-symm}
X \to X^\prime = X^\prime (X) \; \in \; {\rm Diff}(M)
\ee
and the corresponding transformations of target fields $B$ and $A$
including also a target metric (hidden) in the first term of
Eq.\eq{open-action}.

(II) $\Lambda$-symmetry:
\be \la{lambda-symm}
(B,\; A) \to (B - d\Lambda, \; A + \Lambda)
\ee
where the gauge parameter $\Lambda$ is a one-form in $M$. A simple
application of Stokes' theorem immediately verifies the symmetry
\eq{lambda-symm}. Note that the $\Lambda$-symmetry is present only
when $B \neq 0$. When $B=0$, the symmetry \eq{lambda-symm} is
reduced to $A \to A + d\lambda$, which is the ordinary $U(1)$ gauge
symmetry.

The above two local symmetries in string theory must also be
realized as the symmetries in low energy effective theory. We well
understand the root of the symmetry \eq{diff-symm} since the string
action \eq{open-action} describes a gravitational theory in target
spacetime. The diffeomorphism symmetry \eq{diff-symm} certainly
signifies the emergence of gravity in the target space $M$. A
natural question is then what is a root of the $\Lambda$-symmetry
\eq{lambda-symm}.

Unfortunately, as far as we know, there has been no serious
investigation about a physical consequence of the symmetry
\eq{lambda-symm}. As a provoking comment, let us
first point out that the $\Lambda$-symmetry \eq{lambda-symm} is as
large as the Diff(M)-symmetry \eq{diff-symm} (supposing that the
rank of $B$ is equal to the dimension of $M$) and is present only
when $B \neq 0$, so a stringy symmetry by nature. Indeed this is a
broad hint that there will be a radical change of physics when $B
\neq 0$ -- the new physics due to the $\theta$-deformation in the
Table 1.

To proceed with a general context, let us first discuss a
geometrical interpretation of the $\Lambda$-symmetry without
specifying low energy effective theories. Suppose that the two-form
$B \in \Lambda^2(M)$ is closed in $M$, i.e., $dB=0$, and
nondegenerate, i.e., nowhere vanishing in $M$.\footnote{In string
theory, $H = dB \in \Lambda^3(M)$ is not necessarily zero. We don't
know much about this case, so we will restrict to the symplectic
case. But the connection with the generalized geometry, to be
shortly discussed later, might be helpful to understand more general
cases.} One can then regard the two-form $B$ as a symplectic
structure on $M$ and the pair $(M, B)$ as a symplectic manifold. The
symplectic geometry is a less intuitive type of geometry but it
should be familiar with classical mechanics, especially, the
Hamiltonian mechanics
\ct{mechanics} and, more prominently, quantum mechanics.

The symplectic geometry respects an important property, known as the
Darboux theorem \ct{darboux}, stating that every symplectic manifold
of the same dimension is locally indistinguishable. More precisely,
let $(M, \omega)$ be a symplectic manifold. Then in a neighborhood
of each $P \in M$, there is a local coordinate chart in which
$\omega$ is a constant, i.e., $(M, \omega)
\cong ({\bf R}^{2n}, \sum dq^i \wedge dp_i)$. For our purpose, we
will use its refined version - the Moser lemma \ct{moser} -
describing a cohomological condition for two symplectic structures
to be equivalent. Given two-forms $\omega$ and $\omega^\prime$ such
that $[\omega] = [\omega^\prime] \in H^2(M)$ and $\omega_t = \omega
+ t(\omega^\prime -\omega)$ is symplectic $\forall t \in [0, 1]$,
then there exists a diffeomorphism $\phi_t : M \to M$ such that
$\phi^*_t(\omega_t) = \omega$. This implies that all $\omega_t$ are
related by coordinate transformations generated by a vector field
$X_t$ satisfying
\be \la{moser-eq}
\iota_{X_t} \omega_t + A = 0
\ee
where $\omega^\prime - \omega = dA$. In terms of local coordinates,
there always exists a coordinate transformation $\phi_1$ whose
pullback maps $\omega^\prime = \omega + dA$ to $\omega$, i.e.,
$\phi_1 : y \mapsto x = x(y)$ so that
\be \la{darboux-tr}
\frac{\p x^\alpha}{\p y^a}\frac{\p x^\beta}{\p y^b} \omega^\prime_{\alpha\beta}(x)
= \omega_{ab}(y).
\ee

The Moser lemma \eq{darboux-tr} stating that the symplectic
manifolds $(M, \omega_0)$ and $(M, \omega_1)$ are strongly isotopic
is a global statement and will be applied to our problem as follows.
For a symplectic manifold $(M, \omega_1 = B + F)$ where $F=dA$, by
the Darboux theorem, one can always find a local coordinate chart
$(U; y^1, \cdots, y^{2n})$ centered at $p \in M$ and valid on the
neighborhood $U$ such that $\omega_0(p) = \frac{1}{2} B_{ab} dy^a
\wedge dy^b$ where $B_{ab}$ is a constant symplectic matrix of rank
$2n$. Then there are two symplectic structures on $U$; the given
$\omega_1 = B + F$ and $\omega_0 = B$. Consider a smooth family
$\omega_t = \omega_0 + t(\omega_1 - \omega_0)$ of symplectic forms
joining $\omega_0$ to $\omega_1$. Now the Moser lemma
\eq{darboux-tr} implies that there exists a global diffeomorphism
$\phi: M \times {\bf R} \to M$ such that $\phi^*_t(\omega_t) =
\omega_0, \; 0 \leq t \leq 1$. If there exists such a diffeomorphism,
in terms of the associated time-dependent vector field $X_t \equiv
\frac{d \phi_t}{dt} \circ \phi_t^{-1}$, one would have for all $0 \leq t \leq
1$ that ${\cal L}_{X_t} \omega_t + \frac{d \omega_t}{dt} = 0$ which
can be reduced to Eq.\eq{moser-eq}. One can pointwise solve the
Moser's equation \eq{moser-eq} to obtain a unique smooth family of
vector fields $X_t, \; 0 \leq t \leq 1$, generating the global
diffeomorphism $\phi_t$ satisfying $\frac{d \phi_t}{dt} = X_t \circ
\phi_t$. So everything boils down to solving the Moser's equation
\eq{moser-eq} for $X_t$.

First one may solve the equation \eq{moser-eq} at $t \to 0$ to
determine $X_0 = X_0(y)$ on $U$ in terms of the Darboux coordinates
$y^a$ and extend to all $0 \leq t \leq 1$ by integration
\ct{jurco-schupp}. After integration, one can find a local isotopy
$\phi: U \times [0,1] \to M$ with $\phi_t^*(\omega_t) = \omega_0$
for all $t \in [0,1]$. Let us denote the resulting coordinate
transformation $\phi_1(y)$ on $U$ generated by the vector field
$X_1$ as $x^a(y) = y^a + X_1^a(y)$. (Compare the result with
Eq.\eq{cov-coordinate} where $X_1^a(y) := \theta^{ab}
\widehat{A}_b(y)$.) This is the result we want to get from the data
$(M, \omega_1 = B + F)$ by performing a coordinate transformation
\eq{darboux-tr} onto a local Darboux chart. Therefore sometimes we
will simply refer the Darboux theorem to Eq.\eq{darboux-tr} in a
loose sense as long as the physical meaning is clear.

The string action \eq{open-action} indicates that, when $B \neq 0$,
its natural group of symmetries includes not only the diffeomorphism
\eq{diff-symm} in Riemannian geometry but also the $\Lambda$-symmetry \eq{lambda-symm}
in symplectic geometry. According to the Darboux theorem (precisely,
the Moser lemma stated above), the local change of symplectic
structure due to the $\Lambda$-symmetry \eq{lambda-symm} (or the
$B$-field transformation) can always be translated into a
diffeomorphism symmetry as in Eq.\eq{darboux-tr}. This fact implies
that the $\Lambda$-symmetry \eq{lambda-symm} should be considered as
a par with diffeomorphisms. It turns out \ct{hsy2} that the Darboux
theorem in symplectic geometry plays the same role as the
equivalence principle in general relativity for the geometrization
of the electromagnetic force. These geometrical structures inherent
in the string action \eq{open-action} are summarized below.

\begin{center}
\begin{tabular}{|c|c|}
  \hline
  (I) Riemannian geometry & (II) Symplectic geometry \\
  \hline
  Riemannian manifold $(M, g)$: &  Symplectic manifold $(M, \omega)$: \\
$M$ a smooth manifold & $M$ a smooth manifold \\
and $g: TM \otimes TM \to {\bf R}$ & and $\omega \in \Lambda^2(M)$ \\
a nondegenerate symmetric bilinear form &
a nondegenerate closed 2-form, i.e., $d \omega =0 $ \\
\hline
  Equivalence principle: & Darboux theorem: \\
Locally, $(M, g) \cong ({\bf R}^{2n}, \sum dx^\mu \otimes dx_\mu)$ &
Locally, $(M, \omega) \cong ({\bf R}^{2n}, \sum dq^i \wedge dp_i)$ \\
  \hline
\end{tabular} \\
\vspace{0.5cm}
Table 2. Riemannian geometry vs. Symplectic geometry
\end{center}

Therefore we need a generalized geometry when $B \neq 0$ which
treats both Riemannian geometry and symplectic geometry on equal
footing.\footnote{\label{rvs}A Riemannian geometry is defined by a
pair $(M,g)$ where the metric $g$ encodes all geometric informations
while a symplectic geometry is defined by a pair $(M, \omega)$ where
the 2-form $\omega$ encodes all. See the Table 2. A basic concept in
Riemannian geometry is a distance defined by the metric. One may
identify this distance with a geodesic worldline of a ``particle"
moving in $M$. On the contrary, a basic concept in symplectic
geometry is an area defined by the symplectic structure. One may
regard this area as a minimal worldsheet swept by a ``string" moving
in $M$. Amusingly, the Riemannian geometry is probed by particles
while the symplectic geometry would be probed by strings. But we
know that a Riemannian geometry (or gravity) is emergent from
strings ! This argument, though naive, glimpses the reason why the
$\theta$-deformation in the Table 1 goes parallel to the
$\alpha^\prime$-deformation.} Such kind of generalized geometry was
introduced by N. Hitchin \ct{gcg-hitchin} in 2002 and further
developed by M. Gualtieri and G. R. Cavalcanti \ct{gcg-gualtieri}.
Generalized complex geometry unites complex and symplectic
geometries such that it interpolates between a complex structure $J$
and a symplectic structure $\omega$ by viewing each as a complex (or
symplectic) structure ${\cal J}$ on the direct sum of the tangent
and cotangent bundle $E = TM \oplus T^*M$. A generalized complex
structure ${\cal J}: E \to E$ is a generalized almost complex
structure, satisfying ${\cal J}^2 = -1$ and ${\cal J}^* = - {\cal
J}$, whose sections are closed under the Courant bracket
\footnote{When $H = dB$ is not zero, the Courant bracket on $E$ is
`twisted' by the real, closed 3-form $H$ in the following way
$$[X + \xi, Y + \eta]_H = [X + \xi, Y + \eta]_C  + \iota_Y \iota_X
H. $$ See \ct{gcg-gualtieri} for more details, in particular, a
relation to gerbes.}
\be \la{courant}
[X + \xi, Y + \eta]_C = [X,Y] + {\cal L}_X \eta - {\cal L}_Y \xi -
\frac{1}{2} d\bigl(\iota_X \eta - \iota_Y \xi \bigr),
\ee
where ${\cal L}_X$ is the Lie derivative along the vector field $X$
and $d \, (\iota)$ is the exterior (interior) product.

An important point in generalized geometry is that the symmetries of
$E$, i.e., the endomorphisms of $E$ (the group of orthogonal Courant
automorphisms of $E$), are the composition of a diffeomorphism of
$M$ and a $B$-field transformation defined by $e^B(X+\xi) = X +
\xi + \iota_X B$ for any $X + \xi \in E$, where $B$ is an arbitrary
closed 2-form. This $B$-field transformation can be identified with
the $\Lambda$-symmetry \eq{lambda-symm} as follows. Let $(M, B)$ be
a symplectic manifold where $B = d \xi$, locally, by the Poincar\'e
lemma. The $\Lambda$-symmetry \eq{lambda-symm} can then be
understood as a shift of the canonical 1-form, $\xi
\to \xi - \Lambda$, which is the $B$-field transformation with the
identification $\Lambda = - \iota_X B$. With this notation, the
$B$-field transformation is equivalent to $B \to B + {\cal L}_X B$
since $dB =0$. We thus see that the generalized complex geometry
provides a natural geometric framework to incorporate simultaneously
the two local symmetries in Eq.\eq{diff-symm} and
Eq.\eq{lambda-symm}. That is,
\be \la{courant-auto}
{\rm Courant \; automorphism =  Diff(M) \oplus \Lambda-symmetry}.
\ee

One can introduce a generalized metric on $TM \oplus T^*M$ by
reducing the structure group $U(n,n)$ to $U(n) \times U(n)$. It
turns out \ct{gcg-gualtieri} that the metric on $TM
\oplus T^*M$ compatible with the natural pairing $\langle X + \xi, Y
+ \eta \rangle = \half \bigl(\xi(Y) + \eta(X) \bigr)$ is equivalent
to a choice of metric $g$ on $TM$ and 2-form $B$. \footnote{A
reduction to $U(n) \times U(n)$ is equivalent to the existence of
two generalized almost complex structures ${\cal J}_1, {\cal J}_2$
where ${\cal J}_1$ and ${\cal J}_2$ commute and a generalized
K\"ahler metric $G = -{\cal J}_1 {\cal J}_2$ is positive definite.
This structure is known as the generalized K\"ahler or bi-Hermitian
structure \ct{gcg-gualtieri}. Any generalized K\"ahler metric $G$
takes the form
$$G = \left(
         \begin{array}{cc}
           -g^{-1}B & g^{-1} \\
           g - B g^{-1}B  & B g^{-1}  \\
         \end{array}
       \right)
= \left(
    \begin{array}{cc}
      1 & 0 \\
      B & 1 \\
    \end{array}
  \right)
\left(
  \begin{array}{cc}
    0 & g^{-1} \\
    g & 0 \\
  \end{array}
\right)
\left(
  \begin{array}{cc}
    1 & 0 \\
    -B & 1 \\
  \end{array}
\right),$$
which is the $B$-field transformation of a bare Riemannian metric
$g$ as long as the 2-form $B$ is closed. Interestingly the metric
part $g - B g^{-1}B : TM \to T^* M$ in the generalized K\"ahler
metric $G$ is exactly of the same form as the open string metric in
a $B$-field \ct{sw}.} We now introduce a DBI ``metric" $g +
\kappa B : TM \to T^* M$ which maps $X$ to $\xi = (g + \kappa
B)(X)$. Consider the Courant automorphism
\eq{courant-auto} which is a combination of a $B$-field
transformation followed by a diffeomorphism $\phi: M \to M$
\be \la{courant-tr}
X + \xi \to  \phi^{-1}_* X +  \phi^* (\xi + \iota_X B).
\ee
The above action transforms the DBI metric $g + \kappa B$ according
to
\be \la{dbi-tr}
g + \kappa B \to  \phi^* \Bigl(g + \kappa (B + {\cal L}_X B) \Bigr).
\ee
The Moser lemma \eq{darboux-tr} then implies that there always
exists a diffeomorphism $\phi$ such that $\phi^* (B + {\cal L}_X B)
= B$. In terms of local coordinates $\phi : y
\to x = x(y)$, Eq.\eq{dbi-tr} then reads as
\be \la{dbi-iso}
(g + \kappa B^\prime)_{\alpha\beta} (x) = \frac{\p y^a}{\p x^\alpha}
\Bigl(g^\prime_{ab}(y)+  \kappa B_{ab}(y) \Bigr) \frac{\p y^b}{\p x^\beta}
\ee
where $B^\prime = B + {\cal L}_X B$ and
\be \la{gauge-metric}
g^\prime _{ab}(y) = \frac{\p x^\alpha}{\p y^a} \frac{\p x^\beta}{\p
y^b} g_{\alpha\beta}(x).
\ee
One can immediately see that the diffeomorphism \eq{dbi-iso} between
two different DBI metrics is a direct result of the Moser lemma
\eq{darboux-tr}. We will see that the identity \eq{dbi-iso} leads to
a remarkable relation between symplectic (or Poisson) geometry and
complex (or Riemannian) geometry.

\subsection{DBI action as a generalized geometry}

We observed that the presence of a nowhere vanishing (closed) 2-form
$B$ in spacetime $M$ calls for a generalized geometry, where the two
local symmetries in Eq.\eq{courant-auto} are treated on equal
footing. A crucial point in the generalized geometry is that the
space $\Lambda^2(M)$ of closed 2-forms in $M$ appears as a part of
spacetime geometry, as embodied in Eq.\eq{dbi-iso}, in addition to
the Diff(M) symmetry being a local isometry of Riemannian geometry.
This suggests that, when $B \neq 0$, it is possible to realize a
completely new geometrization of a physical force based on
symplectic geometry rather than Riemannian geometry. So a natural
question is: What is the force ?

We will show that the force is indeed the electromagnetic force and
there exists a novel form of the equivalence principle, i.e., the
Darboux theorem, for the geometrization of the electromagnetism. In
other words, Eq.\eq{darboux-tr} implies that there always exists a
coordinate transformation to locally eliminate the electromagnetic
force as long as the D-brane worldvolume $M$ supports a symplectic
structure $B$, i.e., $M$ becomes a NC space. Furthermore, $U(1)$
gauge transformations in NC spacetime become a `spacetime' symmetry
rather than an `internal' symmetry, which already suggests that the
electromagnetism in NC spacetime can be realized as a geometrical
property of spacetime like gravity.

Let us now discuss the physical consequences of the generalized
geometry, especially, the implications of the $\Lambda$-symmetry
\eq{lambda-symm} in the context of the low energy effective theory
of open strings in the background of an NS-NS 2-form $B$. We will
use the effective field theory description in order to broadly
illuminate what kind of new physics arises from a field theory in
the B-field background, i.e., a NC field theory. It will provide a
clear-cut picture about the new physics though it is not quite
rigorous. In the next section we will put the arguments here on a
firm foundation using the background independent formulation of NC
gauge theory.

A low energy effective field theory deduced from the open string
action \eq{open-action} describes an open string dynamics on a
$(p+1)$-dimensional D-brane worldvolume. The dynamics of D-branes is
described by open string field theory whose low energy effective
action is obtained by integrating out all the massive modes, keeping
only massless fields which are slowly varying at the string scale
$\kappa \equiv 2 \pi
\alpha^\prime$. For a $Dp$-brane in closed string background fields,
the action describing the resulting low energy dynamics is given by
\begin{equation} \label{dbi-general}
S = \frac{2\pi}{g_s (2\pi \kappa)^{\frac{p+1}{2}}}
\int d^{p+1} x \sqrt{\det(g + \kappa (B + F))}
+ {\cal O}(\sqrt{\kappa} \partial F, \cdots),
\end{equation}
where $F=dA$ is the field strength of $U(1)$ gauge fields. The DBI
action \eq{dbi-general} respects the two local symmetries,
\eq{diff-symm} and \eq{lambda-symm}, as expected.

(I) Diff(M)-symmetry: Under a local coordinate transformation
$\phi^{-1}: x^\alpha \mapsto {x^\prime}^\alpha$ where worldvolume
fields also transform in usual way
\be \la{bf-transform}
(B^\prime + F^\prime)_{ab} (x^\prime) =
 \frac{\p x^\alpha}{\p {x^\prime}^a} \frac{\p x^\beta}{\p {x^\prime}^b}
 (B + F)_{\alpha\beta}(x)
\ee
together with the metric transformation \eq{metric-tr}, the action
\eq{dbi-general} is invariant.

(II) $\Lambda$-symmetry: One can easily see that the action
\eq{dbi-general} is invariant under the transformation
\eq{lambda-symm} with any 1-form $\Lambda$.

Note that ordinary $U(1)$ gauge symmetry is a special case of
Eq.\eq{lambda-symm} where the gauge parameter $\Lambda$ is exact,
namely, $\Lambda = d \lambda$, so that $B \to B, \; A \to A +
d\lambda$. Indeed the $U(1)$ gauge symmetry is a diffeomorphism
(known as a symplectomorphism) generated by a vector field $X$
satisfying ${\cal L}_X B = 0$. We see here that the gauge symmetry
becomes a `spacetime' symmetry rather than an `internal' symmetry,
as well as an infinite-dimensional and non-Abelian symmetry when $B$
is nowhere vanishing. This fact unveils a connection between NC
gauge fields and spacetime geometry.

The geometrical data of D-branes, that is a derived category in
mathematics, are specified by the triple $(M, g, B)$ where $M$ is a
smooth manifold equipped with a Riemannian metric $g$ and a
symplectic structure $B$. One can see from the action
\eq{dbi-general} that the data come only into the combination $(M, g, B)
= (M, g + \kappa B)$, which is the DBI metric \eq{dbi-tr} to embody
a generalized geometry. In fact the `D-manifold' defined by the
triple $(M, g, B)$ describes the generalized geometry
\ct{gcg-hitchin,gcg-gualtieri} which continuously interpolates
between a symplectic geometry $(|\kappa Bg^{-1}| \gg 1)$ and a
Riemannian geometry $(|\kappa Bg^{-1}| \ll 1)$. An important point
is that the electromagnetic force $F$ should appear in the gauge
invariant combination $\Omega = B + F$ due to the $\Lambda$-symmetry
\eq{lambda-symm}, as shown in Eq.\eq{dbi-general}. Then the
Darboux theorem \eq{darboux-tr} with the identification
$\omega^\prime = \Omega$ and $\omega = B$ states that one can
``always" eliminate the electromagnetic force $F$ by a suitable
local coordinate transformation as far as the 2-form $B$ is
nondegenerate. Therefore the Darboux theorem  in symplectic goemetry
bears an analogy with the equivalence principle in Section 2.1.

Let us represent the local coordinate transform $\phi: y \mapsto x =
x(y)$ in Eq.\eq{darboux-tr} as follows
\be \la{cov-coordinate}
x^a(y) \equiv y^a + \theta^{ab} \widehat{A}_b(y),
\ee
where $\theta^{ab}$ is a Poisson structure on $M$, i.e.,
$\theta^{ab}= \bigl(\frac{1}{B} \bigr)^{ab}$.
\footnote{\la{poisson}A Poisson structure is a skew-symmetric, contravariant 2-tensor
$\theta = \theta^{ab} \p_a \wedge \p_b \in \bigwedge^2 TM$ which
defines a skew-symmetric bilinear map $\{f,g\}_{\theta} = \langle
\theta, df \otimes dg \rangle = \theta^{ab} \p_a f \p_b g$ for $f, g \in C^\infty(M)$,
so-called, a Poisson bracket. So we get $\theta^{ab}(y) =
\{y^a,y^b\}_{\theta}$.} This particular form of expression has been motivated
by the fact that $\omega^\prime_{ab}(x) =
\omega_{ab}(y)$ in the case of $F = dA =0$, so the second term in
Eq.\eq{cov-coordinate} should take care of the deformation of the
symplectic structure coming from $F=dA$. As was shown above, $U(1)$
gauge transformations are generated by a Hamiltonian vector field
$X_\lambda$ satisfying $\iota_{X_\lambda}B + d \lambda =0$ and the
action of $X_\lambda$ on $x^a(y)$ is given by
\bea \la{semi-gauge-tr}
\delta x^a(y) &\equiv& X_\lambda (x^a) = \{ x^a, \lambda \}_{\theta} \xx
&=& \theta^{ab} \bigl( \p_b \lambda + \{\widehat{A}_b, \lambda
\}_{\theta} \bigr),
\eea
where the last expression presumes a constant $\theta^{ab}$. The
above transformation will be identified with the NC $U(1)$ gauge
transformation after a NC deformation, so $\widehat{A}_a(y)$ turns
out to be NC gauge fields. The coordinates $x^a(y)$ in
\eq{cov-coordinate} will play a special role, since they are
background independent \ct{seiberg} as well as gauge covariant
\ct{madore}.

We showed before that the local equivalence \eq{darboux-tr} between
symplectic structures brings in the diffeomorphic equivalence
\eq{dbi-iso} between two different DBI metrics, which in turn
leads to a remarkable identity between DBI actions \ct{cornalba}:
\begin{equation} \label{mirror}
\int d^{p+1} x \sqrt{\det \bigl(g(x) + \kappa (B + F)(x)\bigr)}
= \int d^{p+1} y \sqrt{\det\bigl(h(y) + \kappa B(y)\bigr)}.
\end{equation}
Note that gauge field fluctuations now appear as an induced metric
on the brane given by
\begin{equation} \label{induced-metric}
h_{ab}(y) =  \frac{\partial x^\alpha}{\partial y^a}
\frac{\partial x^\beta}{\partial y^b} g_{\alpha\beta}(x).
\end{equation}
The identity \eq{mirror} can also be obtained by considering the
coordinate transformations \eq{metric-tr} and \eq{bf-transform}
satisfying $(B^\prime + F^\prime)_{ab} (x^\prime) =
B_{ab}(x^\prime)$. This kind of coordinate transformation always
exists thanks to the Darboux theorem \eq{darboux-tr}. Note that all
these underlying structures are very parallel to general relativity
(see Section 2.1). For instance, considering the fact that a
diffeomorphism $\phi
\in$ Diff(M) acts on $E$ as $X + \xi \mapsto \phi^{-1}_* X + \phi^*
\xi$, we see that the covariant coordinates $x^a(y)$ in
Eq.\eq{cov-coordinate} correspond to the locally inertial
coordinates $\xi^\alpha(x)$ in Eq.\eq{eq-free-spacetime} while the
coordinates $y^a$ play the same role as the laboratory Cartesian
coordinates $x^\mu$ in Eq.\eq{eom-lab}.

We will now discuss important physical consequences we can
get from the identity \eq{mirror}.

(1) The identity \eq{mirror} says that gauge field fluctuations on a
rigid D-brane are equivalent to dynamical fluctuations of the
D-brane itself without gauge fields. Indeed this picture is
omnipresent in string theory with the name of open-closed string
duality although it is not formulated in this way.

(2) The identity \eq{mirror} cannot be true when $B=0$, i.e.,
spacetime is commutative. In this case the $\Lambda$-symmetry is
reduced to ordinary $U(1)$ gauge symmetry. The gauge symmetry has no
relation to a diffeomorphism symmetry and it is just an internal
symmetry rather than a spacetime symmetry.

(3) Let us consider a curved D-brane in a constant B-field
background whose shape is described by an induced metric $h_{ab}$.
We may consider the right-hand side of Eq.\eq{mirror} with a
constant $B_{\rm{const}}$ as the corresponding DBI action. The
induced metric $h_{ab}$ can be represented as in
Eq.\eq{induced-metric} with a flat metric $g_{\alpha\beta}(x) =
\delta_{\alpha\beta}$. The nontrivial shape of the curved D-brane
described by the metric $h_{ab}$ can then be translated in the
left-hand side of Eq.\eq{mirror} into a nontrivial condensate of
gauge fields on a flat D-brane given by
\be \la{inhomo-b}
B_{ab}(x) = \bigl(B_{\rm{const}} + F_{\rm{back}}(x)\bigr)_{ab}.
\ee
The converse is also suggestive. Any symplectic 2-form on a
noncompact space can be written as the form \eq{inhomo-b} where
$B_{\rm{const}}$ is an asymptotic value of the 2-form $B_{ab}(x)$,
i.e., $F_{\rm{back}}(x)
\to 0$ at $|x| \to \infty$. And the gauge field configuration
$F_{\rm{back}}(x)$ can be interpreted as a curved D-brane manifold
in the $B_{\rm{const}}$ background. Thus we get an intriguing result
that a curved D-brane with a canonical symplectic 2-form (or a
constant Poisson structure) is equivalently represented as a flat
D-brane with an inhomogeneous symplectic 2-form (or a nonconstant
Poisson structure). Our argument here also implies a fascinating
result that $B_{\rm{const}}$, a uniform condensation of gauge fields
in a vacuum, would be a `source' of flat spacetime. Later we will
return to this point with an elaborated viewpoint.

(4) One can expand the right-hand side of Eq.\eq{mirror} around the background
$B$, arriving at the following result \ct{cornalba}
\bea \label{semi-expansion}
&& \int d^{p+1} y \sqrt{\det\bigl(h(y) + \kappa B(y)\bigr)} \xx
&& \qquad =
\int d^{p+1} y \sqrt{\det\bigl(\kappa B \bigr)}
\Bigl(1 + \frac{1}{4 \kappa^2}  g_{ac}g_{bd}
\{x^a, x^b\}_{\theta} \{x^c, x^d\}_{\theta} + \cdots \Bigr)
\eea
where $\{x^a, x^b\}_{\theta}$ is a Poisson bracket (defined in
footnote \ref{poisson}) between the covariant coordinates
\eq{cov-coordinate}. For constant $B$ and $g$,
Eq.\eq{semi-expansion} is equivalent to the IKKT matrix model
\ct{ikkt} after a quantization {\it \`a la} Dirac, i.e.,
$\{x^a, x^b\}_{\theta} \Rightarrow -i [\widehat{x}^a,
\widehat{x}^b]_\star$, which is believed to describe the
nonperturbative dynamics of the type IIB string theory. Furthermore
one can show that Eq.\eq{semi-expansion} reduces to a NC gauge
theory, using the relation
\be \la{poisson-bracket}
[\widehat{x}^a, \widehat{x}^b]_\star = - i \Bigl( \theta(\widehat{F}
- B)\theta
\Bigr)^{ab}
\ee
where the NC field strength is given by
\be \la{semi-f}
\widehat{F}_{ab} = \p_a \widehat{A}_b - \p_b \widehat{A}_a - i
[\widehat{A}_a, \widehat{A}_b ]_\star.
\ee
Therefore the identity \eq{mirror} is, in fact, the Seiberg-Witten
equivalence between commutative and NC DBI actions \ct{sw}.

(5) It was explicitly demonstrated in \ct{hsy1,hsy2} how NC gauge
fields manifest themselves as a spacetime geometry, as
Eq.\eq{semi-expansion} glimpses this geometrization of the
electromagnetic force. Surprisingly it turns out \ct{hsy1} that
self-dual electromagnetism in NC spacetime is equivalent to
self-dual Einstein gravity. (We rigorously show this equivalence in
Appendix A.) For example, $U(1)$ instantons in NC spacetime are
actually gravitational instantons
\ct{sty-ys}. This picture also reveals a beautiful geometrical
structure that self-dual NC electromagnetism perfectly fits with the
twistor space describing curved self-dual spacetime. The deformation
of symplectic (or K\"ahler) structure of a self-dual spacetime due
to the fluctuation of gauge fields appears as that of complex
structure of the twistor space.

(6) All these properties appearing in the geometrization of
electromagnetism may be summarized in the context of derived
category. More closely, if $M$ is a complex manifold whose complex
structure is given by $J$, we see that dynamical fields in the
left-hand side of Eq.\eq{mirror} act only as the deformation of
symplectic structure $\Omega (x) = B + F(x)$ in the triple $(M, J,
\Omega)$, while those in the right-hand side of Eq.\eq{mirror} appear only as
the deformation of complex structure $J^\prime(y)$ in the triple
$(M^\prime, J^\prime, B)$ through the metric \eq{induced-metric}. In
this notation, the identity \eq{mirror} can thus be written as
follows
\be \la{h-mirror}
(M, J, \Omega) \cong  (M^\prime, J^\prime, B).
\ee
The equivalence \eq{h-mirror} is very reminiscent of the homological
mirror symmetry \ct{kont-mirror}, stating the equivalence between
the category of A-branes (derived Fukaya category corresponding to
the triple $(M, J, \Omega)$) and the category of B-branes (derived
category of coherent sheaves corresponding to the triple $(M^\prime,
J^\prime, B)$).

There is a subtle but important difference between the Riemannian
geometry and symplectic geometry. Strictly speaking, the equivalence
principle in general relativity is a point-wise statement at any
given point $P$ while the Darboux theorem in symplectic geometry is
defined in an entire neighborhood around $P$. This is the reason why
there exist local invariants, e.g., curvature tensors, in Riemannian
geometry while there is no such kind of local invariant in
symplectic geometry.\footnote{If the equivalence principle held over
an entire neighborhood of a point $P$, curvature tensors would
identically vanish. Indeed the existence of local invariants such as
Riemann curvature tensors results from the implicit assumption that
it is always possible to discriminate total gravitational fields
between two arbitrary nearby spacetime points (see Sec. 2.1). This
exhibits a sign that there will be a serious conflict between the
equivalence principle and the Heisenberg's uncertainty principle. In
this perspective, it seems like a vain attempt to mix with water and
oil to try to quantize Einstein gravity itself, which is based on
Riemann curvature tensors of which the equivalence principle is in
the heart.} This raises a keen puzzle about how Riemannian geometry
is emergent from symplectic geometry though their local geometries
are in sharp contrast to each other.

We suggest a following resolution. A symplectic structure $B$ is
nowhere vanishing. In terms of physicist language, this means that
there is an (inhomogeneous in general) condensation of gauge fields
in a vacuum, i.e.,
\be \la{vacuum-condense}
\langle B_{ab}(x) \rangle_{{\rm vac}} = \theta^{-1}_{ab}(x).
\ee
Let us consider a constant symplectic structure for simplicity. The
background \eq{vacuum-condense} then corresponds to a uniform
condensation of gauge fields in a vacuum given by $\langle A_a^0
\rangle_{{\rm vac}} = - B_{ab} y^b$. It will be suggestive to
rewrite the covariant coordinates
\eq{cov-coordinate} as (actually to invoke a renowned Goldstone
boson $\varphi = \langle \varphi \rangle + h$)\footnote{\la{zee}In
this respect, it would be interesting to quote a recent comment of
A. Zee \ct{zee}: ``The basic equation for the graviton field has the
same form $g_{\mu\nu} = \eta_{\mu\nu} + h_{\mu\nu}$. This naturally
suggests that $\eta_{\mu\nu} = \langle g_{\mu\nu} \rangle$ and
perhaps some sort of spontaneous symmetry breaking." We will show
later that this pattern is not an accidental happening.}
\be \la{vacuum-coordinate}
x^a(y) = \theta^{ab} \Bigl( - \langle A_b^0 \rangle_{{\rm vac}} +
\widehat{A}_b(y) \Bigr).
\ee
This naturally suggests some sort of spontaneous symmetry breaking
where $y^a$ are vacuum expectation values of $x^a(y)$, specifying
the background \eq{vacuum-condense} as usual, and $\widehat{A}_b(y)$
are fluctuating (dynamical) coordinates (fields).

Note that the vacuum \eq{vacuum-condense} picks up a particular
symplectic structure, introducing a typical length scale $||\theta||
= l^2_{nc}$. This means that the $\Lambda$-symmetry $G$ in
Eq.\eq{lambda-symm} is spontaneously broken to the symplectomorphism
$H$ preserving the vacuum \eq{vacuum-condense} \ct{hsy2}. The
$\Lambda$-symmetry is the local equivalence between two symplectic
structures belonging to the same cohomology class. But the
transformation in Eq.\eq{lambda-symm} will not preserve the vacuum
\eq{vacuum-condense} except its subgroup generated by the gauge
parameter $\Lambda = d \lambda$ which is equal to the NC $U(1)$
gauge symmetry \eq{semi-gauge-tr}.\footnote{\label{b-shift}We will show later that a
constant shift of the symplectic structure, $B \to B^\prime = B +
\delta B$, does not affect any physics, so a symmetry of the theory,
although it readjusts the vacuum \eq{vacuum-condense}.} So the
deformations of the vacuum manifold \eq{vacuum-condense} by NC gauge
fields take values in the coset space $G/H$, which is equivalent to
the gauge orbit space of NC gauge fields or the physical
configuration space of NC electromagnetism
\ct{hsy2}. The spontaneous symmetry breaking also explains why only
ordinary $U(1)$ gauge symmetry is observed at large scales $\gg
l_{nc}$. We argued in \ct{hsy2} that the spontaneous symmetry
breaking \eq{vacuum-condense} will explain why Einstein gravity,
carrying local curvature invariants, can emerge from symplectic
geometry.\footnote{Here we are not saying that symplectic geometry
is missing an important ingredient. Instead our physics simply
requires to distinguish the background (nondynamical) and
fluctuating (dynamical) parts of a symplectic structure. This will
be a typical feature appearing in a background independent theory.}
In other words, Riemannian geometry would simply be a result of
coarse-graining of symplectic geometry at the scales $\gtrsim
l_{nc}$ like as the Einstein gravity in string theory where the
former simply corresponds to the limit $\alpha' \to 0$.

\section{Emergent Gravity}

Sometimes a naive reasoning also suggests a road in mist. What is
quantum gravity ? Quantum gravity means to quantize gravity.
Gravity, according to Einstein's general relativity, is the dynamics
of spacetime geometry which is usually described by a Hausdorff
space $M$ while quantization {\it \`a la} Dirac will require a phase
space structure of spacetime as a prequantization. The phase space
structure of spacetime $M$ can be specified by introducing a
symplectic structure $\omega$ on $M$. Therefore our naive reasoning
implies that the pair $(M,\omega)$, a symplectic manifold, might be
a proper starting point for quantum gravity, where fluctuations of
spacetime geometry would be fluctuations of the symplectic structure
$\omega$ and the quantization of symplectic manifold $(M,\omega)$
could be performed via the deformation quantization {\it \`a la}
Kontsevich \ct{kontsevich}.\footnote{This quantization scheme is
different from the usual canonical quantization of gravity where
metrics $g$ and their conjugates $\pi_g$ constitute fundamental
variables for quantization, i.e., a phase space $(g,
\pi_g)$. We believe that the conventional quantization scheme is much
like an escapade to quantize an elasticity of solid (e.g., sound
waves) or hydrodynamics and it is supposed to be failed due to the
choice of wrong variables for quantization, since it turns out that
Riemannian metrics are not fundamental variables but collective (or
composite) variables.} This state of art is precisely the situation
we have encountered in the previous section for the generalized
geometry emerging from the string theory
\eq{open-action} when $B \neq 0$.

A symplectic structure $B = \half B_{ab} dy^a \wedge dy^b$ defines a
Poisson structure $\theta^{ab} \equiv (B^{-1})^{ab}$ on $M$ (see
footnote \ref{poisson}) where $a, b=1, \dots, 2n$. From now on, we
will refer to a constant symplectic structure unless otherwise
specified. The Dirac quantization with respect to the Poisson
structure $\theta^{ab}$ then leads to a quantum phase space
\eq{nc-spacetime}. And the argument in Section 2.3 also explains
why a condensation of gauge fields in a vacuum,
Eq.\eq{vacuum-condense}, gives rise to the NC spacetime
\eq{nc-spacetime}, i.e.,
\begin{equation} \label{vacuum-spacetime}
\langle B_{ab} \rangle_{\rm vac} = (\theta^{-1})_{ab} \;\; \Leftrightarrow
\;\; [y^a, y^b]_\star = i \theta^{ab} \;\; \Leftrightarrow \;\; [a_i, a_j^\dagger] =
\delta_{ij},
\end{equation}
where $a_i$ and $a_j^\dagger$ with $i,j = 1, \cdots, n$ are
annihilation and creation operators, respectively, in the Heisenberg
algebra of an $n$-dimensional harmonic oscillator.

It is a well-known fact from quantum mechanics that the
representation space of NC ${\bf R}^{2n}$ is given by an
infinite-dimensional, separable Hilbert space
\be \la{fock}
\CH = \{ |\vec{n} \rangle \equiv |n_1,\cdots,n_n \rangle,
\; n_i=0,1, \cdots \}
\ee
which is orthonormal, i.e., $\langle \vec{n}| \vec{m} \rangle
= \delta_{\vec{n}\vec{m}}$ and complete, i.e.,  $
\sum_{\vec{n} = 0}^\infty |\vec{n} \rangle \langle \vec{n}| = 1$.
Note that every NC space can be represented as a theory of operators
in the Hilbert space ${\cal H}$, which consists of NC
$\star$-algebra ${\cal A}_\theta$ like as a set of observables in
quantum mechanics. Therefore any field $\widehat{\Phi} \in
\CA_\theta $ in the NC space
\eq{vacuum-spacetime} becomes an operator acting on ${\cal H}$ and
can be expanded in terms of the complete operator basis
\begin{equation}\label{matrix-basis}
\CA_\theta = \{ |\vec{n} \rangle \langle \vec{m}|, \; n_i, m_j = 0,1, \cdots \},
\end{equation}
that is,
\begin{equation}\label{op-matrix}
    \widehat{\Phi}(y) = \sum_{\vec{n},\vec{m}} \Phi_{\vec{n}\vec{m}}
    |\vec{n} \rangle \langle \vec{m}|.
\end{equation}
One may use the `Cantor diagonal method' to put the $n$-dimensional
positive integer lattice in $\CH$ into a one-to-one correspondence
with the infinite set of natural numbers (i.e., $1$-dimensional
positive integer lattice): $|\vec{n} \rangle
\leftrightarrow |n \rangle, \; n=1,\cdots,N
\to \infty $. In this one-dimensional basis, Eq.\eq{op-matrix} can
be relabeled as the following form
\be \la{matrix-expansion}
\widehat{\Phi}(y) = \sum_{n,m=1}^\infty \Phi_{nm} \;|n \rangle \langle m |.
\ee
One can regard $\Phi_{nm}$ in Eq.\eq{matrix-expansion} as components
of an $N \times N$ matrix $\Phi$ in the $N \to \infty$ limit. We
then get the following relation \ct{nc-review,szabo,hsy3}:
\be \la{sun-sdiff}
\mathrm{Any \; field \; on \; NC} \; {\bf R}^{2n} \; \cong \; N \times N \;
\mathrm{matrix} \; \mathrm{at} \; N \to \infty.
\ee
If $\widehat{\Phi}$ is a real field, then $\Phi$ should be a
Hermitian matrix. The relation \eq{sun-sdiff} means that a NC field
can be regarded as a master field of a large $N$ matrix.

We have to point out that our statements in the previous section
about emergent geometries should be understood in the
`semi-classical' limit where the Moyal-Weyl commutator,
$-i[\widehat{f},\widehat{g}]_\star$, can be reduced to the Poisson
bracket $\{f,g\}_{\theta}$. Now the very notion of a point in NC
spaces such as Eq.\eq{vacuum-spacetime} is doomed but replaced by a
state in $\CH$. So the usual concept of geometry based on smooth
manifolds would be replaced by a theory of operator algebra, e.g.,
NC geometry {\it \`a la} Connes \ct{connes-ncg}, or a theory of
deformation quantization {\it \`a la} Kontsevich \ct{kontsevich}.
Thus our next mission is how to lift our previous `semi-classical'
arguments to the full NC world. A nice observation to do this is
that a NC algebra ${\cal A}_\theta$ generated by the NC coordinates
\eq{nc-spacetime} is mathematically equivalent to the one generated
by the NC phase space \eq{nc-phase}.

In classical mechanics, the set of possible states of a system forms
a Poisson manifold and the observables that we want to measure are
smooth functions in $C^\infty(M)$, forming a commutative (Poisson)
algebra. In quantum mechanics, the set of possible states is a
Hilbert space $\CH$ and the observables are self-adjoint operators
acting on $\CH$, forming a NC $\star$-algebra. Pleasingly, there are
two paths to represent the NC algebra. One is the matrix mechanics
where the observables are represented by matrices in an arbitrary
basis in $\CH$. The other is the deformation quantization where,
instead of building a Hilbert space from a Poisson manifold and
associating an algebra of operators to it, the quantization is
understood as a deformation of the algebra of classical observables.
We are only concerned with the algebra to deform the commutative
product in $C^\infty(M)$ to a NC, associative product. Two
approaches have one to one correspondence through the Weyl-Moyal map
\ct{nc-review}.

Similarly, there are two different realizations of the NC algebra
${\cal A}_\theta$. One is the ``matrix representation" we already
introduced in Eq.\eq{sun-sdiff}. The other is to map the NC
$\star$-algebra ${\cal A}_\theta$ to a differential algebra using
the inner automorphism, a normal subgroup of the full automorphism
group, in ${\cal A}_\theta$. We call it ``geometric representation",
which will be used in Sec.3.2. The geometric representation is quite
similar to the dynamical evolution of a system in the Heisenberg
picture in which the time-evolution of dynamical variables is
defined by the inner automorphism of the NC $\star$-algebra
generated by the coordinates in Eq.\eq{nc-phase}. Of course, the two
representations of a NC field theory should describe an equivalent
physics. Now we will apply these two pictures to NC field theories
to see what the equivalence between them implies.

\subsection{Matrix representation}

First we apply the matrix representation \eq{sun-sdiff} to NC $U(1)$
gauge theory on $\mathbf{R}^D =
\mathbf{R}^d_C \times \mathbf{R}^{2n}_{NC}$ where the
$d$-dimensional commutative spacetime $\mathbf{R}^d_C$ will be taken
with either Lorentzian or Euclidean signature.\footnote{The
generalized Darboux theorem was proved in \ct{gcg-gualtieri},
stating that any $m$-dimensional generalized complex manifold, via a
diffeomorphism and a B-field transformation, looks locally like the
product of an open set in ${\bf C}^k$ with an open set in the
standard symplectic space $({\bf R}^{2m-2k}, \sum dq^i \wedge
dp_i)$. The integer $k$ is called the type of the generalized
complex structure, which is not necessarily constant but may rather
vary throughout the manifold -- the jumping phenomenon. The type can
jump up, but always by an even number. Here we will consider the
situation where the type $k$ is constant over the manifold.} We will
be brief since most technical details could be found in
\ct{hsy3}. We decompose $D$-dimensional coordinates $X^M \;
(M=1,\cdots, D)$ into $d$-dimensional commutative ones, denoted as
$z^\mu \; (\mu=1,
\cdots, d)$, and $2n$-dimensional NC ones, denoted as $y^a \; (a =
1, \cdots, 2n)$, satisfying the relation
\eq{vacuum-spacetime}. Likewise, $D$-dimensional gauge fields
$\widehat{A}_M(z,y)$ are also decomposed in a similar way
\bea \la{decomp-cov}
\widehat{D}_M &=& \p_M - i \widehat{A}_M(z,y) \equiv
(\widehat{D}_\mu, \widehat{D}_a)(z,y) \xx &=& (\widehat{D}_\mu,
-i\kappa B_{ab} \widehat{\Phi}^b)(z,y)
\eea
where $\widehat{D}_\mu = \p_\mu - i \widehat{A}_\mu(z,y)$ are
covariant derivatives along $\mathbf{R}^d_C$ and $\widehat{\Psi}_a
(z,y) \equiv \kappa B_{ab} \widehat{\Phi}^b (z,y) = B_{ab}
\widehat{x}^b(z,y)$ are adjoint Higgs fields of mass
dimension defined by the covariant coordinates \eq{cov-coordinate}.

Here, the matrix representation means that NC $U(1)$ gauge fields
$\widehat{\Xi}_M (z,y) \equiv (\widehat{A}_\mu,
\widehat{\Psi}_a)(z,y)$ are represented as $N
\times N$ matrices in the $N \to \infty$ limit as Eq.\eq{matrix-expansion}, i.e.,
\be \la{gauge-higgs-matrix}
\widehat{\Xi}_M (z,y) = \sum_{n,m=1}^\infty (\Xi_M)_{nm}(z) \;|n \rangle \langle m |.
\ee
Note that the $N \times N$ matrices $\Xi_M(z)= (A_\mu, \Psi_a)(z)$
in Eq.\eq{gauge-higgs-matrix} are now regarded as gauge and Higgs
fields in $U(N \to \infty)$ gauge theory on $d$-dimensional
commutative spacetime $\mathbf{R}^d_C$. One can then show that,
adopting the matrix representation
\eq{gauge-higgs-matrix}, the NC $U(1)$ gauge theory on ${\bf R}^d_C
\times {\bf R}^{2n}_{NC}$ is ``exactly" mapped to the $U(N  \to
\infty)$ Yang-Mills theory on $d$-dimensional spacetime ${\bf
R}^d_C$
\begin{eqnarray} \label{matrix-action}
S_B &=& -\frac{1}{4 g^2_{YM}} \int d^D X
(\widehat{F}_{MN} - B_{MN}) \star (\widehat{F}^{MN} - B^{MN}) \nonumber \\
&=& - \frac{(2\pi\kappa)^{\frac{4-d}{2}}}{2\pi g_s} \int d^d z {\rm
Tr} \left(\frac{1}{4} F_{\mu\nu} F^{\mu\nu}  +
\frac{1}{2} D_\mu \Phi^a D^\mu \Phi^a - \frac{1}{4} [\Phi^a, \Phi^b]^2
\right)
\end{eqnarray}
where the matrix $B_{MN}= \left(
                            \begin{array}{cc}
                              0 & 0 \\
                              0 & B_{ab} \\
                            \end{array}
                          \right)$
is the background symplectic 2-form \eq{vacuum-spacetime} of rank
$2n$. For notational simplicity, we have hidden all constant metrics
in Eq.\eq{matrix-action}. Otherwise, we refer
\ct{hsy3} for the general expression.

We showed before that $U(1)$ gauge symmetry in NC spaces is actually
a spacetime symmetry (diffeomorphisms generated by $X$ vector fields
satisfying ${\cal L}_X B=0$) where the NC $U(1)$ gauge
transformation acts on the covariant derivatives in \eq{decomp-cov}
as
\be \la{gauge-tr-cov-der}
\widehat{D}_M \to \widehat{D}_M^\prime = \widehat{U}(X) \star
\widehat{D}_M \star \widehat{U}(X)^{-1}
\ee
for any NC group element $\widehat{U}(X) \in U(1)$. Indeed the idea
that NC gauge symmetries are spacetime symmetries was discussed long
ago by many people. An exposition of these works can be found in
\ct{szabo}. The gauge transformation \eq{gauge-tr-cov-der} can be represented in the matrix
representation \eq{matrix-expansion}. The gauge symmetry now acts as
unitary transformations on the Fock space ${\cal H}$ which is
denoted as $U_{\rm cpt}({\cal H})$. This NC gauge symmetry $U_{\rm
cpt}({\cal H})$ is so large that $U_{\rm cpt}({\cal H})
\supset U(N)\;(N \to \infty)$ \ct{nc-group}. The NC $U(1)$ gauge
transformations in Eq.\eq{gauge-tr-cov-der} are now transformed into
$U(N)$ gauge transformations on ${\bf R}_C^d$ (where we complete
$U_{\rm cpt}({\cal H})$ with $U(N)$ in the limit $N \to \infty$)
given by
\be \la{factor-gauge-tr}
(D_\mu, \Psi_a) \to (D_\mu, \Psi_a)^\prime = U(z) (D_\mu, \Psi_a)
U(z)^{-1}
\ee
for any group element $U(z) \in U(N)$. Thus a NC gauge theory in the
matrix representation can be regarded as a large $N$ gauge theory.

As was explained above, the equivalence bewteen a NC $U(1)$ gauge
theory in higher dimensions and a large $N$ gauge theory in lower
dimensions is an exact map. What is the physical consequence of this
exact equivalence ?

Indeed one can get a series of matrix models from the NC $U(1)$
gauge theory \eq{matrix-action}. For instance, the IKKT matrix model
for $d=0$ \ct{ikkt}, the BFSS matrix model for $d=1$ \ct{bfss} and
the matrix string theory for $d=2$ \ct{matrix-string}. The most
interesting case is that the 10-dimensional NC $U(1)$ gauge theory
on ${\bf R}^{4}_C \times {\bf R}^{6}_{NC}$ is equivalent to the
bosonic part of 4-dimensional ${\cal N} =4$ supersymmetric $U(N)$
Yang-Mills theory, which is the large $N$ gauge theory of the
AdS/CFT duality \ct{ads-cft}. Note that all these matrix models or
large $N$ gauge theories are a nonperturbative formulation of string
or M theories. Therefore it should not be so surprising that a
$D$-dimensional gravity could be emergent from the $d$-dimensional
$U(N  \to \infty)$ gauge theory, according to the large $N$ duality
or AdS/CFT correspondence and thus from the $D$-dimensional NC gauge
theory in Eq.\eq{matrix-action}. We will show further evidences that
the action \eq{matrix-action} describes a theory of (quantum)
gravity.

A few remarks are in order.

(1) The equivalence \eq{matrix-action} raises a far-reaching
question about the renormalization property of NC field theory. If
we look at the first action in Eq.\eq{matrix-action}, the theory
superficially seems to be non-renormalizable for $D>4$ since the
coupling constant $g^2_{YM} \sim m^{4-D}$ has a negative mass
dimension. But this non-renormalizability appears as a fake if we
use the second action in Eq.\eq{matrix-action}. The resulting
coupling constant, denoted as $g^2_d \sim m^{4-d}$, in the matrix
action \eq{matrix-action} depends only on the dimension of the
commutative spacetime rather than the entire spacetime \ct{hsy3}.

The change of dimensionality is resulted from the relationship
\eq{sun-sdiff} where all dependence of NC coordinates appears as
matrix degrees of freedom. An important point is that the NC space
\eq{nc-spacetime} now becomes an $n$-dimensional positive integer
lattice (fibered $n$-torus $\mathbf{T}^n$, but whose explicit
dependence is mysteriously not appearing in the matrix action
\eq{matrix-action}). Thus the transition from commutative to NC
spaces accompanies the mysterious cardinality transition {\it \`a
la} Cantor from aleph-one (real numbers) to aleph-null (natural
numbers). Of course this transition is akin to that from classical
to quantum world in quantum mechanics. The transition from a
continuum space to a discrete space should be radical even affecting
the renormalization property \ct{harold-extra}.

Actually the matrix regularization of a continuum theory is an old
story, for instance, a relativistic membrane theory in light-front
coordinates (see, for example, a review \ct{m-review} and references
therein). The matrix regularization of the membrane theory on a
Riemann surface of any genus is based on the fact that the symmetry
group of area-preserving diffeomorphisms can be approximated by
$U(N)$. This fact in turn alludes that adjoint fields in $U(N)$
gauge theory should contain multiple branes with arbitrary
topologies. In this sense it is natural to think of the matrix
theory \eq{matrix-action} as a second quantized theory from the
point of view of the target space \ct{m-review}.

(2) From the above construction, we know that the number of adjoint
Higgs fields $\Phi^a$ is equal to the rank of the B-field
\eq{vacuum-spacetime}. Therefore the matrix theory
in Eq.\eq{matrix-action} can be defined in different dimensions by
changing the rank of the $B$-field. This change of dimensionality
appears in the matrix theory as the `matrix T-duality' (see Sec.
VI.A in \ct{m-review}) defined by\footnote{\label{odd}One can change
the dimensionality of the matrix model by any integer number by the
matrix T-duality \eq{matrix-t} while the rank of the $B$-field can
be changed only by an even number. Hence it is not obvious what kind
of background can explain the NC field theory with an odd number of
adjoint Higgs fields. A plausible guess is that there is a 3-form
$C_{\mu\nu\rho}$ which reduces to the 2-form $B$ in
Eq.\eq{vacuum-spacetime} by a circle compactification, so may be of
M-theory origin. Unfortunately, we don't know how to construct a
corresponding NC field theory with the 3-form background, although
very recent developments seem to go toward that direction.}
\be \la{matrix-t}
iD_\mu \rightleftarrows \Phi^a.
\ee
Applying the matrix T-duality \eq{matrix-t} to the action
\eq{matrix-action}, on one hand, one can arrive at the 0-dimensional
IKKT matrix model (in the case of Euclidean signature) or the
1-dimensional BFSS matrix model (in the case of Lorentzian
signature). On the other hand, one can also go up to $D$-dimensional
pure $U(N)$ Yang-Mills theory given by
\be \la{pure-ym}
S_C = -\frac{1}{4 g^2_{YM}} \int d^D X \Tr F_{MN} F^{MN}.
\ee
Note that the $B$-field is now completely disappeared, i.e., the
spacetime is commutative. In fact the T-duality between
Eq.\eq{matrix-action} and Eq.\eq{pure-ym} is an analogue of the
Morita equivalence on a NC torus stating that NC $U(1)$ gauge theory
with rational $\theta = M/N$ is equivalent to an ordinary $U(N)$
gauge theory \ct{sw}.

(3) One may notice that the second action in Eq.\eq{matrix-action}
can also be obtained by a dimensional reduction of the action
\eq{pure-ym} from $D$-dimensions to $d$-dimensions. However there is a
subtle but important difference between these two.

A usual boundary condition for NC gauge fields in
Eq.\eq{matrix-action} is that $\widehat{F}_{MN} \to 0$ at $|X| \to
\infty$. So the following maximally commuting matrices
\be \la{max-comm-matrix}
[\Phi^a, \Phi^b] = 0 \;\;\; \cong \;\;\; \Phi^a = {\rm
diag}(\phi_1^a, \cdots, \phi_N^a), \;\;\; \forall a
\ee
could not be a vacuum solution of Eq.\eq{matrix-action} (see
Eq.\eq{poisson-bracket}), while they could be for the Yang-Mills
theory dimensionally reduced from Eq.\eq{pure-ym}. The vacuum
solution of Eq.\eq{matrix-action} is rather
Eq.\eq{vacuum-spacetime}.

A proper interpretation for the contrast will be that the flat space
${\bf R}^{2n}$ in Eq.\eq{matrix-action} is not {\it a priori} given
but defined by (or emergent from) the background
\eq{vacuum-spacetime}. (We will show this fact later.)
But, in Eq.\eq{pure-ym}, a flat $D$-dimensional spacetime ${\bf
R}^D$ already exists, so it is no longer needed to specify a
background for the spacetime, contrary to Eq.\eq{matrix-action}. It
was shown by Witten
\ct{witten-d} that the low-energy theory describing a system of $N$
parallel D$p$-branes in flat spacetime is the dimensional reduction
of ${\cal N}=1$, (9+1)-dimensional super Yang-Mills theory to
$(p+1)$ dimensions. The vacuum solution describing a condensation of
$N$ parallel D$p$-branes in flat spacetime is then given by
Eq.\eq{max-comm-matrix}. So a natural inference is that the
condensation of $N$ parallel D$p$-branes in Eq.\eq{max-comm-matrix}
is described by a different class of vacua from the background
\eq{vacuum-spacetime}.

\subsection{Geometric representation}

Now we move onto the geometric representation of a NC field theory.
A crux is that translations in NC directions are an inner
automorphism of the NC $\star$-algebra ${\cal A}_\theta$ generated
by the coordinates in Eq.\eq{vacuum-spacetime},
\be \la{nc*-algebra}
e^{-i k^a B_{ab}y^b} \star \widehat{f}(z,y) \star e^{i k^a
B_{ab}y^b} = \widehat{f}(z, y + k)
\ee
for any $\widehat{f}(z,y)\in {\cal A}_\theta$. Its infinitesimal
form defines the inner derivation \eq{inner-der} of the algebra
${\cal A}_\theta$. It might be worthwhile to point out that the
inner automorphism \eq{nc*-algebra} is nontrivial only in the case
of a NC algebra. In other words, commutative algebras do not possess
any inner automorphism. In addition, Eq.\eq{nc*-algebra} clearly
shows that (finite) space translations are equal to (large) gauge
transformations.\footnote{\la{hbar-derivation}It may be interesting
to compare with a similar relation on a commutative space
\be \la{outer}
e^{l^\mu \p_\mu} f(z,y) e^{-l^\mu \p_\mu} = f(z + l, y).
\ee
A crucial difference is that translations in commutative space are
an outer automorphism since $e^{l^\mu \p_\mu}$ is not an element of
the underlying $\star$-algebra. So every points in commutative space
are distinguishable, i.e., unitarily inequivalent while every
``points" in NC space are indistinguishable, i.e., unitarily
equivalent. As a result, one loses the meaning of ``points" in NC
spacetime. This is a consequence of the fact that the set of prime
ideals defining the spectrum of the algebra $\CA_\theta$ is rather
small for $\theta \neq 0$ contrary to the commutative case. Note
that, after turning on $\hbar$, the relation \eq{outer} turns into
an inner automorphism of NC algebra generated by the NC phase space
\eq{nc-phase} since $e^{l^\mu \p_\mu}= e^{\frac{i}{\hbar} l^\mu
p_\mu}$ is now an algebra element. Another intriguing difference is
that the translation in \eq{outer} is parallel to the generator
$\p_\mu$ while the translation in \eq{nc*-algebra} is transverse to
the generator $y^a$ due to the antisymmetry of $B_{ab}$. It would be
interesting to contemplate this fact from the perspective in the
footnote \ref{rvs}.} It is a generic feature in NC spaces that an
internal symmetry of physics turns into a spacetime symmetry, as we
already observed in Eq.\eq{semi-gauge-tr}.

If electromagnetic fields are present in the NC space
\eq{vacuum-spacetime}, covariant objects, e.g., Eq.\eq{decomp-cov},
under the NC $U(1)$ gauge transformation should be introduced. As an
innocent generalization of the inner automorphism \eq{nc*-algebra},
let us consider the following ``dynamical" inner automorphism
\be \la{dynamical-inner}
e^{k^M \widehat{D}_M} \star \widehat{f}(X) \star
e^{-k^M\widehat{D}_M} =
\widehat{W}(X, C_k) \star \widehat{f}(X + k)\star \widehat{W}(X, C_k)^{-1}
\ee
where
\be \la{co-open-wilson}
e^{k^M \widehat{D}_M} \equiv \widehat{W}(X, C_k) \star e^{k^M \p_M}
\ee
with $\p_M \equiv (\p_\mu, -iB_{ab}y^b)$ and we used
Eqs.\eq{nc*-algebra} and \eq{outer} which can be summarized with a
compact form
\be \la{translation}
e^{k^M \p_M} \star \widehat{f}(X) \star e^{-k^M \p_M} =
\widehat{f}(X + k).
\ee
To understand Eq.\eq{dynamical-inner}, first notice that $e^{k^M
\widehat{D}_M}$ is a covariant object under NC $U(1)$ gauge
transformations according to Eq.\eq{gauge-tr-cov-der} and so one can
get
\bea \la{open-wilson-tr}
e^{k^M \widehat{D}_M} \to e^{k^M \widehat{D}_M^\prime} &=&
\widehat{U}(X) \star e^{k^M \widehat{D}_M} \star \widehat{U}(X)^{-1}
\xx
&=& \widehat{U}(X)  \star \widehat{W}(X, C_k) \star
\widehat{U}(X+k)^{-1} \star e^{k^M \p_M}
\eea
where Eq.\eq{translation} was used. Eq.\eq{open-wilson-tr} indicates
that $\widehat{W}(X, C_k)$ is an extended object whose extension is
proportional to the momentum $k^M$. Indeed $\widehat{W}(X, C_k)$ is
the open Wilson line, well-known in NC gauge theories, defined by
\be \la{open-wilson}
\widehat{W}(X, C_k) = P_\star \exp\Bigl(i \int_0^1 d
\sigma
\p_\sigma \xi^M(\sigma) \widehat{A}_M (X + \xi(\sigma)) \Bigr),
\ee
where $P_\star$ denotes path ordering with respect to the
$\star$-product along the path $C_k$ parameterized by
\begin{equation}\label{path}
\xi^M (\sigma) = k^M \sigma.
\end{equation}

The most interesting feature in NC gauge theories is that there do
not exist local gauge invariant observables in position space as
Eq.\eq{nc*-algebra} shows that the `locality' and the `gauge
invariance' cannot be compatible simultaneously in NC space. Instead
NC gauge theories allow a new type of gauge invariant observables
which are nonlocal in position space but localized in momentum
space. These are the open Wilson lines in Eq.\eq{open-wilson} and
their descendants with arbitrary local operators attached at their
endpoints. It turns out
\ct{gross-hi} that these nonlocal gauge invariant operators behave very much
like strings ! Indeed this behavior might be expected from the
outset since both theories carry their own non-locality scales set
by $\alpha^\prime$ (string theory) and $\theta$ (NC gauge theories)
which are equally of dimension of (length)$^2$, as advertised in the
Table 1.

The inner derivation \eq{inner-der} in the presence of gauge fields
is naturally covariantized by considering an infinitesimal version
of the dynamical inner automorphism
\eq{dynamical-inner}\footnote{\la{flat-index}From now on, for our later purpose,
we denote the indices carried by the covariant objects in
Eq.\eq{decomp-cov} with $A, B, \cdots$ to distinguish them from
those in the local coordinates $X^M$. The indices $A, B, \cdots$
will be raised and lowered using the flat Lorentzian metric
$\eta^{AB}$ and $\eta_{AB}$.}
\bea \la{co-inner}
ad_{\widehat{D}_A}[\widehat{f}] (X) &\equiv& [\widehat{D}_A,
\widehat{f}]_\star (X) = D_A^M(z,y)\frac{\p f(X)}{\p X^M} + \cdots
\xx &\equiv& D_A[f](X) + {\cal O}(\theta^3),
\eea
where $D_A^\mu = \delta^\mu_A$ since we define $[\p_\mu,
\widehat{f}(X)]_\star = \frac{\p f(X)}{\p z^\mu}$. It is easy to
check that the covariant inner derivation \eq{co-inner} satisfies
the Leibniz rule and the Jacobi identity, i.e.,
\bea \la{leibniz}
&& ad_{\widehat{D}_A}[\widehat{f} \star \widehat{g}] =
ad_{\widehat{D}_A}[\widehat{f}]
\star \widehat{g} + \widehat{f}
\star ad_{\widehat{D}_A}[\widehat{g}],
\\
\la{jacobi}
&& \bigl(ad_{\widehat{D}_A} \star ad_{\widehat{D}_B} -
ad_{\widehat{D}_B}
\star ad_{\widehat{D}_A}
\bigr) [\widehat{f}] = ad_{[\widehat{D}_A, \widehat{D}_B]_\star}[\widehat{f}].
\eea
In particular, one can derive from Eq.\eq{jacobi} the following
identities
\bea \la{map-ncem}
&& ad_{[\widehat{D}_A, \widehat{D}_B]_\star}[\widehat{f}] (X) =
-i [\widehat{F}_{AB}, \widehat{f}]_\star (X) = [D_A, D_B][f] (X) + \cdots \\
\la{fund-identity}
&& [ad_{\widehat{D}_A}, [ad_{\widehat{D}_B},
ad_{\widehat{D}_C}]_\star ]_\star [\widehat{f}](X) = - i
[\widehat{D}_A
\widehat{F}_{BC}, \widehat{f}]_\star (X) \equiv {\mathfrak{R}_{ABC}}^M (X) \p_M
f(X) + \cdots.
\eea
Note that the ellipses in the above equations correspond to higher
order derivative corrections generated by generalized vector fields
$\widehat{D}_A$.

We want to emphasize that the leading order of the map \eq{co-inner}
is nothing but the Poisson algebra. It is well-known \ct{mechanics}
that, for a given Poisson algebra $(C^\infty(M),
\{\cdot,\cdot\}_\theta)$, there exists a natural map $C^\infty (M) \to
TM: f \mapsto X_f$ between smooth functions in $C^\infty(M)$ and
vector fields in $TM$ such that
\be \la{duality}
X_f (g) = \{g, f \}_{\theta}
\ee
for any $g \in C^\infty(M)$. Indeed the assignment between a
Hamiltonian function $f$ and the corresponding Hamiltonian vector
field $X_f$ is the Lie algebra homomophism in the sense
\be \la{poisson-lie}
X_{\{f,g\}_{\theta}} = - [X_f,X_g]
\ee
where the right-hand side represents the Lie bracket between the
Hamiltonian vector fields. One can see that the Hamiltonian vector
fields on $M$ are the limit where the star-commutator
$-i[\widehat{D}_A, \widehat{f}]_\star$ is replaced by the Poisson
bracket $\{D_A, f \}_{\theta}$ or the Lie derivative ${\cal L}_{D_A}
(f)$.

The properties, \eq{leibniz} and \eq{jacobi}, show that the adjoint
action \eq{co-inner} can be identified with the derivations of the
NC algebra $\mathcal{A}_\theta$, which naturally generalizes the
notion of vector fields. In addition their dual space will
generalize that of 1-forms. Noting that the above NC differential
algebra recovers the ordinary differential algebra at the leading
order of NC deformations, it should be obvious that almost all
objects known from the ordinary differential geometry find their
counterparts in the NC case; e.g., a metric, connection, curvature
and Lie derivatives, and so forth. Actually, according to the Lie
algebra homomorphism \eq{poisson-lie}, $D_A(X)= D_A^M(X)
\frac{\p}{\p X^M}$ in the leading order of the map \eq{co-inner} can
be identified with ordinary vector fields in $TM$ where $M$ is any
D-dimensional (pseudo-)Riemannian manifold. More precisely, the
$D$-dimensional NC $U(1)$ gauge fields $\widehat{D}_M (X)
=(\widehat{D}_\mu,
\widehat{D}_a) (X)$ at the leading order appear as vector fields
(frames in tangent bundle) on a $D$-dimensional manifold $M$ given
by
\be \la{local-vector}
D_\mu(X) = \p_\mu + A_\mu^a(X) \frac{\p}{\p y^a}, \qquad D_a(X) =
D_a^b(X) \frac{\p}{\p y^b},
\ee
where
\be \la{com-vec}
A_\mu^a \equiv - \theta^{ab} \frac{\p \widehat{A}_\mu}{\p y^b},
\qquad  D_a^b \equiv \delta^b_a - \theta^{bc} \frac{\p \widehat{A}_a}{\p
y^c}.
\ee

Thus the map in Eq.\eq{co-inner} definitely leads to the vector
fields
\be \la{vec-d}
D_A (X) = (\p_\mu + A_\mu^a \p_a,  D_a^b \p_b)
\ee
or with matrix notation\footnote{We notice that this structure
shares a striking similarity with the Kaluza-Klein construction of
non-Abelian gauge fields from a higher dimensional Einstein gravity
\ct{cho-freund}. (Our matrix convention is swapping the row and column
in \ct{cho-freund}.) We will discuss in Sec. 5 a possible origin of
the similarity between the Kaluza-Klein theory and the emergent
gravity. A very similar Kaluza-Klein type origin of gravity from NC
gauge theory was also noticed in the earlier work \ct{langman-szabo}
where it was shown that a particular reduction of NC gauge theory
captures the qualitative manner in which NC gauge transformations
realize general covariance.}
\be \la{vec-matrix}
D_A^M(X) = \left(
  \begin{array}{cc}
    \delta^\nu_\mu & A_\mu^a  \\
    0 &  D_a^b \\
  \end{array}
\right).
\ee
One can easily check from Eq.\eq{com-vec} that $D_A$'s in
Eq.\eq{vec-d} take values in the Lie algebra of volume-preserving
vector fields, i.e., $\p_M D_A^M = 0$. One can also determine the
dual basis $D^A = D^A_M dX^M \in T^*M$ defined by
Eq.\eq{dual-vector} which is given by
\be \la{form-d}
D^A (X) = \bigl(dz^\mu,  V^a_b(dy^b- A_\mu^b dz^\mu)\bigr)
\ee
or with matrix notation
\be \la{form-matrix}
D_M^A (X) = \left(
  \begin{array}{cc}
    \delta^\nu_\mu & - V_b^a A_\mu^b  \\
    0 &  V_b^a \\
  \end{array}
\right)
\ee
where $V_a^c D_c^b = \delta_a^b$.

Through the dynamical inner automorphism \eq{dynamical-inner}, NC
$U(1)$ gauge fields $\widehat{A}_M(X)$ or $U(N \to \infty)$
gauge-Higgs system $(A_\mu, \Phi^a)$ in the action
\eq{matrix-action} are mapped to vector fields in $TM$ (to be
general ``a NC tangent bundle" $TM_\theta$) defined by
Eq.\eq{co-inner}. This is a remarkably transparent way to get a
$D$-dimensional gravity emergent from NC gauge fields or large $N$
gauge fields. We provide in Appendix A a rigorous proof of the
equivalence between self-dual NC electromagnetism and self-dual
Einstein gravity, originally first shown in \ct{hsy1}, to illuminate
how the map
\eq{co-inner} achieves the duality between NC gauge fields and
Riemannian geometry.

Now our next goal is obvious; the emergent gravity in general. Since
the equation of motion \eq{sde-ncem} for self-dual NC gauge fields
is mapped to the Einstein equation \eq{g-instanton} for self-dual
four-manifolds, one may anticipate that the equations of motion for
arbitrary NC gauge fields would be mapped to the vacuum Einstein
equations, in other words,
\be \la{maxwell-einstein}
\widehat{D}^A \widehat{F}_{AB} = 0 \quad \stackrel{?}{\Longleftrightarrow} \quad
E_{MN} \equiv R_{MN} - \half g_{MN} R = 0
\ee
together with the Bianchi identities
\be \la{bianchi}
\widehat{D}_{[A} \widehat{F}_{BC]} = 0 \quad \stackrel{?}{\Longleftrightarrow} \quad
R_{M[ABC]} = 0.
\ee
(We will often use the notation $\Gamma_{[ABC]} =  \Gamma_{ABC} +
\Gamma_{BCA} + \Gamma_{CAB}$ for the cyclic permutation of
indices.) After some thought one may find that the guess
\eq{maxwell-einstein} is not a sound reasoning since it should be
implausible if arbitrary NC gauge fields allow only Ricci flat
manifolds. Furthermore we know well that the NC $U(1)$ gauge theory
\eq{matrix-action} will recover the usual Maxwell theory in the commutative limit.
But if Eq.\eq{maxwell-einstein} is true, the Maxwell has been lost
in the limit. Therefore we conclude that the guess
\eq{maxwell-einstein} must be something wrong.

We need a more careful musing about the physical meaning of emergent
gravity. The emergent gravity proposes to take Einstein gravity as a
collective phenomenon of gauge fields living in NC spacetime much
like the superconductivity in condensed matter physics where it is
understood as a collective phenomenon of Cooper pairs (spin-0 bound
states of two electrons). It means that the origin of gravity is the
collective excitations of NC gauge fields at scales $\sim l^2_{nc} =
|\theta|$ which are described by a new order parameter, probably of
spin-2, and they should be responsible to gravity even at large
scales $\gg l_{nc}$, like as the classical physics emerges as a
coarse graining of quantum phenomena when $\hbar \ll 1$ (the
correspondence principle). Therefore the emergent gravity
presupposes a spontaneous symmetry breaking of some big symmetry
(see the Table 3) to trigger a spin-2 order parameter (graviton as a
Cooper pair of two gauge fields). If any, ``the correspondence
principle" for the emergent gravity will be that it should recover
the Maxwell theory (possibly with some other fields) coupling to the
Einstein gravity in commutative limit $|\theta| \to 0$ or at large
distance scales $\gg l_{nc}$.\footnote{This is not to say that the
electromagnetism is only relevant to the emergent gravity. The weak
and the strong forces should play a role in some way which we don't
know yet. But we guess that they will affect only a microscopic
structure of spacetime since they are short range forces. See
Section 4 for some related discussion.} Then the Maxwell theory will
appear in the right-hand side of the Einstein equation as an
energy-momentum tensor, i.e.,
\be \la{einstein-energy}
E_{MN} = \frac{8 \pi G_D}{c^4} T_{MN}
\ee
where $G_D$ is the gravitational Newton constant in $D$ dimensions.

Let us first discuss the consequence of the gravitational
correspondence principle postponing to Section 5 the question about
the existence of spin-2 bound states in NC spacetime. According to
the above scheme, we are regarding the NC $U(1)$ gauge theory in
Eq.\eq{matrix-action} as a theory of gravity. Hence the parameters,
$g_{YM}$ and $|\theta|$, defining the NC gauge theory should be
related to the gravitational Newton constant $G_D$ defining the
emergent gravity in $D$ dimensions. A dimensional analysis
(recovering $\hbar$ and $c$ too) simply shows that
\be \la{newton-constant}
\frac{G_D \hbar^2}{c^2} \sim g_{YM}^2 |{\rm Pf} \theta|^{\frac{1}{n}}
\ee
where $2n$ is the rank of $\theta^{ab}$. Suppose that $g_{YM}$ is
nonzero and always $c=1$ in Eq.\eq{newton-constant}. One can take a
limit $|\theta| \to 0$ and $\hbar \to 0$ simultaneously such that
$G_D$ is nonzero. In this limit we will get the classical Einstein
gravity coupling with the Maxwell theory which we are interested in.
Instead one may take a limit $|\theta|  \to 0$ and $G_D \to 0$
simultaneously, but $\hbar \neq 0$. This limit will correspond to
quantum electrodynamics. On the other hand, the classical Maxwell
theory will correspond to the limit, $\frac{G_D \hbar^2}{|{\rm Pf}
\theta|^{\frac{1}{n}}} \sim g_{YM}^2$ = constant, when $G_D \to 0,
\; \hbar \to 0$ and $|\theta| \to 0$.\footnote{As a completely different limit,
one may keep $|\theta|$ nonzero while $g_{YM} \to 0$. Note that this
limit does not necessarily mean that NC gauge theories are
non-interacting since, for an adjoint scalar field $\widehat{\phi}$
as an example, $\widehat{D}_a \widehat{\phi} =
\p_a \widehat{\phi} - i \frac{g_{YM}}{\hbar c} [\widehat{A}_a, \widehat{\phi}]_\star =
\p_a \widehat{\phi} + \frac{g_{YM}\theta^{bc}}{\hbar c} \frac{\p \widehat{A}_a}{\p y^b}
\frac{\p \widehat{\phi}}{\p y^c} + \cdots$, recovering
the original form of gauge coupling. $\frac{g_{YM}\theta^{bc}}{\hbar
c}$ can be nonzero depending on the limit under control. The
relation \eq{newton-constant} implies that there exist gravitational
($G_D \neq 0$) and non-gravitational ($G_D = 0$) theories for the
case at hand. Unfortunately we did not understand what they are.}

We will check the above speculation by showing that
Eq.\eq{einstein-energy} is correct equations of motion for emergent
gravity. Indeed we will find the Einstein gravity with the
energy-momentum tensor given by Maxwell fields and a ``Liouville"
field related to the volume factor in Eq.\eq{D-volume}. But we will
see that the guess
\eq{bianchi} is generally true. Note that self-dual gauge fields
have a vanishing energy-momentum tensor that is the reason why the
self-dual NC gauge fields simply satisfy the relation in
Eq.\eq{maxwell-einstein}.

We will use the notation in Appendix A with obvious minor changes
for a D-dimensional Lorentzian manifold. Define structure equations
of the vectors $D_A \in TM$ as
\be \la{structure-d}
[D_A, D_B] = - {\mathfrak{f}_{AB}}^C D_C
\ee
where ${\mathfrak{f}_{AB}}^\mu = 0, \; \forall A,B$ for the basis
\eq{vec-d}. From the experience of the self-dual case, we know
that the vector fields $D_A$ are related to the orthonormal frames
(vielbeins) $E_A$ by $D_A =
\lambda E_A$ where the conformal factor $\lambda$ will be determined
later. (This situation is reminiscent of the string frame ($D_A$)
and the Einstein frame ($E_A$) in string theory.) Hence the
D-dimensional metric is given by
\bea \la{D-metric}
ds^2 &=& \eta_{AB} E^A \otimes E^B \xx &=& \lambda^2
\eta_{AB} D^A \otimes D^B =
\lambda^2 \eta_{AB} D^A_M D^B_N
\; dX^M \otimes dX^N
\eea
where $E^A = \lambda D^A$. In particular, the dual basis \eq{form-d}
determines its explicit form up to a conformal factor as \ct{ward}
\be \la{ward-metric}
ds^2 = \lambda^2 \Bigl(\eta_{\mu\nu} dz^\mu dz^\nu +
\delta_{ab} V^a_c V^b_d (dy^c - \mathbf{A}^c)(dy^d - \mathbf{A}^d)
\Bigr)
\ee
where $\mathbf{A}^a = A_\mu^a dz^\mu$. The structure function
${\mathfrak{f}_{AB}}^C$ is also conformally mapped to
Eq.\eq{structure-fn} with
\be \la{2-structure}
{\mathfrak{f}_{AB}}^C = \lambda {f_{AB}}^C - D_A \log \lambda
\delta^C_B + D_B \log \lambda
\delta^C_A.
\ee

In the case of $D = 4$, Eq.\eq{map-ncem} immediately shows that the
leading order of self-dual NC gauge fields described by
Eq.\eq{sde-ncem} reduces to the following self-duality equation
\be \la{sde-structure}
{\mathfrak{f}_{AB}}^E = \pm \half {\varepsilon_{AB}}^{CD}
{\mathfrak{f}_{CD}}^E.
\ee
We proved in Appendix A that the metric \eq{ward-metric} satisfying
Eq.\eq{sde-structure} describes self-dual Einstein manifolds where
the conformal factor $\lambda^2$ is given by Eq.\eq{lambda}.

Now let us fix the conformal factor $\lambda^2$ in the metric
\eq{D-metric}. By an $SO(d-1,1) \times SO(2n)$ rotation of basis vectors $E_A$,
we can impose the condition that
\be \la{structure-f}
{f_{BA}}^B \equiv \phi_A = (3-D) E_A \log \lambda
\ee
and Eq.\eq{2-structure} in turn implies
\be \la{D-structure-f}
{\mathfrak{f}_{BA}}^B \equiv \rho_A = 2 D_A \log \lambda.
\ee
Note that ${\mathfrak{f}_{AB}}^\mu = 0, \; \forall A,B$ which is the
reason why one has to use only $SO(d-1,1) \times SO(2n)$ rotations
to achieve the condition \eq{structure-f} (see the footnote
\ref{so4-rotation} for a similar argument for self-dual gauge
fields). Eq.\eq{structure-f} means that the vector fields $E_A$ are
volume preserving with respect to a D-dimensional volume form
$\mathfrak{v} = \lambda^{(3-D)} \mathfrak{v}_g$ where
\be \la{D-volume-form}
\mathfrak{v}_g = E^1 \wedge \cdots \wedge E^D
\ee
and then the vector fields $D_A$ are volume preserving with respect
to the volume form $\mathfrak{v}_D = \lambda^{(2-D)}
\mathfrak{v}_g$. (See Eq.\eq{volume-preserving} for its proof.)
Therefore we get\footnote{\la{gradient-volume}One can directly check
Eq.\eq{D-structure-f} as follows. Acting ${\cal L}_{D_A}$ on both
sides of Eq.\eq{D-volume}, we get ${\cal L}_{D_A}
\Bigl( \mathfrak{v}_D (D_1, \cdots, D_D) \Bigr) = ({\cal L}_{D_A}
\mathfrak{v}_D) (D_1, \cdots, D_D) + \sum_{B=1}^D
\mathfrak{v}_D (D_1, \cdots, {\cal L}_{D_A} D_B, \cdots,  D_D)
= ({\cal L}_{D_A} \mathfrak{v}_D)(D_1, \cdots, D_D) +
\sum_{B=1}^D \mathfrak{v}_D (D_1, \cdots, [D_A, D_B], \cdots,  D_D)
= (\nabla \cdot D_A + {\mathfrak{f}_{BA}}^B ) \mathfrak{v}_D (D_1,
\cdots,  D_D) = (2 D_A \log \lambda) \mathfrak{v}_D (D_1, \cdots, D_D)$.
Since ${\cal L}_{D_A} \mathfrak{v}_D = (\nabla \cdot D_A)
\mathfrak{v}_D = 0$, Eq.\eq{D-structure-f} is deduced. Conversely,
if ${\mathfrak{f}_{BA}}^B  = 2 D_A \log \lambda$, $D_A$'s all
preserve the volume form $\mathfrak{v}_D$, i.e., ${\cal L}_{D_A}
\mathfrak{v}_D = (\nabla \cdot D_A) \mathfrak{v}_D = 0$.}
\be \la{D-volume}
\lambda^2 = \mathfrak{v}_D (D_1, \cdots, D_D).
\ee
Since $\p_M D_A^M = 0$, we know that the invariant volume is given
by $\mathfrak{v}_D = dz^1 \wedge \cdots \wedge dz^d \wedge dy^1
\wedge \cdots \wedge dy^{2n}$. Therefore we finally get
\be \la{D-lambda}
\lambda^2 = \det^{-1} V^a_b.
\ee

In terms of the structure functions one can get the map in
Eq.\eq{fund-identity}
\be \la{nc-map}
-i[\widehat{D}_A \widehat{F}_{BC}, \widehat{f}]_\star =
\big(D_A {\mathfrak{f}_{BC}}^D -
{\mathfrak{f}_{BC}}^E{\mathfrak{f}_{AE}}^D \big)D_D[f] + \cdots.
\ee
In other words, one can get the following maps for the equations of
motion and the Bianchi identities
\bea \la{map-eom}
&& \widehat{D}^A \widehat{F}_{AB} = 0 \quad \Longleftrightarrow
\quad \eta^{AB} \Bigl(D_A {\mathfrak{f}_{BC}}^D -
{\mathfrak{f}_{BC}}^E{\mathfrak{f}_{AE}}^D \Bigr) = 0, \\
\la{map-bianchi}
&& \widehat{D}_{[A} \widehat{F}_{BC]} = 0 \quad
\Longleftrightarrow} \quad D_{[A} {\mathfrak{f}_{BC]}}^D -
{\mathfrak{f}_{[BC}}^E {{\mathfrak{f}_{A]E}}^D = 0.
\eea

The spacetime geometry described by the metric \eq{D-metric} or
\eq{ward-metric} is an emergent gravity arising from NC gauge
fields whose underlying theory is defined by the action
\eq{matrix-action}. The fundamental variables in our approach are of course
gauge fields which should be subject to Eqs.\eq{map-eom} and
\eq{map-bianchi}. A spacetime metric is defined by NC (or non-Abelian) gauge fields
and regarded as a collective variable (a composite or bilinear of
gauge fields). Therefore our goal is to show that the equations of
motion \eq{map-eom} for NC gauge fields together with the Bianchi
identity \eq{map-bianchi} can be rewritten using the map
\eq{co-inner} as the Einstein equation for the metric
\eq{D-metric}. In other words, the Einstein equation
$E_{MN} = 8 \pi G_D T_{MN}$ is nothing but the equation of motion
for NC gauge fields represented from the (emergent) spacetime point
of view.

Our strategy is the following. First note that the Riemann curvature
tensors defined by Eq.\eq{D-riemann} have been expressed with the
orthonormal basis $E_A$. Since we will impose on them
Eqs.\eq{map-eom} and \eq{map-bianchi}, it will be useful to
represent them with the gauge theory basis $D_A$. As a consequence,
it will be shown that Einstein manifolds emerge from NC gauge fields
after imposing Eqs.\eq{map-eom} and \eq{map-bianchi}. All
calculations can straightforwardly be done using the relations
\eq{2-structure} and \eq{spin-conformal}. All the details show up in Appendix B.

The result is very surprising. The emergent gravity derived from NC
gauge fields predicts a new form of energy which we call the
``Liouville" energy-momentum tensor. Indeed this form of energy was
also noticed in \ct{madore-poisson} with a nonvanishing Ricci
scalar. The terminology is attributed to the following fact. The
vector fields $D_A$ are volume preserving with respect to the
symplectic volume $\mathfrak{v}_D$ (see the footnote
\ref{gradient-volume}). Thus $\mathfrak{v}_D$ is constant along
integral curves of $D_A$, in which case $D_A$ are called
incompressible with respect to $\mathfrak{v}_D$ and which is known
as the {\it Liouville theorem} in Hamiltonian mechanics
\ct{mechanics}. (See \ct{big-gravity} for the Liouville theorem in
curved spacetime.) Superficially this seems to imply that spacetime
behaves like an incompressible fluid so that spacetime volume does
not change along the flow generated by the vector field $D_A$. But
we have to be careful to interpret the geometrical meaning of the
Liouville theorem because the symplectic volume $\mathfrak{v}_D$ is
different from the Riemannian volume $\mathfrak{v}_g =
\lambda^{(D-2)} \mathfrak{v}_D$ in Eq.\eq{D-volume-form}. Furthermore, as we showed
in Appendix B, the vector field $D_A$ contributes to both sides of
the Einstein equation \eq{einstein-energy}. So the spacetime volume
given by $\mathfrak{v}_g$ can change along the flow described by the
vector field $D_A$ and its shape may also change in very complicated
ways. But this kind of a local expansion, distortion and twisting of
spacetime manifold will spend some energy, which should be supplied
from the right-hand side. This picture may be clarified by looking
at the so-called Raychaudhuri equation \ct{raycha,hawking-ellis}.

The Raychaudhuri equation is evolution equations of the expansion,
shear and rotation of flow lines along the flow generated by a
vector field in a background spacetime. Here we introduce an affine
parameter $\tau$ labeling points on the curves of the flow. Given a
timelike unit vector field $u^M$, i.e., $u^M u_M = -1$, the
Raychaudhuri equation in $D$ dimensions is given by
\be \la{raychaudhuri}
\dot{\Theta} -  {\dot{u}^M}_{;M} + \Sigma_{MN} \Sigma^{MN} -
\Omega_{MN} \Omega^{MN} + \frac{1}{D-1} \Theta^2 = - R_{MN} u^M u^N.
\ee
$\Theta ={u^M}_{;M}$ represents the expansion/contraction of volume
and $\dot{\Theta} = \frac{d \Theta}{d \tau}$ while $\dot{u}^M =
{u^M}_{;N} u^N$ represents the acceleration due to nongravitational
forces, e.g., the Lorentz force. $\Sigma_{MN}$ and $\Omega_{MN}$ are
the shear tensor and the vorticity tensor, respectively, which are
all orthogonal to $u^M$, i.e., $\Sigma_{MN} u^N = \Omega_{MN} u^N
=0$. The Einstein equation \eq{einstein-energy} can be rewritten as
\be \la{einstein-eq-energy}
R_{MN} = 8 \pi G_D \big( T_{MN} - \half g_{MN} {T_P}^P \big)
\ee
where $T_{MN} = E_M^A E_N^B T_{AB}$. In four dimensions, one can see
from Eq.\eq{einstein-eq-energy} that the right-hand side of
Eq.\eq{raychaudhuri} is given by
\be \la{ray-scalar}
 - R_{MN} u^M u^N = - \frac{1}{2\lambda^2} u^M u^N (\rho_M \rho_N  + \Psi_M \Psi_N)
- 8 \pi G_4 T^{(M)}_{MN} u^M u^N
\ee
where the Lorentzian energy-momentum tensor in
Eq.\eq{einstein-eq-energy} can be read off from
Eq.\eq{em-tensor-maxwell} and Eq.\eq{dark-energy} having in mind the
footnote \ref{wick-rotation}.

Suppose that all the terms except the expansion evolution
$\dot{\Theta}$ on the left-hand side of Eq.\eq{raychaudhuri} as well
as the Maxwell term $T^{(M)}_{MN}$ in Eq.\eq{ray-scalar} vanish or
become negligible. In this case the Raychaudhuri equation reduces to
\be \la{reduce-raychaudhuri}
\dot{\Theta} = - \frac{1}{2\lambda^2} u^M u^N
(\rho_M \rho_N + \Psi_M \Psi_N).
\ee
Note that the Ricci scalar is given by $R = \frac{1}{2\lambda^2}
g^{MN} (\rho_M \rho_N + \Psi_M \Psi_N)$. Therefore $R < 0$ when
$\rho_M$ and $\Psi_M$ are timelike while $R > 0$ when $\rho_M$ and
$\Psi_M$ are spacelike. Remember that our metric signature is
$(-+++)$. So, for the timelike perturbations, $\dot{\Theta} < 0$
which means that the volume of a three dimensional spacelike
hypersurface orthogonal to $u_M$ decreases. However, if spacelike
perturbations are dominant, the volume of the three dimensional
spacelike hypersurface can expand. For example, consider the most
symmetric perturbations as in Eq.\eq{so4-invariant}, i.e.,
\be \la{so31-invariant}
\langle \rho_A \rho_B \rangle = \frac{1}{4} \eta_{AB} \rho_C^2, \quad
\langle \Psi_A \Psi_B  \rangle = \frac{1}{4} \eta_{AB} \Psi_C^2.
\ee
More precisely, one can decompose the perturbation
\eq{reduce-raychaudhuri} into trace (scalar), anti-symmetric (vector)
and symmetric-traceless (tensor) parts. Since we look at only the
scalar perturbation in Eq.\eq{raychaudhuri}, simply assume that the
vector and tensor modes are negligible for some reasons, e.g., the
cosmological principle. In this case, Eq.\eq{reduce-raychaudhuri}
becomes
\be \la{raychaudhuri-desitter}
\dot{\Theta} = \frac{1}{8\lambda^2} g^{MN}
(\rho_M \rho_N + \Psi_M \Psi_N) > 0.
\ee
The perturbation \eq{so31-invariant} does not violate the energy
condition since $u^M u^N T_{MN}^{(L)} = \frac{1}{64 \pi G_4
\lambda^2} g^{MN}(\rho_M \rho_N + \Psi_M \Psi_N) > 0$. See
Eq.\eq{dark-energy-4d}. This means that the spacetime geometry is in
a de Sitter phase. Thus we see that the Liouville energy-momentum
tensor can act as a source of gravitational repulsion. We will
further discuss in Section 3.4 this energy as a plausible candidate
of dark energy.

Up to now we have considered fluctuations around the vacuum
\eq{vacuum-spacetime} corresponding to a uniform condensation of
gauge fields. In this case if we turn off all fluctuations, i.e.,
$\widehat{A}_M = 0$ in Eq.\eq{co-inner}, the metric
\eq{D-metric} or \eq{ward-metric} simply reduces to a flat spacetime.
We have to point out that the fluctuations need not be small. Our
ignorance of the next leading order, ${\cal O}(\theta^3)$, in
Eq.\eq{co-inner} corresponds to the limit of slowly varying fields,
$\sqrt{2\pi \alpha^\prime} |\frac{\p F}{F}| \ll 1$, in the sense
keeping field strengths (without restriction on their size) but not
their derivatives \ct{hsy2}. Since the Ricci curvature
\eq{ricci-eom} is purely determined by $\mathfrak{f}_{ABC} \sim
F_{AB}$ (see Eq.\eq{nc-vec-identity}), this approximation
corresponds to the limit of slowly varying curvatures compared to
the NC scale $|\theta| \sim l_{nc}^2$ but without restriction on
their size. This implies that NC effects should be important for a
violently varying spacetime, e.g., near the curvature singularity,
as expected.

\subsection{General NC spacetime}

Now the question is how to generalize the emergent gravity picture
to the case of a nontrivial vacuum, e.g., Eq.\eq{inhomo-b},
describing an inhomogeneous condensate of gauge fields. The Poisson
structure $\Theta^{ab}(x) = (\frac{1}{B})^{ab}(x)$ is nonconstant in
this case, so the corresponding NC field theory is defined by a
nontrivial star-product
\be \la{general-nc}
[Y^a, Y^b]_{\widetilde{\star}} = i \Theta^{ab}(Y)
\ee
where $Y^a$ denote vacuum coordinates which are designed with the
capital letters to distinguish them from $y^a$ for the constant
vacuum \eq{vacuum-spacetime}. The star product $[\widehat{f},
\widehat{g}]_{\widetilde{\star}}$ for $\widehat{f},
\widehat{g} \in {\cal A}_\Theta$ can be perturbatively computed via the
deformation quantization \ct{kontsevich}. There are excellent
earlier works \ct{cornalba-gen} especially relevant for the analysis
of the DBI action as a generalized geometry though a concrete
formulation of NC field theories for a general NC spacetime is still
out of reach.

Recall that we are interested in the commutative limit so that
\bea \la{general-c-limit}
-i [\widehat{f}, \widehat{g}]_{\widetilde{\star}} &=& \Theta^{ab}(Y)
\frac{\p f(Y)}{\p Y^a} \frac{\p g(Y)}{\p Y^b} + \cdots \xx
&\equiv& \{f,g \}_\Theta + \cdots
\eea
for $\widehat{f}, \widehat{g} \in {\cal A}_\Theta$. Using the
Poisson bracket \eq{general-c-limit}, we can similarly realize the
Lie algebra homomophism $C^\infty (M) \to TM: f \mapsto X_f$ between
a Hamiltonian function $f$ and the corresponding Hamiltonian vector
field $X_f$. To be specific, for any given function $f
\in C^\infty (M)$, we can always assign a Hamiltonian vector field $X_f$
defined by $X_f (g) = \{g,f\}_{\Theta}$ with some function $g
\in C^\infty (M)$. Then the following Lie algebra homomophism holds
\be \la{lie-homo-gen}
X_{\{f,g\}_{\Theta}} = - [X_f, X_g]
\ee
as long as the Jacobi identity for the Poisson bracket
$\{f,g\}_{\Theta}$ holds or, equivalently, the Schouten-Nijenhuis
bracket for the Poisson structure $\Theta^{ab}$ vanishes
\ct{kontsevich}.

Furthermore there is a natural automorphism $D(\hbar)$ which acts on
star-products \ct{kontsevich}:
\be \la{star-auto}
f \, {\widetilde{\star}} \, g = D(\hbar) \Big( D(\hbar)^{-1} (f)
\star D(\hbar)^{-1} (g) \Big).
\ee
In the commutative limit where $D(\hbar) \approx 1$,
Eq.\eq{star-auto} reduces to the following condition
\be \la{poisson-equiv}
{\{f,g\}_{\Theta}} = {\{f,g\}_{\theta}}.
\ee
Let us explain what Eq.\eq{poisson-equiv} means. For $f=Y^a(y)$ and
$g=Y^b(y)$, Eq.\eq{poisson-equiv} implies that
\be \la{two-theta}
\Theta^{ab}(Y) = \theta^{cd} \frac{\p Y^a}{\p y^c} \frac{\p Y^b}{\p
y^d}
\ee
whose statement is, of course, equivalent to the Moser lemma
\eq{darboux-tr}. Also notice that Eq.\eq{poisson-equiv} defines
diffeomorphisms between vector fields $X_f^\prime (g) \equiv
\{g,f\}_{\Theta}$ and $X_f (g) \equiv \{g,f\}_{\theta}$ such that
\be \la{poisson-diffeo}
{X^\prime_f}^a = \frac{\p Y^a}{\p y^b} X^b_f.
\ee
Indeed the automorphism \eq{star-auto} corresponds to a global
statement that the two star-products involved are cohomologically
equivalent in the sense that they generate the same Hochschild
cohomology \ct{kontsevich}.

It is still premature to know the precise form of the full NC field
theory defined by the star product \eq{general-c-limit}. Even the
commutative limit where the star commutator reduces to the Poisson
bracket in Eq.\eq{general-c-limit} still bears some difficulty since
the derivatives of $\Theta^{ab}$ appear here and there. For example,
\be \la{deriv-gener}
\{ B_{ab}(Y) Y^b, f \}_{\Theta} = \frac{\p f}{\p Y^a} + \Theta^{bc}
\frac{\p B_{ad}}{\p Y^b} Y^d \frac{\p f}{\p Y^c}.
\ee
In particular, $\{ B_{ab}(Y) Y^b, f \}_{\Theta} \neq \frac{\p f}{\p
Y^a}$. There is no simple way to realize the derivative
$\frac{\p}{\p Y^a}$ as an inner derivation.\footnote{To be precise,
we have to point out that the extra term in Eq.\eq{deriv-gener} can
be ignored under the limit of our consideration. We are considering
the limit of slowly varying fields where the derivative of field
strengths is ignored (see the last paragraph in Section 3.2). Then
Eq.\eq{deriv-gener} defines the inner derivation in this limit. We
expect the analysis in this limit will be very straightforward. But
we will not push to this direction because the coming new approach
seems to provide a more clear insight for the emergent geometry.}
Now we will suggest an interesting new approach for the nontrivial
background \eq{inhomo-b} based on the remark (3) in Section 2.3.

Let us return to the remark (3). Denote the nontrivial B-field in
Eq.\eq{inhomo-b} as
\be \la{back-b-bar}
B_{ab}(x) = (\bar{B} + \bar{F}(x))_{ab}
\ee
where $\bar{B}_{ab}= \big({\theta}^{-1}\big)_{ab}$ describes a
constant background such as Eq.\eq{vacuum-spacetime} while $
\bar{F}(x) = d \bar{A}(x)$ describes an inhomogeneous condensate of gauge fields.
Then the left-hand side of Eq.\eq{mirror} is of the form $g + \kappa
(\bar{B} + \mathcal{F})$ where $\mathcal{F} = d \mathcal{A}$ with
$\mathcal{A}(x) = \bar{A}(x) + A(x)$. It should be completely
conceivable that it can be mapped to the NC gauge theory of the
gauge field $\mathcal{A}(x)$ in the constant $\bar{B}$-field
background according to the Seiberg-Witten equivalence \ct{sw}. Let
us denote the corresponding NC gauge field as $\widehat{A}_a \equiv
\widehat{B}_a + \widehat{C}_a$. The only notable point is that the
gauge field $\widehat{A}_a$ has an inhomogeneous background part
$\widehat{B}_a$ and $\widehat{C}_a$ describes fluctuations around
this background. This situation should be familiar, for example,
with a gauge theory in an instanton (or soliton) background.

So everything goes parallel to the previous case. We will suppose a
general situation so that the background gauge fields
$\overline{\widehat{A}}_\mu (z,y)$ as well as $\widehat{B}_b (z,y)$
depend on $z^\mu$. Let us introduce the following covariant
coordinates
\bea \la{cov-coord-gen}
\widehat{X}^a(z, y) &=& y^a + \theta^{ab} \widehat{A}_b (z, y) = y^a +
\theta^{ab} \widehat{B}_b (z,y) + \theta^{ab} \widehat{C}_b (z, y) \xx
& \equiv & Y^a (z,y) + \theta^{ab} \widehat{C}_b (z, y)
\eea
where we identified the vacuum coordinates $Y^a$ in
Eq.\eq{general-nc} because we have to recover them after completely
turning off the fluctuation $\widehat{C}_b$. Now the covariant
derivative $\widehat{D}_M$ in Eq.\eq{decomp-cov} can be defined in
the exactly same way
\be \la{cov-der-gen}
\widehat{D}_M = \p_M - i \widehat{A}_M(z, y) = (\widehat{D}_\mu, - i \bar{B}_{ab}
\widehat{X}^b)(z, y)
\ee
where $\p_M = (\p_\mu, - i \bar{B}_{ab} y^b)$. In addition the NC
fields $\widehat{D}_A$ in Eq.\eq{cov-der-gen} (see the footnote
\ref{flat-index}) can be mapped to vector fields in the same way as
Eq.\eq{co-inner}.

Since the results in Section 3.2 can be applied to arbitrary NC
gauge fields in the constant $B$-field, the same formulae can be
applied to the present case at hand with the understanding that the
vector fields $D_A$ in Eq.\eq{co-inner} refer to total gauge fields
including the background. This means that the vector fields $D_A =
\lambda E_A \in TM$ reduce to $\bar{D}_A = \bar{\lambda}
\bar{E}_A$ after completely turning off the fluctuations
where $\bar{D}_A$ is determined by the background $(\p_\mu - i
\overline{\widehat{A}}_\mu (z,y), - i
\bar{B}_{ab} Y^b(z,y))$ and $\bar{\lambda}$ satisfies the relation
\be \la{back-lambda}
{\bar{\lambda}}^2 = \mathfrak{v}_D(\bar{D}_1, \cdots, \bar{D}_D).
\ee
Therefore the metric for the background is given by
\bea \la{D-back-metric}
ds^2 &=& \eta_{AB} \bar{E}^A \otimes \bar{E}^B \xx &=&
{\bar{\lambda}}^2
\eta_{AB} \bar{D}^A \otimes \bar{D}^B =
{\bar{\lambda}}^2 \eta_{AB} \bar{D}^A_M \bar{D}^B_N
\; dX^M \otimes dX^N.
\eea
Of course we have implicitly assumed that the background $\bar{D}_A$
also satisfies Eqs.\eq{map-eom}-\eq{map-bianchi}. In four
dimensions, for instance, we know that the metric \eq{D-back-metric}
describes Ricci-flat four manifolds if $\bar{D}_A$ satisfies the
self-duality equation \eq{sde-structure}.

Now let us look at the picture of the right-hand side of
Eq.\eq{mirror}. After applying the Darboux transform
\eq{darboux-tr} for the symplectic structure \eq{back-b-bar}, the right-hand side becomes
of the form $h_{ab}(y) + \kappa (\bar{B}_{ab} +
\mathfrak{F}_{ab}(y))$ where
\be \la{f-darboux}
\mathfrak{F}_{ab}(y) = \frac{\p x^\alpha}{\p y^a} \frac{\p x^\beta}{\p y^b}
F_{\alpha\beta}(x) \equiv \p_a \mathfrak{A}_b (y) - \p_b
\mathfrak{A}_a (y)
\ee
and the metric $h_{ab}(y)$ is given by Eq.\eq{induced-metric}. Note
that in this picture the gauge fields $\mathfrak{A}_a (y)$ are
regarded as fluctuations propagating in the background $h_{ab}(y)$
and $\bar{B}_{ab}$. Therefore it would be reasonable to interpret
the right-hand side of Eq.\eq{mirror} as a NC gauge theory of the
gauge field $\mathfrak{A}_a (y)$ defined by the canonical NC space
\eq{vacuum-spacetime} but in curved space described by the metric
$h_{ab}(y)$.

Although the formulation of NC field theory in a generic curved
spacetime is still a challenging problem, we want to speculate on
how to formulate the emergent gravity within this picture since the
underlying picture for the identity \eq{mirror} is rather
transparent. In this regard, the results in \ct{cornalba-gen} would
be useful. In this approach the inhomogeneous condensate of gauge
fields in the vacuum \eq{back-b-bar} appears as an explicit
background metric, which implies that the metric \eq{D-metric} in
this picture will be replaced by
\bea \la{D-darboux-metric}
ds^2 &=& g_{AB} E^A \otimes E^B \xx &=& \Lambda^2 g_{AB} D^A
\otimes D^B =  \Lambda^2 g_{AB}
D^A_M D^B_N \; dX^M \otimes dX^N
\eea
where $g_{AB}$ is the metric in the space spanned by the
noncoordinate bases $E^A = \Lambda D^A$ \ct{cho-freund}. Since the
anholonomic basis $D^A$ in Eq.\eq{D-darboux-metric} is supposed to
be flat when the fluctuations are turned off, i.e.,
$\mathfrak{F}_{ab} = 0$, the metric $\Lambda^2 g_{AB}$ will
correspond to the background metric $h_{ab}(y)$ in the DBI action
\eq{mirror}. Since the metric \eq{D-darboux-metric} has
the Riemannian volume form $\mathfrak{v}_g = \sqrt{-g}  E^1 \wedge
\cdots \wedge E^D$ instead of Eq.\eq{D-volume-form},
the volume form $\mathfrak{v}_D = \Lambda^{(2-D)}\mathfrak{v}_g$ in
Eq.\eq{D-volume} will be given by
\be \la{D-curved-volume-form}
\mathfrak{v}_D =  \sqrt{-g} \Lambda^2 D^1 \wedge
\cdots \wedge D^D.
\ee
So the function $\Lambda$ in Eq.\eq{D-darboux-metric} will satisfy
the condition
\be \la{D-volume-curved}
\sqrt{-g} \Lambda^2 = \mathfrak{v}_D(D_1, \cdots, D_D).
\ee
And it is easy to infer that $\sqrt{-g} \Lambda^2 \to 1$ for
vanishing fluctuations since $D_A$ becomes flat for that case.

According to the metric \eq{D-darboux-metric}, the indices $A,B,
\cdots$ will be raised and lowered using the metric $g_{AB}$. As
usual, the torsion free condition \eq{torsion-free} for the metric
\eq{D-darboux-metric} will be imposed to get the relation
\eq{spin-structure} where $\omega_{ABC} = g_{BD}{{\omega_A}^D}_C$
and $f_{ABC} = g_{CD}{f_{AB}}^D$. Since $g_{AB}$ is not a flat
metric, ${{\omega_A}^B}_C$ in Eq.\eq{D-spin1} or Eq.\eq{D-spin2}
will actually be the Levi-Civita connections in noncoordinate bases
rather than the spin connections, but we will keep the notation for
convenience. And the condition that the metric
\eq{D-darboux-metric} is covariantly constant, i.e., $\nabla_C
\Bigl( g_{AB} E^A \otimes E^B
\Bigr) = 0$, leads to the relation \ct{cho-freund}
\be \la{metric-condition-curved}
\omega_{ABC} = \half \big(E_A g_{BC} - E_B g_{CA} + E_C g_{AB} \big)
+ \half \big(f_{ABC} - f_{BCA} + f_{CAB} \big).
\ee
The curvature tensors have exactly the same form as
Eq.\eq{D-riemann}.

All the calculations in Appendix B can be repeated in this case
although the details will be much more complicated. We will not
perform this calculation since it seems to be superfluous at this
stage. But we want to draw some interesting consequences from the
natural requirement that the metric
\eq{D-darboux-metric} must be equivalent to the metric \eq{D-metric} or
\eq{ward-metric} in general, not only for backgrounds.

Let us summarize the two pictures we have employed. Let us indicate
the first picture with (L) and the second picture with (R). When all
fluctuations are vanishing, we have the following results:
\bea \la{L-1}
\mathrm{(L)}: ds^2 &=& {\bar{\lambda}}^2 \eta_{AB} \bar{D}^A_M \bar{D}^B_N \;
dX^M \otimes dX^N \xx
 &=& {\bar{\lambda}}^2 \Bigl(\eta_{\mu\nu} dz^\mu dz^\nu +
\delta_{ab} V^a_c V^b_d (dy^c - \mathbf{A}^c)(dy^d - \mathbf{A}^d)
\Bigr) \\
\la{L-2}
\mathfrak{v}_D &=& dz^1 \wedge \cdots \wedge dz^d \wedge dy^1
\wedge \cdots \wedge dy^{2n} \\
\la{L-3}
{\bar{\lambda}}^2 &=& \det^{-1} V^a_b \\
\la{R-1}
\mathrm{(R)}: ds^2 &=& \Lambda^2 g_{MN} dX^M \otimes dX^N \\
\la{R-2}
\mathfrak{v}_D &=& dz^1 \wedge \cdots \wedge dz^d \wedge dy^1
\wedge \cdots \wedge dy^{2n} \\
\la{R-3}
\Lambda^2 &=& \frac{1}{\sqrt{-g}}.
\eea
One can immediately see that (L) and (R) are equal each other if
$g_{MN} = \eta_{AB} \bar{D}^A_M \bar{D}^B_N$. Indeed, this
equivalence is nothing but the geometric manifestation of the
equivalence \eq{mirror}. Therefore we conjecture that the
equivalence between the two pictures (L) and (R) remains true even
after including all fluctuations.

Now let us examine whether the action \eq{matrix-action} allows a
conformally flat metric as a solution. First we point out that
$\Lambda^2 = 1$ for the flat metric $g_{MN} = \eta_{MN}$ as
Eq.\eq{R-3} immediately shows. This can also be seen from the
picture (L). Since we put $\mathbf{A}^c = 0$, $g_{MN} =
\eta_{MN}$ corresponds to a coordinate transformation $y^a \to
\tilde{y}^a$ such that $V^a_b dy^b = d \tilde{y}^a$.
This coordinate transformation can be expressed as $D_a^b =
\frac{\p y^b}{\p \tilde{y}^a}$ using Eq.\eq{com-vec}.
That is, the coordinate $\tilde{y}^a$ is a solution of the equation
$D_a \tilde{y}^b \equiv \frac{\p \tilde{y}^b}{\p y^a} +
\{ \widehat{A}_a, \tilde{y}^b \}_\theta = \delta^b_a$.
Thus we can replace the vector field $D_a \in TM$ by $\frac{\p}{\p
\tilde{y}^a}$ in the space described by the coordinates $(z^\mu,
\tilde{y}^a)$. Then Eq.\eq{back-lambda} is automatically satisfied
since the volume form \eq{L-2} is equal to $\mathfrak{v}_D =
\det^{-1} V^a_b \; dz^1
\wedge \cdots \wedge dz^d \wedge d\tilde{y}^1
\wedge \cdots \wedge d\tilde{y}^{2n} = {\bar{\lambda}}^2 dz^1
\wedge \cdots \wedge dz^d \wedge d\tilde{y}^1 \wedge \cdots \wedge d\tilde{y}^{2n}$.
Because we already put $\widehat{A}_\mu = 0 $, the vector fields in
$TM$ are now represented by $D_A[f](z^\mu, \tilde{y}^a) =
\big( \frac{\p}{\p z^\mu}, \frac{\p}{\p \tilde{y}^a})[f]$, which implies
$\forall \;{\mathfrak{f}_{AB}}^C = 0$. Therefore $\bar{\lambda}$
should be a constant due to the relation \eq{D-structure-f}.

Thereby we see that the conformally flat metric is instead given by
the vector field $\bar{D}_A = \phi(z,y) \p_A$, which corresponds to
the coordinate transformations $z^\mu \to \tilde{z}^\mu, \; y^a
\to \tilde{y}^a$ such that $dz^\mu = \phi^{-1} d \tilde{z}^\mu$ and
$V^a_b dy^b = \phi^{-1} d \tilde{y}^a$. In this case the metric
\eq{L-1} and the volume form \eq{L-2} are given by
\bea \la{conformal-metric}
ds^2 &=& \phi^{D-2} \big(\eta_{\mu\nu} d\tilde{z}^\mu
d\tilde{z}^\nu + d\tilde{y}^a d\tilde{y}^a \big) \\
\la{conformal-volume}
\mathfrak{v}_D &=& d\tilde{z}^1 \wedge \cdots
\wedge d\tilde{z}^d \wedge d\tilde{y}^1 \wedge \cdots \wedge d\tilde{y}^{2n}
\eea
where we used Eq.\eq{R-3}, i.e., $\Lambda^2 = {\bar{\lambda}}^2 =
\phi^D$. For the vector field $\bar{D}_A = \phi(\tilde{z},
\tilde{y}) \p_A$, the equation of motion \eq{map-eom} becomes
\be \la{ads-eom}
0 = \{ \widehat{D}^A \widehat{F}_{AB}, f \}_\theta =
\phi (\p^A \phi \p_A \phi + \phi \p^A \p_A \phi) \p_B f
- \phi (\p^A \phi \p_B \phi + \phi \p^A \p_B \phi) \p_A f
\ee
for any reference function $f = f(\tilde{z}, \tilde{y})$.

We will try two kinds of simple ansatz
\bea \la{ansatz-time}
(I): && \phi = \phi(\tau) \;\; \mathrm{where} \;\; \tau =
\tilde{z}^0,
\\
(II): && \phi = \phi(\rho) \;\; \mathrm{where} \;\; \rho^2 =
\sum_{a=1}^{2n} \tilde{y}^a \tilde{y}^a.
\eea
One can find for the ansatz (I) that Eq.\eq{ads-eom} leads to the
equation $\frac{d}{d\tau} \big( \phi \frac{d\phi}{d\tau} \big)=0$
and so $\phi(\tau) = \gamma \sqrt{\tau + \tau_0}$. In four
dimensions, this solution describes an expanding cosmological
solution \ct{big-gravity,hawking-ellis}. It is interesting that the
expanding cosmological solution comes out from ``pure" NC
electromagnetism \eq{matrix-action} without any source
term.\footnote{In comoving coordinates, the metric
\eq{conformal-metric} is of the form $ds^2 = -dt^2 + a(t)^2 d
\mathbf{x}^2$ where $t = \frac{2}{3} \gamma
\tau^{\frac{3}{2}}$ and $a(t)^2 = \gamma^2 \tau \equiv
\alpha t^{\frac{2}{3}}$. Since $a(t) \propto t^{\frac{2}{3(1+w)}}$,
we see that this metric corresponds to a universe characterized by
the equation of state $p = \rho$, i.e., $w=1$. It has been argued in
\ct{banks-fishler} that the $p = \rho$ cosmology
corresponds to the most holographic background and the most entropic
initial condition for the universe. We thank Qing-Guo Huang for
drawing our attention to \ct{banks-fishler}.}

However, for the ansatz (II), we found that only $\phi =
\rm{constant}$ can be a solution. This seems to be true in general.
Hence we claim that a conformally flat metric for the ansatz (II) is
trivial. A source term might be added to the action
\eq{matrix-action} to realize a nontrivial solution. The solution for the
ansatz (II) should be interesting because the $AdS_p \times {\bf
S}^q$ space with $q+1 = 2n$ belongs to this class and it can be
described by Eq.\eq{conformal-metric} by choosing
\be \la{radius}
\phi^{D-2} = \frac{L^2}{\rho^2}.
\ee
In particular, $AdS_5 \times {\bf S}^5$ space is given by the case,
$d=4, \; n=3$, that is,
\be \la{ads5}
ds^2 = \frac{L^2}{\rho^2} \big(\eta_{\mu\nu} d\tilde{z}^\mu
d\tilde{z}^\nu + d\tilde{y}^a d\tilde{y}^a \big) =
\frac{L^2}{\rho^2}
\big(\eta_{\mu\nu} d\tilde{z}^\mu d\tilde{z}^\nu + d \rho^2 \big) + L^2 d
\Omega_5^2.
\ee
We hope to address in the near future what kind of source term
should be added to get the conformal factor \eq{radius}.
Eq.\eq{radius} looks like a potential of codimension-$2n$ Coulomb
sources in $D$ dimensions when we identify the harmonic function
$H(\rho)^{\frac{1}{n-1}} = \phi^{D-2} = L^2/\rho^2$, which
presumably corresponds to the vacuum \eq{max-comm-matrix}.

\subsection{Hindsights}

We want to ponder on the spacetime picture revealed from NC gauge
fields and the emergent gravity we have explored so far.

The most remarkable picture emerging from NC gauge fields is about
the origin of flat spacetime, which is absent in Einstein gravity.
Of course the notorious problem for emergent time is elusive as
ever. We will refer to the emergence of spaces only here, but we
will discuss in Section 4 how ``Emergent Time" would be defined in
the context of emergent gravity.

Note that the flat spacetime is a geometry of special relativity
rather than general relativity and the special relativity is a
theory about kinematics rather than dynamics. Hence the general
relativity says nothing about the dynamical origin of flat spacetime
since the flat spacetime defining a local inertial frame is assumed
to be {\it a priori} given without reference to its dynamical
origin. So there is a blind point about the dynamical origin of
spacetime in general relativity.

Our scheme for the emergent gravity implies that the uniform
condensation of gauge fields in a vacuum \eq{vacuum-spacetime} will
be a source of flat spacetime. Now we will clarify the dynamical
origin of flat spacetime based on the geometric representation in
Section 3.2. We will equally refer to the commutative spacetime
${\bf R}_C^d$ with the understanding that it has been T-dualized
from a fully NC space (except time) in the sense of Eq.\eq{matrix-t}
although the transition from  NC to commutative ones is mysterious
(see the remark (1) in Section 3.1). Therefore we will regard
$\p_\mu$ in Eq.\eq{co-inner} as a background part since it is
related to $y^a/\kappa$ via the matrix T-duality
\eq{matrix-t}.

The basic principle for the emergent gravity is the map
\eq{co-inner} or the correspondence \eq{duality}
between NC fields in ${\cal A}_\theta$ and vector fields in $TM$.
The most notable point is that we necessarily need a Poisson (or
symplectic) structure on $M$, viz., NC spacetime, to achieve the
correspondence between ${\cal A}_\theta$ and $\Gamma(TM)$, sections
of tangent bundle $TM \to M$. Basically the $\theta$-deformation
\eq{nc-spacetime} introduces the duality between NC gauge fields and
spacetime geometry. The crux is that there exists a novel form of
the equivalence principle, guaranteed by the global Moser lemma, for
the electromagnetism in the context of symplectic geometry. In this
correspondence a flat spacetime is coming from the constant
background itself defining the NC spacetime
\eq{vacuum-spacetime}. This observation, trivial at the first
glance, was the crucial point for the proposal in \ct{hsy-cc} to
resolve the cosmological constant problem.

We know that the uniform condensation of stress-energy in a vacuum
will appear as a cosmological constant in Einstein gravity. For
example, if we shift a matter Lagrangian $\mathcal{L}_M$ by a
constant $\Lambda$, that is,
\be \la{shift-lagrangian}
\mathcal{L}_M \to \mathcal{L}_M - 2 \Lambda,
\ee
this shift results in the change of the energy-momentum tensor of
matter by $T_{MN} \to T_{MN} - \Lambda g_{MN}$ in the Einstein
equation \eq{einstein-energy} although the equations of motion for
matters are invariant under the shift \ct{pad}. Definitely this
$\Lambda$-term will appear as a cosmological constant in Einstein
gravity and it has an observable physical effect. For example, a
flat spacetime is no longer a solution of the Einstein equation in
the case of $\Lambda \neq 0$.

The emergent gravity defined by the action \eq{matrix-action}
responds completely differently to the constant shift
\eq{shift-lagrangian}. To be specific, let us consider a constant
shift of the background $B_{MN} \to B_{MN} + \delta B_{MN}$. Then
the action
\eq{matrix-action} in the new background becomes
\begin{equation} \label{shift-action}
S_{B + \delta B} = S_{B} + \frac{1}{2 g^2_{YM}} \int d^D X
\widehat{F}_{MN} \delta B_{MN} - \frac{1}{4 g^2_{YM}} \int d^D X
\Big( \delta B_{MN}^2 - 2 B^{MN} \delta B_{MN} \Big).
\end{equation}
The last term in Eq.\eq{shift-action} is simply a constant and thus
it will not affect the equations of motion \eq{map-eom}. The second
term is a total derivative and so it will vanish if
$\widehat{F}_{MN}$ well behaves at infinity. (It is a defining
property in the definition of a star product that $\int d^D X
\widehat{f} \star \widehat{g} = \int d^D X \widehat{f} \cdot \widehat{g}$.
Then the second term should vanish as far as $\widehat{A}_M \to 0$
at infinity.) If spacetime has a nontrivial boundary, the second
term could be nonvanishing at the boundary which will change the
theory under the shift. We will not consider a nontrivial spacetime
boundary since the boundary term is not an essential issue in the
cosmological constant problem, though there would be an interesting
physics at the boundary. Then we get the result $S_{B + \delta B}
\cong S_{B}$. Indeed this is the Seiberg-Witten equivalence between
NC field theories defined by the noncommutativity $\theta'=
\frac{1}{B+ \delta B}$ and $\theta = \frac{1}{B}$ \ct{sw}. Although
the vacuum \eq{vacuum-spacetime} readjusts itself under the shift,
the Hilbert spaces ${\cal H}_{\theta'}$ and ${\cal H}_{\theta}$ in
Eq.\eq{fock} are completely isomorphic if and only if $\theta$ and
$\theta'$ are nondegenerate constants. Furthermore the vector fields
in Eq.\eq{co-inner} generated by $B + \delta B$ and $B$ backgrounds
are equally flat as long as they are constant. We also observed in
Eq.\eq{em-maxwell} that the background gauge field does not
contribute to the energy-momentum tensor.

Therefore we conclude that the constant shift of energy density such
as Eq.\eq{shift-lagrangian} is a symmetry of the theory
\eq{matrix-action} although the action \eq{matrix-action} defines a
theory of gravity in the sense of emergent gravity. Thus the
emergent gravity is completely immune from the vacuum energy. In
other words, {\it the vacuum energy does not gravitate} unlike as
Einstein gravity. This was an underlying logic in \ct{hsy-cc} why
the emergent gravity can resolve the cosmological constant problem.

One has realized that the cosmological constant can be interpreted
as a measure of the energy density of the vacuum. One finds that the
resulting energy density is of the form
\be \la{vacuum-energy}
\rho_{\mathrm{vac}} = \frac{1}{V} \sum_{\mathbf{k}} \half \hbar
\omega_{\mathbf{k}} \sim \hbar k^4_{max}
\ee
where $k_{max}$ is a certain momentum cutoff below which an
underlying theory can be trusted. Thus the vacuum energy
\eq{vacuum-energy} may be understood as a vast accumulation of
harmonic oscillators in space. Note that the vacuum
\eq{vacuum-spacetime} is also the uniform condensation of
harmonic oscillators in space. The immune difference is that the
harmonic oscillator in Eq.\eq{vacuum-energy} is defined by the NC
phase space \eq{nc-phase} while the harmonic oscillator in
Eq.\eq{vacuum-spacetime} is defined by the NC space
\eq{nc-spacetime}.

The current framework of quantum field theory, which has been
confirmed by extremely sophisticated experiments, mostly predicts
the vacuum energy of the order $\rho_{\mathrm{vac}} \sim (10^{18}
GeV)^4$. The real problem is that this huge energy couples to
gravity in the framework of Einstein gravity and so results in a
bizarre contradiction with contemporary astronomical observations.
This is the notorious cosmological constant problem.

But we have observed that the emergent gravity shows a completely
different picture about the vacuum energy. The vacuum energy
\eq{vacuum-energy} does not gravitate regardless of how large it is
as we explained above. So there is no cosmological constant problem
in emergent gravity. More remarkable picture in emergent gravity is
that the huge energy $M_{Pl} = (8 \pi G)^{-1/2} \sim 10^{18} GeV$ is
actually the origin of the flat spacetime. Here the estimation of
the vacuum energy for the condensate \eq{vacuum-spacetime}, for
example, $\rho_{\mathrm{vac}} \sim |B_{ab}|^2 \sim M_P^4$ in four
dimensions, is coming from our identification of the Newton constant
\eq{newton-constant}. In other words, the emergent gravity says that
a flat spacetime is not free gratis but a result of the Planck
energy condensation in a vacuum.

An important point is that the vacuum \eq{vacuum-spacetime}
triggered by the Planck energy condensation causes the spacetime to
be NC and the NC spacetime is the essence of emergent gravity. Since
the flat spacetime is emergent from the uniform vacuum
\eq{vacuum-spacetime} and the Lorentz symmetry is its spacetime
symmetry, the dynamical origin of flat spacetime implies that the
Lorentz symmetry is also emergent from the NC spacetime
\eq{vacuum-spacetime}. In addition, if the vacuum
\eq{vacuum-spacetime} was triggered by the Planck energy
condensation, the flat spacetime as well as the Lorentz symmetry
should be very robust against any perturbations since the Planck
energy is the maximum energy in Nature.

Furthermore the noble picture about the dynamical origin of the flat
spacetime may explain why gravity is so weak compared to other
forces. Let us look at Eq.\eq{cov-coordinate}.  As we know, $y^a$ is
a background part defining a flat spacetime and the gauge field
$\widehat{A}_a$ describes dynamical fluctuations around the flat
spacetime. (As we mentioned at the beginning of this section, the
commutative space in Eq.\eq{decomp-cov} can also be incorporated
into this picture using the T-duality \eq{matrix-t}.) One may
imagine these fluctuations as shaking the background spacetime
lattice defined by the Fock space \eq{fock}, which generates
gravitational fields. But the background lattice is very solid since
the stiffness of the lattice is supposed to be the Planck scale. In
other words, the gravity generated by the deformations of the
spacetime lattice \eq{fock} will be very weak since it is suppressed
by the background stiffness of the Planck scale. So, ironically, the
weakness of gravitational force may be due to the fact that the flat
spacetime is originated from the Planck energy.

The emergent gravity thus reveals a remarkably beautiful and
consistent picture about the origin of flat spacetime. Does it also
say something about dark energy ?

Over the past ten or twenty years, several magnificent astronomical
observations have confirmed that our Universe is composed of 5 \%
ordinary matters and radiations while 23 \% dark matter and 72 \%
dark energy. The observed value of the dark energy turned out to be
very very tiny, say,
\be \la{dark-energy-obs}
\Delta \rho^{obs} \leq (10^{-12} GeV)^4
\ee
which is desperately different from the theoretical estimation
\eq{vacuum-energy} by the order of $10^{120}$. What is the origin of
the tiny dark energy \eq{dark-energy-obs} ?

We suggested in \ct{hsy-cc} that the dark energy
\eq{dark-energy-obs} is originated from vacuum fluctuations around
the primary background \eq{vacuum-spacetime}. Since the background
spacetime \eq{vacuum-spacetime} is NC, any UV fluctuations of the
Planck scale $L_P$ in the NC spacetime will be necessarily paired
with IR fluctuations of a typical scale $L_H$ related to the size of
cosmic horizon in our Universe due to the UV/IR mixing \ct{uv-ir}. A
simple dimensional analysis shows that the energy density of the
vacuum fluctuation is of the order
\be \la{vac-fluctuation}
\Delta \rho \sim \frac{1}{L_P^2 L_H^2}
\ee
which is numerically in agreement with the observed value
\eq{dark-energy-obs} up to a factor \ct{hsy-cc}.
It should be remarked that the vacuum fluctuation
\eq{vac-fluctuation} will be an inevitable consequence if our
picture about the dynamical origin of flat spacetime is correct. If
the vacuum \eq{vacuum-spacetime} or equivalently the flat spacetime
is originated from the Planck energy condensation (it should be the
case if the identification \eq{newton-constant} is correct), the
energy density of the vacuum \eq{vacuum-spacetime} will be
$\rho_{{\rm vac}} \sim M_{Pl}^4$ which is the conventionally
identified vacuum energy predicted by quantum field theories. Thus
it is natural to expect that cosmological fluctuations around the
vacuum \eq{vacuum-spacetime} or the flat spacetime will add a tiny
energy $\Delta \rho$ to the vacuum so that the total energy density
is equal to $\rho \sim M_{Pl}^4 \big(1 + \frac{L_P^2}{L_H^2} \big)$
since $L^2_P \equiv 8 \pi G_4$ and $L_H^2  \equiv 1/\Lambda$ are
only the relevant scales in the Einstein equation
\eq{einstein-energy} with $T_{MN} = - \frac{\Lambda}{8\pi G_4} g_{MN}
= - M_{Pl}^4 \big( \frac{L_P}{L_H} \big)^2 g_{MN}$ \ct{pad}. Since
the first term does not gravitate, the second term
\eq{vac-fluctuation} will be the leading contribution to the
deformation of spacetime curvature, leading to possibly a de Sitter
phase. It should be remarked that the fluctuation
\eq{vac-fluctuation} is of the finite size $L_H$. So one cannot
apply the argument \eq{shift-action} since $\Delta \rho$ is not
constant over the entire spacetime even if it is constant over a
Hubble patch.

Now we will argue that the Liouville energy \eq{dark-energy} may (or
can) explain the dark energy \eq{vac-fluctuation}. First let us
perform the Wick rotation for the energy-momentum tensor
\eq{dark-energy} using the rule in the footnote \ref{wick-rotation}
to get the Lorentzian energy-momentum tensor in the 4-dimensional
spacetime. It is then given by
\be \la{dark-energy-4d}
T_{MN}^{(L)} = \frac{1}{16 \pi G_4 \lambda^2} \Big( \rho_M \rho_N +
\Psi_M \Psi_N - \frac{1}{2} g_{MN}(\rho_P^2 + \Psi_P^2) \Big)
\ee
where $\rho_M = 2 \p_M \lambda$ and $\Psi_M = E_M^A \Psi_A$.
First of all we emphasize that we already checked in Eq.\eq{reduce-raychaudhuri}
that it can exert a negative pressure causing an expansion of
universe, possibly leading to a de Sitter phase. We also pointed out
below Eq.\eq{average-dark-energy} that it can behave like a cosmological
constant, i.e., $\rho = - p$, in a constant (or almost constant)
curvature spacetime. Another important property is that the
Liouville energy \eq{dark-energy-4d} is vanishing for the flat
spacetime. So it should be small if the spacetime is not so curved.

To be more quantitative, let us consider the fluctuation \eq{so31-invariant}
and look at the energy density $u^M u^N T_{MN}^{(L)}$ along the flow
represented by a timelike unit vector $u^M$ as in Eq.\eq{ray-scalar}. Note that the
Riemannian volume is given by $\mathfrak{v}_g =
\lambda^2 \mathfrak{v}_4 = \lambda^2 d^4 x$. Also
recall that $\Psi_M$ is the Hodge-dual to the 3-form $H$ in
Eq.\eq{3-form-h}. Thus $u^M \rho_M$ and $u^M \Psi_M$ refer to the
volume change of a three dimensional spacelike hypersurface
orthogonal to $u^M$. Assume that the radius of the three dimensional
hypersurface is $R(\tau)$ at time $\tau$, where $\tau$ is an affine
parameter labeling the curve of the flow. Then it is reasonable to
expect that $u^M \rho_M \approx u^M \Psi_M \approx 2 \lambda/
R(\tau)$ where we simply assumed that $u^M \rho_M \approx u^M
\Psi_M$. Then we approximately get
\be \la{dark-energy-cal}
u^M u^N T_{MN}^{(L)} \sim \frac{1}{8 \pi G_4 R^2}.
\ee
If we identify the radius $R$ with the size of cosmic horizon,
$L_H$, the energy density \eq{dark-energy-cal} reproduces the dark
energy \eq{vac-fluctuation} up to a factor.

\section{Electrodynamics as a Symplectic Geometry}

This section does contain mostly speculations. We will not intend
any rigor. Rather we will revisit the $\hbar$-deformation
\eq{nc-phase} to reinterpret the electrodynamics of a
charged particle in terms of symplectic geometry defined in phase
space. We want to point out its beautiful aspects since in our
opinion it has not been well appreciated by physicists. Furthermore
it will provide a unifying view about $U(1)$ gauge theory in terms
of symplectic geometry. Nevertheless our main motivation for the
revival is to get some glimpse on how to introduce matter fields
within the framework of emergent gravity. As a great bonus, it will
also outfit us with a valuable insight about how to define ``Time"
in the sense of emergent spacetime.

\subsection{Hamiltonian dynamics and emergent time}

Let us start to revisit the derivation of the Darboux theorem
\eq{darboux-tr} due to Moser \ct{moser}. A remarkable point in the
Moser's proof is that there always exists a one-parameter family of
diffeomorphisms generated by a smooth time-dependent vector field
$X_t$ satisfying $\iota_{X_t} \omega_t + A = 0$ for the change of a
symplectic structure within the same cohomology class from $\omega$
to $\omega_t = \omega + t (\omega' - \omega)$ for all $0 \leq t \leq
1$ where $\omega' - \omega = dA$. The evolution of the symplectic
structure is locally described by the flow $\phi_t$ of $X_t$
starting at $\phi_0$ = identity. (Of course the ``time" $t$ here is
just an affine parameter labeling the flow. At this stage it does
not necessarily refer to a physical time.) By the Lie derivative
formula, we have
\bea \la{time-flow}
\frac{d}{dt} \big( \phi_t^* \omega_t \big) &=&
\phi_t^* \big( \mathcal{L}_{X_t} \omega_t \big) + \phi_t^* \frac{d \omega_t}{dt} \xx
&=& \phi_t^* d \iota_{X_t} \omega_t + \phi_t^*(\omega' - \omega) =
\phi_t^* \big(\omega' - \omega -dA) = 0.
\eea
Thus $\phi_1^* \omega' =  \phi_0^* \omega = \omega$, so $\phi_1$
provides a chart describing the evolution from $\omega$ to $\omega'
= \omega + dA$.

A whole point of the emergent gravity is the global existence of the
one-parameter family of diffeomorphisms $\phi_t$ describing the
local deformation of a symplectic structure due to the
electromagnetic force. Therefore the electromagnetism in NC
spacetime is nothing but a symplectic geometry (at the leading order
or commutative limit). Now our question is how to understand matter
fields or particles in the context of emergent geometry or
symplectic geometry.

As a first step, we want to point out that the coupling of a charged
particle with $U(1)$ gauge fields is beautifully understood in the
context of symplectic geometry \ct{sternberg,dyson}. This time the
symplectic geometry of matters is involved with the
$\hbar$-deformation \eq{nc-phase} rather than the
$\theta$-deformation \eq{nc-spacetime} which is the symplectic
geometry of gravity. It is rather natural that matters or particles
are described by the symplectic geometry of the phase space since
the particles by definition are prescribed by their positions and
momenta besides their intrinsic charges, e.g., spin, electric
charge, isospin, etc. We will consider only the electric charge
among their internal charges for simplicity. We refer some
interesting works \ct{sternberg, dyson, feynman-qed,
symplectic-technique} addressing this problem.

Let $(M, \omega)$ be a symplectic manifold. One can properly choose
local canonical coordinates $y^a = (q^1, p_1, \cdots, q^n, p_n)$ in $M$
such that the symplectic structure $\omega$ can be written in the form
\be \la{symplectic-phase}
\omega = \sum_{i = 1}^n dq^i \wedge dp_i.
\ee
Then $\omega \in \bigwedge^2 T^*M$ can be thought as a bundle map
$TM \to T^*M$. Since $\omega$ is nondegenerate at any point $y \in
M$, we can invert this map to obtain the map $\vartheta \equiv
\omega^{-1}: T^* M \to TM$. This cosymplectic structure
$\vartheta \in \bigwedge^2 TM$ is called the Poisson structure of
$M$ which defines a Poisson bracket $\{\cdot,\cdot\}_\vartheta$. See
the footnote \ref{poisson}. In a local chart with coordinates $y^a$,
we have
\be \la{poisson-phase}
\{f, g\}_\vartheta = \sum_{a,b=1}^{2n} \vartheta^{ab} \frac{\p f}{\p y^a}
\frac{\p g}{\p y^b}.
\ee

Let $H: M \to \mathbf{R}$ be a smooth function on a Poisson manifold
$M$. The vector field $X_H$ defined by $\iota_{X_H} \omega = dH$ is
called the Hamiltonian vector field with the energy function $H$. We
define a dynamical flow by the differential equation
\be \la{hamilton-eq}
\frac{df}{dt} = X_H (f) + \frac{\p f}{\p t} = \{f, H\}_\vartheta + \frac{\p f}{\p t}.
\ee
A solution of the above equation is a function $f$ such that for any
path $\gamma: [0,1] \to M$ we have
\be \la{hamilton-path}
\frac{df(\gamma(t))}{dt} = \{f, H\}_\vartheta(\gamma(t)) + \frac{\p f(\gamma(t))}{\p t}.
\ee

The dynamics of a charged particle in an external static magnetic
field is described by the Hamiltonian
\be \la{particle-hamiltonian}
H = \frac{1}{2m}\big(\mathbf{p}- e \mathbf{A} \big)^2
\ee
which is obtained by the free Hamiltonian $H_0 =
\frac{\mathbf{p}^2}{2m}$ with the replacement
\be \la{minimal-coupling}
\mathbf{p}' = \mathbf{p}- e \mathbf{A}.
\ee
Here the electric charge of an electron is $q_e = - e$ and $e$ is a
coupling constant identified with $g_{YM}$. The symplectic structure
\eq{symplectic-phase} leads to the Hamiltonian vector field $X_H$ given by
\be \la{ham-vec-x}
X_H = \frac{\p H}{\p p_i}\frac{\p}{\p q^i} - \frac{\p H}{\p
q^i}\frac{\p}{\p p_i}.
\ee
Then the Hamilton's equation \eq{hamilton-eq} reduces to the
well-known Lorentz force law
\be \la{force-law}
m \frac{d\mathbf{v}}{dt} = e \mathbf{v} \times \mathbf{B}.
\ee

An interesting observation \ct{sternberg} (orginally due to
Jean-Marie Souriau) is that the Lorentz force law \eq{force-law} can
be derived by keeping the Hamiltonian $H = H_0$ but instead shifting
the symplectic structure
\be \la{soriau}
\omega \to \omega' = \omega - e B
\ee
where $B (q) = \half B_{ij} (q) dq^i \wedge dq^j$. In this case the
Hamiltonian vector field $X_H$ defined by $\iota_{X_H} \omega' = dH$
is given by
\be \la{ham-vec-f}
X_H = \frac{\p H}{\p p_i}\frac{\p}{\p q^i} -
\Big( \frac{\p H}{\p q^i} - e B_{ij} \frac{\p H}{\p p_j} \Big)
\frac{\p}{\p p_i}.
\ee
Then one can easily check that the Hamilton's equation
\eq{hamilton-eq} with the vector field \eq{ham-vec-f} reproduces the
Lorentz force law \eq{force-law}. Actually one can show that the
symplectic structure $\omega'$ in Eq.\eq{soriau} introduces a NC
phase space \ct{nc-review} such that the momentum space becomes NC,
i.e., $[p'_i, p'_j] = - i \hbar eB_{ij}$.

If a particle is interacting with electromagnetic fields, the
influence of the magnetic field $B=dA$ is described by the `minimal
coupling' \eq{minimal-coupling} and the new momenta $\mathbf{p}' =
-i\hbar (\nabla - i \frac{e}{\hbar} \mathbf{A})$ are covariant under
$U(1)$ gauge transformations. Let us point out that the minimal
coupling \eq{minimal-coupling} can be understood as the Darboux
transformation \eq{darboux-tr} between $\omega$ and $\omega'$.
Consider the coordinate transformation $y^a \mapsto x^a(y) = (Q^1,
P_1, \cdots, Q^n, P_n)(q,p)$ such that
\be \la{darboux-phase}
\sum_{i = 1}^n dq^i \wedge dp_i =
\sum_{i = 1}^n dQ^i \wedge dP_i - \frac{e}{2} \sum_{i,j=1}^n B_{ij} (Q) dQ^i \wedge dQ^j
\ee
but the Hamiltonian is unchanged, i.e., $H=\frac{\mathbf{P}^2}{2m}$.
The condition \eq{darboux-phase} is equivalent to the following
equations
\bea \la{darboux-123}
&& \frac{\p q^i}{\p Q^j} \frac{\p p_i}{\p Q^k} - \frac{\p q^i}{\p
Q^k} \frac{\p p_i}{\p Q^j} = - e B_{jk}, \xx && \frac{\p q^i}{\p
Q^j} \frac{\p p_i}{\p P_k} - \frac{\p q^i}{\p P_j} \frac{\p p_i}{\p
Q^k} = \delta_j^k, \\
&& \frac{\p q^i}{\p P_j} \frac{\p p_i}{\p P_k} - \frac{\p q^i}{\p
P_k} \frac{\p p_i}{\p P_j} = 0. \nonumber
\eea
The above equations are solved by
\be \la{darboux-sol}
q^i = Q^i, \qquad p_i = P_i + e A_i(Q).
\ee

In summary the dynamics of a charged particle in an electromagnetic
field has two equivalent descriptions:
\be \la{particle-equivalence}
\Big(H=\frac{(\mathbf{p}- e \mathbf{A})^2}{2m}, \omega \Big)(q,p)
\quad \cong \quad \Big(H=\frac{\mathbf{P}^2}{2m}, \omega'= \omega - eB \Big)(Q,P).
\ee

The equivalence \eq{particle-equivalence} can easily be generalized
to a time-dependent background $A^\mu = (A^0, \mathbf{A})(q,t)$ with
the Hamiltonian $H = \frac{1}{2m}\big(\mathbf{p}- e \mathbf{A}
\big)^2 + eA^0$. The Hamilton's equation \eq{hamilton-eq} in this case becomes
\be \la{force-law-time}
m \frac{d\mathbf{v}}{dt} = e \big( \mathbf{E} + \mathbf{v} \times
\mathbf{B} \big).
\ee
The equivalence \eq{particle-equivalence} now means that the Lorentz
force law \eq{force-law-time} can be obtained by the Hamiltonian
vector field \eq{ham-vec-f} with the Hamiltonian $H =
\frac{\mathbf{p}^2}{2m} + eA^0$ by noticing that the time dependence
of the external fields now appears as the explicit $t$-dependence of
momenta $p_i =p_i(t)$. Indeed the electric field $\mathbf{E}$
appears as the combination $\mathbf{E} = - \nabla A^0 +
\frac{1}{e} \frac{\p \mathbf{p}}{\p t}$. But note that the coordinates
$(q^i, p_i)$ in Eq.\eq{ham-vec-f} correspond to $(Q^i, P_i)$ in the
notation \eq{darboux-phase} and so $\frac{\p \mathbf{p}}{\p t} = - e
\frac{\p \mathbf{A}}{\p t}$ by Eq.\eq{darboux-sol}.

In a very charming paper \ct{dyson}, Dyson explains the Feynman's
view about the electrodynamics of a charged particle. Feynman starts
with an assumption that a particle exists with position $q^i$ and
velocity $\dot{q}_i$ satisfying commutation relations
\be \la{feynman-comm}
[q^i, q^j] = 0, \qquad m [q^i, \dot{q}_j] = i \hbar\delta^i_j.
\ee
Then he asks a question: What is the most general form of forces
appearing in the Newton's equation consistent with the commutation
relation \eq{feynman-comm} ? Remarkably he ends up with the
electromagnetic force \eq{force-law-time}. In a sense, the Feynman's
result is a no-go theorem for the consistent interaction of
particles in quantum mechanics. The only room for some modification
to the Feynman's argument seems to introduce internal degrees of
freedom such as spin, isospin, color, etc \ct{feynman-qed}. Then a
particle motion is defined on $\mathbf{R}^3 \times F$ with an
internal space $F$. The dynamics of the particle carrying an
internal charge in $F$ is defined by a symplectic structure on $T^*
\mathbf{R}^3 \times F$. See \ct{feynman-qed} for some details.

The Feynman's approach clearly shows that the electromagnetism is an
inevitable structure in quantum particle dynamics. Furthermore, as
emphasized by Dyson, the Feynman's formulation shows that
nonrelativistic Newtonian mechanics and relativistic Maxwell
equations are coexisting peacefully. This is due to the gauge
symmetry that the Lorentz force \eq{force-law-time} is generated by
the minimal coupling $p_\mu \to \mathfrak{P}_\mu \equiv p_\mu - e
A_\mu$. Moreover, Souriau and Sternberg show that the minimal
coupling can be encoded into the deformation of symplectic
structure, which can be summarized as the relativistic form
\ct{symplectic-technique}: $\omega = - d \xi \to \omega' = \omega -
e F = - d \big( \xi + e A \big) $ where $\xi = \mathfrak{P}_\mu
dQ^\mu$ and $A = A_\mu (Q) dQ^\mu$. Therefore the Maxwell equation
$dF=0$ is simply interpreted as the closedness of the symplectic
structure.

Now we have perceived that the dynamics of a charged particle can be
interpreted as a symplectic geometry in phase space. The evolution
of the system is described by the dynamical flow \eq{hamilton-path}
generated by a Hamiltonian vector field, e.g., Eq.\eq{ham-vec-x},
for a given Hamiltonian $H$. Basically, the time in the Hamilton's
equation \eq{hamilton-eq} is an affine parameter to trace out the
history of a particle and it is operationally defined by the
Hamiltonian. Therefore the time in the Hamiltonian dynamics is
intrinsically assigned to the particle itself. But we have to notice
that, only when the symplectic structure is fixed for a given
Hamiltonian, the evolution of the system is completely determined by
the evolution equation \eq{hamilton-eq}. In this case the dynamics
of the system can be formulated in terms of an evolution with a
single time parameter. In other words, we have a globally
well-defined time for the evolution of the system. This is the usual
situation we consider in classical mechanics.

We observed the equivalence \eq{particle-equivalence} for the
dynamics of a charged particle. Let us consider a dynamical
evolution described by the change of a symplectic structure from
$\omega$ to $\omega_t = \omega + t (\omega' -
\omega)$ for all $0 \leq t \leq 1$ where $\omega' -
\omega = - e dA$. The Moser lemma \eq{time-flow} says that
there always exists a one-parameter family
of diffeomorphisms generated by a smooth time-dependent vector field
$X_t$ satisfying $\iota_{X_t} \omega_t = e A$. Although the vector
field $X_t$ defines a dynamical one-parameter flow, the vector field
$X_t$ is in general not even a locally Hamiltonian since $dA=B
\neq 0$. The evolution of the system in this case is locally
described by the flow $\phi_t$ of $X_t$ starting at $\phi_0$ =
identity but it is no more a (locally) Hamiltonian flow. That is,
there is no well-defined or global time for the particle system. The
flow can be a (locally) Hamiltonian, i.e., $\phi_t$ = identity for
all $0 \leq t \leq 1$, only for $dA = 0$. In other words, the time
flow $\phi_t$ of $X_t$ defined on a local chart describes a local
evolution of the system.

Let us summarize the above situation by looking at the familiar
picture in Eq.\eq{particle-equivalence} by fixing the symplectic
structure but instead changing the Hamiltonian. (Note that the
magnetic field in the Lorentz force \eq{force-law} does not do any
work. So there is no energy flow during the evolution.) At time
$t=0$, the system is described by the free Hamiltonian $H_0$ but it
ends up with the Hamiltonian \eq{particle-hamiltonian} at time
$t=1$. Therefore the dynamics of the system cannot be described with
a single time parameter covering the entire period $0
\leq t \leq 1$. We can introduce at most a local time during $\delta t <
\epsilon$ on a local patch and smoothly adjust to a neighboring
patch. To say, a clock of the particle will tick each time with a
different rate since the Hamiltonian of the particle is changing
during time evolution.

We have faced a similar situation in the $\theta$-deformation
\eq{nc-spacetime} as summarized in Eq.\eq{time-flow}.
Of course one should avoid a confusion between the dynamical
evolution of particle system related to the phase space
\eq{nc-phase} and the dynamical evolution of spacetime geometry
related to the NC space \eq{nc-spacetime}. But we should get an
important lesson from Souriau and Sternberg \ct{sternberg} that the
Hamiltonian dynamics in the presence of electromagnetic fields can
be described by the deformation of symplectic structure of phase
space. More precisely, we observed that the emergent geometry is
defined by a one-parameter family of diffeomorphisms generated by a
smooth vector field $X_t$ satisfying $\iota_{X_t} \omega_t + A = 0$
for the change of a symplectic structure within the same cohomology
class from $\omega$ to $\omega_t = \omega + t (\omega' - \omega)$
for all $0 \leq t \leq 1$ where $\omega' - \omega = dA$. The vector
field $X_t$ is in general not a Hamiltonian flow, so any global time
cannot be assigned to the evolution of the symplectic structure
$\omega_t$. But, if there is no fluctuation of symplectic structure,
i.e., $F=dA = 0$ or $A = - dH$, there can be a globally well-defined
Hamiltonian flow. In this case we can define a global time by
introducing a unique Hamiltonian such that the time evolution is
defined by $df/dt = X_H(f) = \{f, H \}_{\theta= \omega^{-1}}$
everywhere. In particular, when the initial symplectic structure
$\omega$ is constant (homogeneous), a clock will tick everywhere at
the same rate. Note that this situation happens for the constant
background \eq{vacuum-spacetime} from which a flat spacetime emerges
as we observed in Section 3.4. But, if $\omega$ is not constant, the
time evolution will not be uniform over the space and a clock will
tick at the different rate at different places. This is consistent
with Einstein gravity since a nonconstant $\omega$ corresponds to a
curved space in our picture.

We suggest the concept of ``Time" in emergent gravity as a contact
manifold $({\bf R} \times M, \widetilde{\omega})$ where $(M,
\omega)$ is a symplectic manifold and $\widetilde{\omega} = \pi_2^*
\omega$ is defined by the projection $\pi_2: {\bf R} \times M
\to M, \; \pi_2(t,p) =p$. See Section 5.1 in \ct{mechanics}
for time dependent Hamiltonian systems. A question is then how to
recover the (local) Lorentz symmetry in the end. As we pointed out
above, if $(M, \omega)$ is a canonical symplectic manifold, i.e., $M
={\bf R}^{2n}$ and $\omega$=constant, a $(2n+1)$-dimensional Lorentz
symmetry will appear from the contact manifold $({\bf R} \times M,
\widetilde{\omega})$. (So our $(3+1)$-dimensional
Lorentzian world needs a more general argument. See the footnote
\ref{odd}.) Once again, the Darboux theorem says that there always
exists a local coordinate system where the symplectic structure is
of the canonical form. See the Table 2. Then it is quite plausible
that the local Lorentz symmetry would be recovered in the previous
way on a local Darboux chart. Furthermore, the Feynman's argument
\ct{dyson} implies that the Lorentz symmetry is just derived from
the symplectic structure on the contact manifold $({\bf R} \times M,
\widetilde{\omega})$. For example, one can recover the gauge
symmetry along the time direction by defining the Hamiltonian $H =
A_0 + H'$ and the time evolution of a spacetime geometry by the
Hamilton's equation $D_0 f \equiv df/dt + \{A_0, f
\}_{\widetilde{\theta} = \widetilde{\omega}^{-1}} =
\{f, H'\}_{\widetilde{\theta} = \widetilde{\omega}^{-1}}$. And
then one may interpret the Hamilton's equation as the infinitesimal
version of an inner automorphism like Eq.\eq{dynamical-inner}, which
was indeed used to define the vector field $D_0(X)$ in
Eq.\eq{local-vector}.

Our proposal for the emergent time is based on the fact that a
symplectic manifold $(M, \omega)$ always admits a Hamiltonian
dynamical system on $M$ defined by a Hamiltonian vector field $X_H$,
i.e., $\iota_{X_H} \omega = dH$. The purpose to pose the issue of
``Emergent Time" is to initiate and revisit this formidable issue
after a deeper understanding of emergent gravity. We refer here some
related works for future references: Our proposal is closely related
to the picture in \ct{time-gravity}, where the time is basically
defined by a one-parameter group of automorphisms of a von Neumann
algebra. Note that the deformation quantization of a Poisson
manifold \ct{kontsevich} also exhibits a similar automorphism
$D(\hbar)$ in Eq.\eq{star-auto} acting on star-products. Section 5.5
in \ct{mechanics} and Chapter 21 in \ct{big-gravity} (and references
therein) provide an exposition on infinite-dimensional Hamiltonian
systems, especially, the Hamiltonian formulation of Einstein
gravity.

\subsection{Matter fields from NC spacetime}

Now let us pose our original problem about what matters are in
emergent geometry. We will not intend to solve the problem. Instead
we will suggest a plausible picture based on the Fermi-surface
scenario in \ct{horava,volovik}. We will return to this problem with
more details in the next publication.

Particles are by definition characterized by their positions and
momenta besides their intrinsic charges, e.g., spin, isospin and an
electric charge. They should be replaced by a matter field in a
relativistic quantum theory in order to incorporate pair creations
and annihilations. Moreover, in a NC space such as
\eq{vacuum-spacetime}, the very notion of a point is replaced by a
state in the Hilbert space \eq{fock} and thus the concept of
particles (and matter fields too) becomes ambiguous. So a genuine
question is what is the most natural notion of a particle or a
corresponding matter field in the NC $\star$-algebra
\eq{matrix-basis}. We suggest it should be a K-theory object in the
sense of \ct{horava}.

Let us briefly summarize the K-theory picture in \ct{horava}. Ho\v
rava considers nonrelativistic fermions in $(d+1)$-dimensional
spacetime having $N$ complex components. Gapless excitations are
supported on a $(d-p)$-dimensional Fermi surface $\Sigma$ in
$(\mathbf{k}, \omega)$ space. Consider an inverse exact propagator
\be \la{inverse-propagator}
{\mathcal{G}_a}^{a'} = \delta_a^{a'}(i\omega -  \mathbf{k}^2/2m +
\mu) + {\Pi_a}^{a'}(\mathbf{k}, \omega)
\ee
where ${\Pi_a}^{a'}(\mathbf{k}, \omega)$ is the exact self-energy
and $a, a' = 1, \cdots, N$. Assuming that $\mathcal{G}$ has a zero
along a submanifold $\Sigma$ of dimension $d-p$ in the
$(d+1)$-dimensional $(\mathbf{k}, \omega)$ space, the question of
stability of the manifold $\Sigma$ of gapless modes reduces to the
classification of the zeros of the matrix $\mathcal{G}$ that cannot
be lifted by small perturbations ${\Pi_a}^{a'}$. Consider a sphere
$\mathbf{S}^p$ wrapped around $\Sigma$ in the transverse $p+1$
dimensions in order to classify stable zeros. The matrix
$\mathcal{G}$ is nondegenerate along this $\mathbf{S}^p$ and
therefore defines a map
\be \la{homotopy}
\mathcal{G}: \mathbf{S}^p \to GL(N,\mathbf{C})
\ee
from $\mathbf{S}^p$ to the group of nondegenerate complex $N \times
N$ matrices. If this map represents a nontrivial class in the $p$th
homotopy group $\pi_p(GL(N,\mathbf{C}))$, the zero along $\Sigma$
cannot be lifted by a small deformation of the theory. The Fermi
surface is then stable under small perturbations, and the
corresponding nontrivial element of $\pi_p(GL(N,\mathbf{C}))$
represents the topological invariant responsible for the stability
of the Fermi surface. As a premonition, we mention that it is enough
to regard the Fermi surface $\Sigma$ as a (stable) vacuum manifold
with a sharp Fermi momentum $\mathbf{p}_F$ where all small
excitations are supported, regardless of fermions themselves.

A remarkable point is that there is the so-called stable regime at
$N > p/2$ where $\pi_p(GL(N,\mathbf{C}))$ is independent of $N$. In
this stable regime, the homotopy groups of $GL(N,\mathbf{C})$ or
$U(N)$ define a generalized cohomology theory, known as K-theory
\ct{mina-moore,k-theory,harvey-k}.
In K-theory which involves vector bundles and gauge fields, any
smooth manifold $X$ is assigned an Abelian group $K(X)$. Aside from
a deep relation to D-brane charges and RR fields in string theory
\ct{mina-moore,k-theory}, the K-theory is also deeply connected with the theory
of Dirac operators, index theorem, Riemannian geometry, NC geometry,
etc. \ct{connes-ncg}.

Let us look at the action \eq{matrix-action} recalling that it
describes fluctuations around a vacuum, e.g.,
Eq.\eq{vacuum-spacetime}. One may identify the map \eq{homotopy}
with the gauge-Higgs system $(A_\mu, \Phi^a)(z)$ as the maps from
$\mathbf{R}_{C}^{d}$ to $U(N \to \infty)$. More precisely, let us
identify the $(d-p)$-dimensional Fermi surface $\Sigma$ with
$\mathbf{R}_{NC}^{2n}$ described by Eq.\eq{vacuum-spacetime} and the
$(p+1)$-dimensional transverse space with $X = \mathbf{R}_{C}^{d}$.
In this case the Fermi surface $\Sigma$ is defined by the vacuum
\eq{vacuum-spacetime} whose natural energy scale is the Planck
energy $E_{Pl}$ as we observed in Section 3.4, so the Fermi momentum
$p_F$ is basically given by $E_{Pl}$. The magic of Fermi surface
physics is that gapless excitations near the Fermi surface easily
forget the possibly huge background energy.

Now we want to consider gapless fluctuations supported on the Fermi
surface $\Sigma$. The matrix action in Eq.\eq{matrix-action} shows
that $\mathbf{R}_{C}^{d}$ is not only a hypersurface but also
supports a $U(N \to \infty)$ gauge bundle. This is the reason
\ct{k-theory,harvey-k} why $K(X)$  comes into play to classify the
topological class of excitations in the $U(N)$ gauge-Higgs system.
As we observed in Section 3.4, a generic fluctuation in
Eq.\eq{co-inner} will noticeably deform the background spacetime
lattice defined by the Fock space \eq{fock} and it will generate
non-negligible gravitational fields. But our usual concept of
particle is that it does not appreciably disturb the ambient
gravitational field. This means that the gapless excitation should
be a sufficiently localized state in $\mathbf{R}_{NC}^{2n}$. In
other words, the state is described by a compact operator in ${\cal
A}_\theta$, e.g., a Gaussian rapidly vanishing away from $y
\sim y_0$ or the matrix elements for a compact operator $\widehat{\Phi} \in {\cal A}_\theta$
in the representation \eq{matrix-expansion} are mostly vanishing
excepts a few elements. A typical example satisfying these
properties is NC solitons, e.g., GMS solitons \ct{gms}.

Since a gauge invariant observable in NC gauge theory is
characterized by its momentum variables as we discussed in Section
3.2, it will be rather useful to represent the state in momentum
space. Another natural property we impose is that it should be
stable up to pair creations and annihilations. Therefore it must be
generated by the K-theory group of the map \eq{homotopy}
\ct{mina-moore,k-theory,harvey-k}, where we will identify the NC
$\star$-algebra ${\cal A}_\theta$ with $GL(N,
\mathbf{C})$ using the relation \eq{sun-sdiff}. Note that the map
\eq{homotopy} is contractible to the group of maps from $X$ to
$U(N)$.

With the above requirements in mind, let us find an explicit
construction of a topologically non-trivial excitation. It is
well-known \ct{harvey-k} that this can be done using an elegant
construction due to Atiyah, Bott and Shapiro (ABS) \ct{abs}. The
construction uses the gamma matrices of the transverse rotation
group $SO(p,1)$ for $X = \mathbf{R}_{C}^{d}$ to construct explicit
generators of $\pi_{p}(U(N))$ where $d = p+1$. Let $X$ be even
dimensional and $S_\pm$ be two irreducible spinor representations of
$Spin(d)$ Lorentz group and $p_\mu \; (\mu = 0, 1, \cdots, p)$ be
the momenta along $X$, transverse to $\Sigma$ in $(\mathbf{k},
\omega)$. We define the gamma matrices $\Gamma^\mu:  S_+ \to S_-$
of $SO(p,1)$ to satisfy $\{ \Gamma^\mu, \Gamma^\nu \} = 2
\eta^{\mu\nu}$. At present we are considering excitations around the constant
vacuum \eq{vacuum-spacetime} and so the vacuum geometry is flat.
But, if we considered excitations in a nontrivial vacuum such as
Eq.\eq{back-b-bar}, the vacuum manifold might be curved. So the
Clifford algebra in this case would be replaced by $\{ \Gamma^\mu,
\Gamma^\nu \} = 2 g^{\mu\nu}$ where the metric $g^{\mu\nu}$ is given by
Eq.\eq{D-back-metric}. Finally we introduce an operator
$\mathcal{D}: \mathcal{H} \times S_+ \to \mathcal{H} \times S_-$
\ct{horava} such that
\be \la{dirac-op}
\mathcal{D} = \Gamma^\mu p_\mu + \cdots
\ee
which is regarded as a linear operator acting on a Hilbert space
$\mathcal{H}$, possibly much smaller than the Fock space \eq{fock},
as well as the spinor vector space $S_\pm$.

The ABS construction implies \ct{horava,volovik} that the Dirac
operator \eq{dirac-op} is a generator of $\pi_p(U(N))$ as a
nontrivial topology in momentum space $(\mathbf{k}, \omega)$ where
the low lying excitations in Eq.\eq{homotopy} near the Fermi surface
$\Sigma$ carry K-theory charges and so they are stable. Such modes
are described by coarse-grained fermions $\chi^A (\omega,
\mathbf{p}, \theta)$ with $\theta$ denoting collective coordinates
on $\Sigma$ and $\mathbf{p}$ being the spatial momenta normal to
$\Sigma$ \ct{horava}. The ABS construction determines the range
$\widetilde{N}$ of the index $A$ carried by the coarse-grained
fermions $\chi^A$ to be $\widetilde{N} = 2^{[p/2]} n
\leq N$ complex components. The precise form of the fermion $\chi^A$
depends on its K-theory charge whose explicit representation on
$\mathcal{H} \times S_\pm$ will be determined later. And we will
apply the Feynman's approach \ct{dyson} to see what the multiplicity
$n$ means. For a moment, we put $n=1$. At low energies, the
dispersion relation of the fermion $\chi^A$ near the Fermi surface
is given by the relativistic Dirac equation
\be \la{dirac-eq}
i \Gamma^\mu \p_\mu \chi + \cdots =0
\ee
with possible higher order corrections in higher energies. Thus we
get a spinor of the Lorentz group $SO(p,1)$ from the ABS
construction as a topological solution in momentum space. For
example, in four dimensions, i.e., $p=3$, $\chi^A$ has two complex
components and so it describes a chiral Weyl fermion.

Although the emergence of $(p+1)$-dimensional spinors is just a
consequence due to the fact that the ABS construction uses the
Clifford algebra to construct explicit generators of
$\pi_{p}(U(N))$, it is mysterious and difficult to understand its
physical origin. But we believe that the fermionic nature of the
excitation $\chi$ is originated from some unknown Planck scale
physics. For example, if the Dirac operator
\eq{dirac-op} is coming from GMS solitons \ct{gms} in $\mathbf{R}_{NC}^{2n}$,
the GMS solitons correspond to eigenvalues of $N \times N$ matrices
in Eq.\eq{sun-sdiff}. As is well known from $c=1$ matrix models, the
eigenvalues behave like fermions, although it is the
(1+1)-dimensional sense, after integrating out off-diagonal
interactions. Another evidence is the stringy exclusion principle
\ct{stringy-exclusion} that the AdS/CFT correspondence puts a
limit on the number of single particle states propagating on the
compact spherical component of the $AdS_p \times \mathbf{S}^q$
geometry which corresponds to the upper bound on $U(1)$ charged
chiral primaries on the compact space $\mathbf{S}^q$.

It should be important to clearly understand the origin of the
fermionic nature of particles arising from the vacuum
\eq{vacuum-spacetime}. The crux seems to be the mysterious connection
between the Clifford modules and K-theories \ct{abs}. Another
related problem is that we didn't yet understand the dynamical
origin of the particle symplectic structure \eq{symplectic-phase}.
Is it similarly possible to get some insight about the particle mass
and dark matters from the dynamical origin of the symplectic
structure \eq{symplectic-phase} as we did in Section 3.4 for the
dark energy ? If the vacuum \eq{vacuum-spacetime} acts as a Fermi
surface for quarks and leptons, is it a symptom that the local
electroweak symmetry can be broken dynamically without Higgs ?

Now let us address the problem how to determine the multiplicity $n$
of the coarse-grained fermions $\chi^{\alpha a}$ where we decomposed
the index $A = (\alpha a)$ with $\alpha$ the spinor index of the
$SO(d)$ Lorentz group and $a = 1, \cdots, n$ an internal index of an
$n$-dimensional representation of some compact symmetry $G$. One may
address this problem by considering the quantum particle dynamics on
$X \times \Sigma$ and repeating the Feynman's question. To be
specific, we restrict (collective) coordinates of $\Sigma$, denoted
as $Q^I \; (I = 1, \cdots, n^2-1)$, to Lie algebra variables such as
the particle isospins or colors. So the commutation relations we
consider are
\bea \la{sun-alg-1}
&& [Q^I, Q^J] = i f^{IJK} Q^K, \\
\la{sun-alg-2}
&& [q^i, Q^I] = 0
\eea
together with the commutation relations \eq{feynman-comm} determined
by the symplectic structure \eq{symplectic-phase} on $T^*
\mathbf{R}^p$.

Then the question is: What is the most general form of forces
consistent with the commutation relations \eq{feynman-comm},
\eq{sun-alg-1} and \eq{sun-alg-2} ? It was already answered in
\ct{feynman-qed} that the answer is just the non-Abelian version of
the Lorentz force law \eq{force-law-time} with an additional set of
equations coming from the condition that the commutation relation
\eq{sun-alg-2} should be preserved during time evolution, i.e.,
$\frac{d}{dt} [q^i, Q^I] = 0$. This condition can be solved by the
so-called Wong's equations
\be \la{wong}
\dot{Q}^I + f^{IJK} A_i^J Q^K \dot{q}_i = 0.
\ee
The Wong's equations just say that the internal charge $Q^I$ is
parallel-transported along the trajectory of the particle under the
influence of the non-Abelian gauge field $A_i^J$.

Therefore the quantum particle dynamics on $X \times \Sigma$
naturally requires to introduce non-Abelian gauge fields in the
representation of the Lie algebra \eq{sun-alg-1}. And the dynamics
of the particle carrying an internal charge in $\Sigma$ should be
defined by a symplectic structure on $T^* X \times \Sigma$. But note
that we have a natural symplectic structure on $\Sigma$ defined by
Eq.\eq{vacuum-spacetime}. Also note that we have only $U(1)$ gauge
fields on $X \times \Sigma$ in Eq.\eq{decomp-cov}. So the problem is
how to get the Lie algebra generators in Eq.\eq{sun-alg-1} from the
space $\Sigma = \mathbf{R}_{NC}^{2n}$ and how to get the non-Abelian
gauge fields $A_\mu^I(z)$ on $X$ from the $U(1)$ gauge fields on $X
\times \Sigma$ where $z^\mu = (t, q^i)$.

The problem is solved by noting that the $n$-dimensional harmonic
oscillator in quantum mechanics can realize $SU(n)$ symmetries (see
the Chapter 14 in \ct{georgi}). The generators of the $SU(n)$
symmetry on the Fock space \eq{fock} are given by
\be \la{sun}
Q^I = a^\dagger_i T^I_{ik} a_k
\ee
where the creation and annihilation operators are given by
Eq.\eq{vacuum-spacetime} and $T^I$'s are constant $n \times n$
matrices satisfying $[T^I, T^J] = i f^{IJK} T^K$ with the same
structure constants as Eq.\eq{sun-alg-1}. It is easy to check that
the $Q^I$'s satisfy the $SU(n)$ Lie algebra \eq{sun-alg-1}. We
introduce the number operator $Q^0 \equiv a^\dagger_i a_i$ and
identify with a $U(1)$ generator. The operator $\mathfrak{C} =
\sum_I Q^I Q^I$ is the quadratic Casimir operator of the $SU(n)$ Lie
algebra and commutes with all $Q^I$'s. Thus one may identify
$\mathfrak{C}$ with an additional $U(1)$ generator.

Let $\rho(V)$ be a representation of the Lie algebra
\eq{sun-alg-1} in a vector space $V$. We take an $n$-dimensional
representation in $V = \mathbf{C}^n$ or precisely $V =
L^2(\mathbf{C}^n)$, a square integrable Hilbert space. Now we expand
the $U(1)$ gauge field $\widehat{A}_M(z,y)$ in Eq.\eq{decomp-cov} in
terms of the $SU(n)$ basis \eq{sun}
\bea \la{sun-expansion}
\widehat{A}_M(z,y) &=& \sum_{n=0}^\infty \sum_{I_i \in \rho(V)}  A^{I_1 \cdots
I_n}_M (z, \rho, \lambda_n)
\; Q^{I_1} \cdots Q^{I_n} \xx
&=& A_M(z) + A^I_M (z, \rho, \lambda_1)\; Q^I + A^{IJ}_M (z,
\rho, \lambda_2)\; Q^I Q^J + \cdots
\eea
where $\rho$ and $\lambda_n$ are eigenvalues of $Q^0$ and
$\mathfrak{C}$, respectively, in the representation $\rho(V)$. The
expansion \eq{sun-expansion} is formal but it is assumed that each
term in Eq.\eq{sun-expansion} belongs to the irreducible
representation of $\rho(V)$. Thus we get $SU(n)$ gauge fields
$A^I_\mu$ as well as adjoint scalar fields $A^I_a$ in addition to
$U(1)$ gauge fields $A_M(z)$ as low lying excitations.

Note that the coarse-grained fermion $\chi$ in Eq.\eq{dirac-eq}
behaves as a relativistic particle in the spacetime $X = {\bf
R}_C^d$ and a stable excitation as long as the Fermi surface
$\Sigma$ is topologically stable. In addition to these fermionic
excitations, there will also be bosonic excitations arising from
changing the position in $X$ of the surface $\Sigma$ or deformations
of the surface $\Sigma$ itself. But the latter effect (as
gravitational fields in $\Sigma$) will be very small and so can be
ignored since we are interested in the low energy behavior of the
Fermi surface $\Sigma$. Then the gauge fields in
Eq.\eq{sun-expansion} represent collective modes for the change of
the position in $X = {\bf R}_C^d$ of the surface $\Sigma$
\ct{volovik}. They can be regarded as collective dynamical fields in
the vicinity of the Fermi surface $\Sigma$ acting on the fermions in
Eq.\eq{dirac-eq}.

Therefore we regard the Dirac operator \eq{dirac-op} as an operator
$\mathcal{D}: \mathcal{H} \times S_+ \to \mathcal{H} \times S_-$
where $\mathcal{H} = L^2(\mathbf{C}^n)$ and introduce a minimal
coupling with the $U(1)$ and $SU(n)$ gauge fields in
Eq.\eq{sun-expansion} by the replacement $p_\mu \to p_\mu - e A_\mu
- A^I_\mu Q^I$. Then the Dirac equation
\eq{dirac-eq} becomes
\be \la{dirac-eq-gauge}
i \Gamma^\mu (\p_\mu -i e A_\mu - i A^I_\mu Q^I) \chi + \cdots =0.
\ee
Here we see that the coarse-grained fermion $\chi$ in the homotopy
class $\pi_p (U(N))$ is in the fundamental representation of
$SU(n)$. So we identify the multiplicity $n$ in the ABS construction
\eq{dirac-eq} with the number of colors. Unfortunately the role of
the adjoint scalar fields in Eq.\eq{sun-expansion} is not obvious
from the Feynman's approach.

The most interesting case in Eq.\eq{matrix-action} is that $p=3$ and
$n=3$, that is, 10-dimensional NC $U(1)$ gauge theory on
$\mathbf{R}_C^4 \times \mathbf{R}_{NC}^6$. In this case
Eq.\eq{dirac-eq-gauge} is the 4-dimensional Dirac equation where
$\chi$ is a quark, an $SU(3)$ multiplet of chiral Weyl fermions,
coupling with gluons $A^I_\mu(z)$, $SU(3)$ gauge fields for the
color charge $Q^I$, as well as photons $A_\mu(z)$, $U(1)$ gauge
fields for the electric charge $e$. One may consider a similar ABS
construction in the vector space $V = \mathbf{C}^2 \times
\mathbf{C}$, i.e., by breaking the $SU(3)$ symmetry to $SU(2) \times
U(1)$, to get $SU(2)$ gauge fields and chiral Weyl fermions. In this
case $Q^I \;(I=1,2,3)$ in Eq.\eq{sun} are the famous Schwinger
representation of $SU(2)$ Lie algebra.

\section{Musing on Noncommutative Spacetime}

It is a well-accepted consensus that at very short distances, e.g.,
the Planck scale $L_P$, the spacetime is no longer commutating due
to large quantum effects and a NC geometry will play a role at short
distances. In addition, the spacetime geometry at the Planck scale
is not fixed but violently fluctuating, as represented as spacetime
foams. Therefore the NC geometry arising at very short distances has
to be intimately related to quantum gravity. The Moyal space
\eq{nc-spacetime} is the simplest and the most natural example of NC
spacetime. Thus it should be expected that the physical laws defined
in the NC spacetime \eq{nc-spacetime}, for instance, a NC field
theory, essentially refer to a theory of (quantum) gravity. This is
the reason why the $\theta$-deformation in the Table 1 must be
radical as much as the $\hbar$-deformation.

Unfortunately, the NC field theory has not been explored as a theory
of gravity so far. It has been studied as a theory of particles
within the conventional framework of quantum field theory. But we
have to recognize that the NC field theory is a quantum field theory
defined in a highly nontrivial vacuum \eq{vacuum-spacetime}. It
should be different from usual quantum field theories defined in a
trivial vacuum. So we should be careful to correctly identify order
parameters for fluctuations around the vacuum \eq{vacuum-spacetime}.
We may have a wrong choice of the order parameter if we naively
regard the NC field theory as a theory of particles only. As an
illustrating example, in order to describe the superconductivity at
$T \lesssim T_c$, it is important to consider an effect of the
background lattice and phonon exchange with electrons. The
interaction of electrons with the background lattice is resulted in
a new order parameter, the so-called Cooper pairs, and a new
attractive force between them. We know that it is impossible to have
a bound state of two electrons, the Cooper pair, in a trivial
vacuum, i.e., without the background lattice. Thus the
superconductivity is an emergent phenomenon from electrons moving in
a nontrivial background lattice.

We observed that the vacuum \eq{vacuum-spacetime} endows the
spacetime with a symplectic structure whose surprising consequences,
we think, have been considerably explored in this paper. For
example, it brings to the correspondence \eq{sun-sdiff} implying the
large $N$ duality or the gauge/gravity duality. These features do
not arise in ordinary quantum field theories. So it would be
desirable to seriously contemplate about the theoretical structure
of NC field theories from the spacetime point of view.

\subsection{Graviton as a Cooper pair}

Graviton is a spin-2 particle. Therefore the emergent gravity, if
the picture is true, should come from a composite of two spin-1
gauge bosons, not from gauge fields themselves.\footnote{We thank
Piljin Yi for raising this critical issue.} Unfortunately, there is
no rigorous proof that the bound state of two spin-1 gauge bosons
does exist in NC spacetime. But an interesting point is that NC
spacetime is more preferable to the formation of bound states
compared to commutative spacetime. See, for example,
\ct{nc-bound}. Salient examples are GMS solitons \ct{gms} and NC
$U(1)$ instantons \ct{nc-instanton}, which are not allowed in a
commutative spacetime. Furthermore there are many logical evidences
that it will be true, especially inferred from the matrix
formulation of NC gauge theory as we briefly discuss below.

For definiteness, let us consider the case with $d=4$ and $n=3$ for
the action \eq{matrix-action}, that is, 10-dimensional NC $U(1)$
gauge theory on $\mathbf{R}_C^4 \times \mathbf{R}_{NC}^6$. The
matrix representation in the action \eq{matrix-action} is precisely
equal to the bosonic part of 4-dimensional ${\cal N} = 4$
supersymmetric $U(N)$ Yang-Mills theory which is known to be
equivalent to the type IIB string theory on $AdS_5 \times {\bf S}^5$
space \ct{ads-cft}. Therefore the 10-dimensional gravity emergent
from NC gauge theory will essentially be the same as the one in the
AdS/CFT duality. The bulk graviton $g_{\mu\nu}(z,\rho)$ in the
AdS/CFT duality, whose asymptotics at $\rho = 0$ is given by the
metric \eq{ads5}, is defined by the coupling to the energy-momentum
tensor $T_{\mu\nu}(z)$ in the $U(N)$ gauge theory. The
energy-momentum tensor $T_{\mu\nu}(z)$ is a spin-2 composite
operator in the gauge theory rather than a fundamental field. This
means that the bulk graviton is holographically defined as a bound
state of two spin-1 gauge bosons. Schematically, we have the
following relation
\be \la{1+1=2}
(1 \otimes 1)_S \; \rightleftarrows \; 2 \oplus 0 \quad {\rm or}
\quad \subset \otimes \supset \; \rightleftarrows \; \bigcirc.
\ee

Indeed the core relation \eq{1+1=2} has underlain the unification
theories since Kaluza and Klein. In early days people have tried the
scheme $(\leftarrow)$ under the name of the Kaluza-Klein theory. A
basic idea in the Kaluza-Klein theory (including string theory) is
to construct spin-1 gauge fields plus gravity in lower dimensions
from spin-2 gravitons in higher dimensions. An underlying view in
this program is that a ``fundamental" theory exists as a theory of
gravity in higher dimensions and a lower dimensional theory of
spin-1 gauge fields is derived from the higher dimensional
gravitational theory. Though it is mathematically beautiful and
elegant, it seems to be physically unnatural if the higher spin
theory should be regarded as a more fundamental theory.

After the discovery of D-branes in string theory, people have
realized that the scheme $(\rightarrow)$ is also possible, which is
now known as the open-closed string duality or the gauge/gravity
duality. But the scheme $(\rightarrow)$ comes into the world in a
delicate way since there is a general no-go theorem known as the
Weinberg-Witten theorem \ct{ww,florian}, stating that an interacting
graviton cannot emerge from an ordinary quantum field theory in the
same spacetime. One has to notice, however, that Weinberg and Witten
introduced two basic assumptions to prove this theorem. The first
hidden assumption is that gravitons and gauge fields live in the
same spacetime. The second assumption is the existence of a
Lorentz-covariant stress-energy tensor. The AdS/CFT duality
\ct{ads-cft} realizes the emergent gravity by relaxing the first
assumption in the way that gravitons live in a higher dimensional
spacetime than gauge fields. As we observed in Section 3.4, the NC
field theory is even more radical in the sense that the Lorentz
symmetry is not a fundamental symmetry of the theory but emergent
from the vacuum algebra \eq{vacuum-spacetime} defined by a uniform
configuration of NC gauge fields.

Another ingredient supporting the existence of the spin-2 bound
states is that the vacuum \eq{vacuum-spacetime} in NC gauge theory
signifies the spontaneous symmetry breaking of the
$\Lambda$-symmetry \eq{lambda-symm} \ct{hsy2}. If one considers a
small fluctuation around the vacuum \eq{vacuum-spacetime}
parameterized by Eq.\eq{vacuum-coordinate}, the spacetime metric
given by Eq.\eq{ward-metric} looks like
\be \la{small-metric}
g_{MN} = \eta_{MN} + h_{MN}
\ee
where $\eta_{MN} = \langle g_{MN} \rangle$ is the flat metric
determined by the uniform condensation of gauge fields in the
vacuum. As a fluctuating (quantum) field, the existence of the
vacuum expectation value in the metric $\langle g_{MN}
\rangle = \eta_{MN}$ also implies some sort of
spontaneous symmetry breaking as Zee anticipated in \ct{zee} (see
the footnote \ref{zee}). We see here that they indeed have the same
origin. If one look at the Table 2, one can see a common property
that both a Riemannian metric $g$ and a symplectic structure
$\omega$ should be nondegenerate, i.e., nowhere vanishing on $M$. In
the context of physics where $g$ and $\omega$ are regarded as a
field, the nondegeneracy means a nonvanishing vacuum expectation
value. We refer to \ct{hsy2} more discussions about the spontaneous
symmetry breaking.

Instead we will discuss an interesting similarity between the BCS
superconductivity \ct{bcs} and the emergent gravity to get some
insight into the much more complicated spontaneous symmetry breaking
for the $\Lambda$-symmetry \eq{lambda-symm}. A superconductor of any
kind is nothing more or less than a material in which the $G = U(1)$
gauge symmetry is spontaneously broken to $H = \mathbf{Z}_2$ which
is the $180^o$ phase rotation preserved by Cooper pairs
\ct{weinberg-witten}. The spontaneous breakdown of electromagnetic gauge
invariance arises because of attractive forces between electrons via
lattice vibrations near the Fermi surface. In consequence of this
spontaneous symmetry breaking, products of any even number of
electron fields have non-vanishing expectation values in a
superconductor, captured by the relation $\half \otimes \half = 0
\oplus 1$. As we mentioned above, the emergent gravity reveals a similar
pattern of spontaneous symmetry breaking though much more
complicated where the $\Lambda$-symmetry
\eq{lambda-symm}, or equivalently $G =$ Diff(M), is spontaneously
broken to the symplectomorphism \eq{semi-gauge-tr}, or equivalently
$H = U(1)_{NC}$ gauge symmetry. The spontaneous breakdown of the
$\Lambda$-symmetry or $G =$ Diff(M) is induced by the condensate
\eq{vacuum-spacetime} of gauge fields in a vacuum and conceivably the
vacuum \eq{vacuum-spacetime} can act as a Fermi surface for low
energy excitations, as we discussed in Section 4.2.

Then we may find a crude but inciting analogy between the BCS
superconductivity and the emergent gravity:

\begin{center}
\begin{tabular}{|c|c|c|}
  \hline
   Theory  & Superconductivity & Emergent gravity \\
  \hline
  Microscopic degree of freedom & electron & gauge field \\
  \hline
  Order parameter & Cooper pair & graviton \\
  \hline
  G & $U(1)$ &  Diff(M) \\
  \hline
  H  & $\mathbf{Z}_2$  & $U(1)_{NC}$ \\
  \hline
  Control parameter & $ \frac{T_c}{T} - 1 $ & $\theta^{ab}$ \\
  \hline
  Macroscopic description & Laudau-Ginzburg & Einstein gravity \\
  \hline
  Microscopic description & BCS & gauge theory \\
  \hline
\end{tabular} \\
\vspace{0.5cm}
Table 3.  Superconductivity vs. Emergent gravity
\end{center}

The Landau-Ginzburg theory is a phenomenological theory of
superconductivity where the free energy of a superconductor near $T
\approx T_c$ can be expressed in terms of a complex order parameter,
describing Cooper pairs \ct{weinberg-witten}. Of course this
situation is analogous to the emergent gravity in the sense that
Einstein gravity as a macroscopic description of NC gauge fields is
manifest only at the commutative limit, i.e., $|\theta| \to 0$.
Although we should be cautious to employ the analogy in the Table 3,
it may be worthwhile to remark that the flux tubes or Abrikosov
vortices in type II superconductors, realized as a soliton solution
in the Landau-Ginzburg theory, seem to be a counterpart of black
holes in the emergent gravity. We think the Table 3 could serve as a
guidepost more than a plain analogy to understand a detailed
structure of emergent gravity.

\subsection{Fallacies on noncommutative spacetime}

As was remarked before, a NC spacetime arises as a result of large
quantum fluctuations at very short distances. So the conventional
spacetime picture gained from a classical and weak gravity regime
will not be naively extrapolated to the Planck scale. Indeed we
perceived that a NC geometry reveals a novel, radically different
picture about the origin of spacetime.

But the orthodox approach so far has regarded the NC spacetime
described by Eq.\eq{vacuum-spacetime}, for instance, as an
additional background condensed on an already existing spacetime.
For example, field theories defined on the NC spacetime have been
studied from the conventional point of view based on the traditional
spacetime picture. Then the NC field theory is realized with
unpleasant features, breaking the Lorentz symmetry and locality
which are two fundamental principles underlying quantum field
theories. A particle in local quantum field theories is defined as a
state in an irreducible representation of the Poincar\'e symmetry
and internal symmetries. This concept of the particle becomes
ambiguous in the NC field theory due to not only the lack of the
Lorentz symmetry but also the non-Abelian nature of spacetime.
Furthermore the nonlocality in NC field theories appears as a
perplexing UV/IR mixing in nonplanar Feynman diagrams in
perturbative dynamics \ct{uv-ir}. This would appear to spoil the
renormalizability of these theories \ct{nc-review}.

Therefore the NC field theory is not an eligible generalization of
quantum field theory framework as a theory of particles. However,
these unpleasant aspects of the NC field theory turn into a welcome
property or turn out to be a fallacy whenever one realizes it as a
theory of gravity. We believe that the nonperturbative dynamics of
gravity is intrinsically nonlocal. A prominent evidence is coming
from the holographic principle \ct{holography} which states that
physical degrees of freedom in gravitational theories reside on a
lower dimensional screen where gauge fields live. The AdS/CFT
duality \ct{ads-cft} is a thoroughly tested example of the
holographic principle. Recently it was shown in
\ct{steinacker,harald-wien} that the UV/IR mixing in NC gauge
theories can be interpreted as a manifestation of gravitational
nonlocality in the context of emergent gravity. This elegant shift
of wing signifies an internal consistency of emergent gravity.

The basic idea of emergent gravity is to view the gravity as a
collective phenomenon of gauge fields. According to Einstein, the
gravity is nothing but the dynamics of spacetime geometry. This
perspective implies that there is no prescribed notion of spacetime.
The spacetime must also be emergent from or defined by gauge fields
if the picture is anyway correct. We observed in Section 3.4 that
the emergent gravity reveals a novel and consistent picture about
the dynamical origin of spacetime. The most remarkable angle is the
dynamical origin of flat spacetime, which is absent in Einstein
gravity. It turned out that the Lorentz symmetry as well as the flat
spacetime is not {\it a priori} given in the beginning but emergent
from or defined by the uniform condensation
\eq{vacuum-spacetime} of gauge fields. In the prospect,
the Lorentz symmetry is not broken by the background
\eq{vacuum-spacetime} but rather emergent at the cost of huge energy
condensation in the vacuum. Thus the emergent gravity also comes to
the rescue of the Lorentz symmetry breaking in NC field theories.

But we want to point out an intriguing potential relation between
the dark energy \eq{vac-fluctuation} and a possible tiny violation
of the Lorentz symmetry. We observed that the energy density
\eq{vac-fluctuation} is due to the cosmological vacuum fluctuation around the flat
spacetime and does generate an observable effect of spacetime
structure, e.g., an expansion of universe.  Furthermore, since the
tiny energy \eq{vac-fluctuation} represents a deviation from the
flat spacetime over the cosmological scale, it may have another
observable effect of spacetime structure; a very tiny violation of
the Lorentz symmetry. Amusingly, the dark energy scale $ \sim 2
\times 10^{-3} eV$ given by \eq{dark-energy-obs} is of the same order of magnitude
as the neutrino mass. This interesting numerical coincidence may
imply some profound relation between the neutrino mass and the tiny
violation of the Lorentz symmetry \ct{neutrino-mass}.

\section{Discussion}

\begin{quote}
{\it Mathematicians do not study objects, but relations between
objects. Thus, they are free to replace some objects by others so
long as the relations remain unchanged. Content to them is
irrelevant: they are interested in form only.}

-- Henri Poincar\'e (1854-1912)
\end{quote}

Recent developments in string and M theories, especially, after the
discovery of D-branes, have constantly revealed that string and M
theories are not very different from quantum field theories. Indeed
a destination of nonperturbative formulations of string and M
theories has often been quantum field theories again. For instance,
the AdS/CFT duality and the matrix models in string and M theories
are only a few salient examples. It seems to insinuate a message
that quantum field theories already contain `quantum gravity' in
some level. At least we have to contemplate our credulous belief
that the string and M theories should be superordinate to quantum
field theories. Certainly we are missing the first (dynamical)
principle to derive the quantum gravity from quantum field theories.

Throughout the paper, we have emphasized that quantum field theories
in NC spacetime are radically different from their commutative
counterparts and they should be regarded as a theory of gravity
rather than a theory of particles. So the important message we want
to draw is that the $\theta$-deformation in the Table 1 should be
seriously considered as a foundation for quantum gravity. In other
words, the first principle would be the geometrization of gauge
fields based on the symplectic and NC geometry. It my be possible
that the NC geometry also underlies the fundamentals of string
theory.

In this paper, we have mostly focussed on the commutative limit,
$\theta \to 0$, where the Einstein gravity manifests itself as a
macroscopic spacetime geometry of NC $\star$-algebra defined by
gauge fields in NC spacetime. That is, Einstein gravity is just a
low energy effective theory of NC gauge fields or large $N$
matrices. So we naturally wonder what happens in a deep NC space. An
obvious guess is that a usual concept of spacetime based on a smooth
geometry will be doomed. Instead an operator algebra, e.g.,
$\star$-algebra defined by NC gauge fields, will define a relational
fabric between NC gauge fields, whose prototype at macroscopic world
emerges as a smooth spacetime geometry. In a deep NC space, an
algebra between objects is more fundamental. A geometry is a
secondary concept defined by the algebra. Indeed the motto in
emergent gravity is that an algebra defines a geometry. In this
scheme, one has to specify an underlying algebra to talk about a
corresponding geometry. So the Poincar\'e's declaration above may
also refer to physicists who are studying quantum gravity.

\section*{Acknowledgments}

We are grateful to Chiang-Mei Chen, Kuerak Chung, Hoil Kim, Kyung
Kiu Kim, Otto Kong, John Madore, Stefan M\"uller, John J. Oh, Kwon
Park, M. Sivakumar and Harold Steinacker for helpful discussions at
some stage of this work.

\newpage

\appendix

\section{A Proof of the Equivalence between Self-dual NC Electromagnetism
and Self-dual Einstein Gravity}

Here we present a self-contained and friendly proof of the
equivalence between self-dual NC electromagnetism and self-dual
Einstein gravity \ct{hsy1}. Our proof here closely follows the
result in \ct{mason-newman} applying our observation
\eq{co-inner}, of course, decisive for the equivalence, that NC
gauge fields can be mapped to (generalized) vector fields through
the inner automorphism \eq{dynamical-inner}. The self-dual case here
will be a useful guide for deriving the general equivalence between
NC $U(1)$ gauge theories and Einstein gravity presented in Appendix
B.

We introduce at each spacetime point in $M$ a local frame of
reference in the form of 4 linearly independent vectors (vierbeins
or tetrads) $E_A = E_A^M \p_M \in TM$ which are chosen to be
orthonormal, i.e., $E_A \cdot E_B = \delta_{AB}$. The basis $\{ E_A
\}$ determines a dual basis $E^A = E^A_M dX^M \in T^*M$ by
\be \la{dual-vector}
\langle E^A, E_B \rangle = \delta^A_B.
\ee
The above pairing leads to the relation $E^A_M E_B^M = \delta^A_B$.
The metric is the most basic invariant defined by the vectors in
$TM$ or $T^*M$,
\bea \la{inverse-metric}
\Bigl(\frac{\p}{\p s}\Bigr)^2 &=& \delta^{AB} E_A \otimes E_B
= \delta^{AB} E_A^M E_B^N \; \p_M \otimes \p_N \xx &\equiv&
g^{MN}(X) \; \p_M \otimes \p_N
\eea
or
\bea \la{4-metric}
ds^2 &=& \delta_{AB} E^A \otimes E^B = \delta_{AB} E^A_M E^B_N
\; dX^M \otimes dX^N \xx &\equiv& g_{MN}(X) \; dX^M \otimes dX^N.
\eea

Under local frame rotations in $SO(4)$ the vectors transform
according to
\bea \la{frame-rotation}
&& E_A(X) \to E_A^\prime(X) = E_B(X) {\Lambda^B}_A(X), \xx && E^A(X)
\to {E^\prime}^A(X) =   {\Lambda^A}_B(X)E^B(X)
\eea
where ${\Lambda^A}_B(X) \in SO(4)$. The spin connections
$\omega_M(X)$ constitute gauge fields with respect to the local
$SO(4)$ rotations
\be \la{spin-so4}
\omega_M \to \Lambda \omega_M \Lambda^{-1} + \Lambda \p_M \Lambda^{-1}
\ee
and the covariant derivative is defined by
\bea \la{spin-cov}
&& {\cal D}_M E_A = \p_M E_A - {{\omega_M}^B}_A E_B, \xx && {\cal
D}_M E^A = \p_M E^A + {{\omega_M}^A}_B E^B.
\eea

The connection one-form ${\omega^A}_B = {{\omega_M}^A}_B dX^M$
satisfies the Cartan's structure equations \ct{big-gravity},
\bea \la{cartan-torsion}
&& {T_{MN}}^A = \p_M E_N^A - \p_N E_M^A + {{\omega_M}^A}_B E_N^B -
{{\omega_N}^A}_B E_M^B, \\
\la{cartan-curvature}
&& {{R_{MN}}^A}_B = \p_M {{\omega_N}^A}_B - \p_N {{\omega_M}^A}_B+
{{\omega_M}^A}_C {{\omega_N}^C}_B - {{\omega_N}^A}_C
{{\omega_M}^C}_B,
\eea
where we introduced the torsion two-form $T^A = \half {T_{MN}}^A
dX^M \wedge dX^N$ and the curvature two-form ${R^A}_B = \half
{{R_{MN}}^A}_B dX^M \wedge dX^N$. Now we impose the torsion free
condition, ${T_{MN}}^A = {\cal D}_M E_N^A - {\cal D}_N E_M^A = 0$,
to recover the standard content of general relativity, which
eliminates $\omega_M$ as an independent variable, i.e.,
\bea \la{spin-connection}
\omega_{MBC} &=& \half E_M^A(f_{ABC} - f_{BCA} + f_{CAB}) \xx
&=& - \omega_{MCB},
\eea
where
\be \la{f}
f_{ABC} = E_A^M E_B^N (\p_M E_{NC} - \p_N E_{MC}).
\ee
Note that ${f_{AB}}^C$ are the structure functions of the vectors
$E_A \in TM$ defined by
\be \la{structure-fn}
[E_A,E_B] =  - {f_{AB}}^C E_C.
\ee
Here raising and lowering the indices $A,B,\cdots$ are insignificant
with Euclidean signature but we have kept track of the position of
the indices for another use with Lorentzian signature.

Since the spin connection $\omega_{MAB}$ and the curvature tensor
$R_{MNAB}$ are antisymmetric on the $AB$ index pair, one can
decompose them into a self-dual part and an anti-self-dual part as
follows
\bea \la{spin-sd-asd}
&& \omega_{MAB} = \omega_M^{(+)a} \eta^a_{AB} + \omega_M^{(-)a}
\bar{\eta}^a_{AB}, \\
\la{curvature-sd-asd}
&& R_{MNAB} = F_{MN}^{(+)a} \eta^a_{AB} + F_{MN}^{(-)a}
\bar{\eta}^a_{AB}
\eea
where the $4 \times 4$ matrices $\eta^a_{AB}$ and ${\bar
\eta}^a_{AB}$ for $a=1,2,3$ are 't Hooft symbols defined by
\bea \la{tHooft-symbol}
&& {\bar \eta}^a_{ij} = {\eta}^a_{ij} = {\varepsilon}_{aij}, \qquad
i,j \in
\{1,2,3\}, \nonumber\\
&& {\bar \eta}^a_{4i} = {\eta}^a_{i4} = \delta_{ai}.
\eea
We list some identities of the 't Hooft tensors
\bea
\la{self-eta}
&& \eta^{(\pm)a}_{AB} = \pm \half {\varepsilon_{AB}}^{CD}
\eta^{(\pm)a}_{CD}, \\
\la{proj-eta}
&& \eta^{(\pm)a}_{AB}\eta^{(\pm)a}_{CD} =
\delta_{AC}\delta_{BD}
-\delta_{AD}\delta_{BC} \pm
\vare_{ABCD}, \\
\la{self-eigen}
&& \vare_{ABCD} \eta^{(\pm)a}_{DE} = \mp (
\delta_{EC} \eta^{(\pm)a}_{AB} + \delta_{EA} \eta^{(\pm)a}_{BC} -
\delta_{EB} \eta^{(\pm)a}_{AC} ), \\
\la{eta-etabar}
&& \eta^{(\pm)a}_{AB} \eta^{(\mp)b}_{AB}=0, \\
\la{eta^2}
&& \eta^{(\pm)a}_{AC}\eta^{(\pm)b}_{BC} =\delta^{ab}\delta_{AB} +
\vare^{abc}\eta^{(\pm)c}_{AB}, \\
\la{eta-ex}
&& \eta^{(\pm)a}_{AC}\eta^{(\mp)b}_{BC} =
\eta^{(\mp)b}_{AC}\eta^{(\pm)a}_{BC}
\eea
where $\eta^{(+)a}_{AB} = \eta^a_{AB}$ and $\eta^{(-)a}_{AB} = {\bar
\eta}^a_{AB}$. (Since the above 't Hooft tensors are defined in
Euclidean ${\bf R^4}$ where the flat metric $\delta_{AB}$ is used,
we don't concern about raising and lowering the indices.)

Using the identities \eq{eta^2} and \eq{eta-ex}, it is easy to see
that the (anti-)self-dual curvature in Eq.\eq{curvature-sd-asd} is
purely determined by the (anti-)self-dual spin connection without
any mixing, i.e.,
\be \la{su2-curvature}
F_{MN}^{(\pm)a} = \p_M \omega_N^{(\pm)a} - \p_N \omega_M^{(\pm)a} -
2
\vare^{abc} \omega_M^{(\pm)b}\omega_N^{(\pm)c}.
\ee
Of course all these separations are due to the fact, $SO(4)= SU(2)_L
\times SU(2)_R$, stating that any $SO(4)$ rotations can be decomposed
into self-dual and anti-self-dual rotations. Since
$\varepsilon^{abc}$ is the structure constant of $SU(2)$ Lie
algebra, i.e., $[\tau^a, \tau^b] = 2 i \varepsilon^{abc} \tau^c$
where $\tau^a$'s are the Pauli matrices, one may identify
$\omega_M^{(\pm)a}$ with $SU(2)_{L,R}$ gauge fields and
$F_{MN}^{(\pm)a}$ with their field strengths.

In consequence we have arrived at the following important property.
If the spin connection is self-dual, i.e., $\omega_M^{(-)a}=0$, the
curvature tensor is also self-dual, i.e., $F_{MN}^{(-)a}=0$.
Conversely, if the curvature is self-dual, i.e., $F_{MN}^{(-)a}=0$,
one can always choose a self-dual spin connection by a suitable
gauge choice since $F_{MN}^{(-)a}=0$ requires that $\omega_M^{(-)a}$
is a pure gauge. Therefore, in this self-dual gauge, the problem of
finding a self-dual solution to the Einstein equation \ct{gibb-hawk}
\be \la{g-instanton}
R_{MNAB} = \pm \half {\varepsilon_{AB}}^{CD} R_{MNCD}
\ee
is equivalent to one of finding self-dual spin connections
\be \la{sde-spin}
\omega_{EAB} = \pm \half {\varepsilon_{AB}}^{CD} \omega_{ECD}
\ee
where $\omega_{CAB} = E_C^M \omega_{MAB}$. Note that a metric
satisfying the self-duality equation \eq{g-instanton}, known as the
gravitational instanton, is necessarily Ricci-flat because
${R_{MBA}}^B = \pm \frac{1}{6} {\varepsilon_{A}}^{BCD} R_{M[BCD]} =
0$. The gravitational instantons defined by Eq.\eq{g-instanton} are
then obtained by solving the first-order differential equations
given by Eq.\eq{sde-spin}.

Now contracting ${\varepsilon_F}^{EAB}$ on both sides of
Eq.\eq{sde-spin} leads to the relation
\be \la{equiv-1}
\omega_{[ABC]} = \mp {\varepsilon_{ABC}}^D \phi_D
\ee
where $\phi_D = {\omega_{ED}}^E$ and $\omega_{[ABC]} =
\omega_{ABC} + \omega_{BCA} + \omega_{CAB}$.
From Eqs.\eq{spin-connection} and
\eq{f} together with Eq.\eq{equiv-1}, we get
\bea \la{equiv-2}
f_{ABC} &=& \omega_{ABC} - \omega_{BAC} = - \omega_{ACB} -
\omega_{BAC} - \omega_{CBA} + \omega_{CBA} \xx &=& \pm
{\varepsilon_{ACB}}^D \phi_D - \omega_{CAB}
\eea
and so
\be \la{equiv-3}
- \omega_{CAB} = f_{ABC} \pm {\varepsilon_{ABC}}^D \phi_D.
\ee
The self-duality equation \eq{sde-spin} now can be understood as
that of the right-hand side of Eq.\eq{equiv-3} with respect to the
$AB$ index pair. In addition the combination $\phi_{[A} \delta_{B]C}
\mp {\varepsilon_{ABC}}^D \phi_D$ also satisfies the same type of
the self-duality equation with respect to the $AB$ index pair. So we
see that the combination ${f_{AB}}^C + \phi_{[A} \delta_{B]}^C$ also
satisfies the same self-duality equation
\be \la{self-combi}
{f_{AB}}^E + \phi_{[A} \delta_{B]}^E =
\pm \half {\varepsilon_{AB}}^{CD}
\Bigl( {f_{CD}}^E + \phi_{[C} \delta_{D]}^E \Bigr).
\ee

Let us introduce a volume form $\mathfrak{v} = \lambda^{-1}
\mathfrak{v}_g$ for some function $\lambda$ where
\be \la{volume}
\mathfrak{v}_g =  E^1 \wedge E^2 \wedge E^3 \wedge E^4.
\ee
Suppose that $E_A$'s preserve the volume form $\mathfrak{v}$, i.e.,
$\mathcal{L}_{E_A} \mathfrak{v}=0$ which is always possible, as
rigorously proved in \ct{grant2}, by considering an $SO(4)$ rotation
\eq{frame-rotation} of basis vectors and choosing the function
$\lambda$ properly \footnote{\la{so4-rotation}Since we imposed the
vanishing of (anti-)self-dual spin connections, $\omega_M^{(+)a} =
0$ or $\omega_M^{(-)a} = 0$, a remaining symmetry is $SU(2)_{L,R}$
up to a rigid rotation. Together with the function $\lambda$, so
totally four free parameters, it is enough to achieve the condition
$\mathcal{L}_{E_A}
\mathfrak{v}=0$.}. This leads to the relation $\mathcal{L}_{E_A}
\mathfrak{v} = (\nabla \cdot E_A - E_A \log
\lambda ) \mathfrak{v} = 0$. Since $\nabla \cdot E_A = -
{\omega_{BA}}^B = - \phi_A$, we get the identity $\phi_A = - E_A
\log \lambda$ for the volume form $\mathfrak{v}$. Define $D_A \equiv \lambda E_A
\in TM$. Then we have
\bea \la{equiv-4}
[D_A,D_B] &=& \lambda \Bigl( -{f_{AB}}^C + E_A \log \lambda
\delta_B^C - E_B \log \lambda  \delta_A^C \Bigr) D_C \xx &=& -
\lambda \Bigl( {f_{AB}}^C + \phi_{[A} \delta_{B]}^C \Bigr) D_C.
\eea
Finally we get from Eq.\eq{self-combi} the following self-duality
equation \ct{ashtekar,mason-newman}
\be \la{sde-equiv}
[D_A,D_B] = \pm \half {\varepsilon_{AB}}^{CD} [D_C,D_D].
\ee

Conversely one can proceed with precisely reverse order to show that
the vector fields $\{ D_A \}$ satisfying Eq.\eq{sde-equiv} describe
the self-dual spin connections satisfying Eq.\eq{sde-spin}. Note
that the vector fields $D_A$ now preserve a new volume form
$\mathfrak{v}_4 = \lambda^{-2} \mathfrak{v}_g$ which can be seen as
follows
\be \la{volume-preserving}
0 = \mathcal{L}_{E_A} (\lambda^{-1} \mathfrak{v}_g) =
d\bigl(\iota_{E_A} (\lambda^{-1} \mathfrak{v}_g) \bigr) =
d\bigl(\iota_{\lambda E_A} (\lambda^{-2} \mathfrak{v}_g) \bigr) =
d\bigl(\iota_{D_A} \mathfrak{v}_4 \bigr) = \mathcal{L}_{D_A}
\mathfrak{v}_4.
\ee
The function $\lambda$ in terms of $\mathfrak{v}_4$ is therefore
given by
\be \la{lambda}
\lambda^2 = \mathfrak{v}_4  (D_1, D_2, D_3, D_4)
\ee
and the metric is determined by Eq.\eq{4-metric} as
\be \la{sd-metric}
ds^2 = \lambda^2 \delta_{AB} D^A \otimes D^B =
\lambda^2 \delta_{AB} D^A_M D^B_N
\; dX^M \otimes dX^N
\ee
where $E^A = \lambda D^A$.

In summary Eqs.\eq{g-instanton}, \eq{sde-spin} and \eq{sde-equiv}
are equivalent each other (up to a gauge choice) and equally
describe self-dual Einstein gravity.

Now Eq.\eq{sde-equiv} clearly exposes to us that the self-dual
Einstein gravity looks very much like the self-duality equation in
gauge theory. Indeed one can easily see from Eq.\eq{map-ncem} that
the self-dual Einstein gravity in the form of Eq.\eq{sde-equiv}
appears as the leading order of the self-dual NC gauge fields
described by
\be \la{sde-ncem}
\widehat{F}_{AB} = \pm \half {\varepsilon_{AB}}^{CD}
\widehat{F}_{CD}.
\ee
This completes the proof of the equivalence between self-dual NC
electromagnetism on ${\bf R}^4_{NC}$ or ${\bf R}^2_C
\times {\bf R}^2_{NC}$ and self-dual Einstein gravity.

\section{Einstein Equations from Gauge Fields}

In this section we will generalize the equivalence between the
emergent gravity and the Einstein gravity to arbitrary NC gauge
fields. We show that the dynamics of NC $U(1)$ gauge fields at a
commutative limit can be understood as the Einstein gravity
described by Eq.\eq{einstein-energy} where the energy momentum
tensor is given by usual Maxwell fields and by an unusual
``Liouville" field related to the conformal factor (or the size of
spacetime) given by Eq.\eq{D-volume}. In the end, we will find some
remarkable physics regarding to a novel structure of spacetime.

In a non-coordinate (anholonomic) basis $\{E_A\}$ satisfying the
commutation relation \eq{structure-fn}, the spin connections
${{\omega_A}^B}_C$ are defined by
\be \la{D-spin1}
\nabla_A E_C = {{\omega_A}^B}_C  E_B
\ee
where $\nabla_A \equiv \nabla_{E_A}$ is the covariant derivative in
the direction of a vector field $E_A$. Acting on the dual basis
$\{E^A\}$, they are given by
\be \la{D-spin2}
\nabla_A E^B = - {{\omega_A}^B}_C E^C.
\ee
Since we will impose the torsion free condition, i.e.,
\be \la{torsion-free}
T(A,B) = \nabla_{[A} E_{B]} - [E_A,E_B]= 0,
\ee
the spin connections are related to the structure functions
\be \la{spin-structure}
f_{ABC} = -\omega_{ACB} + \omega_{BCA}.
\ee
The Riemann curvature tensors in the basis $\{E_A\}$ are defined by
\be \la{riemann-def}
R(A,B) = [\nabla_A, \nabla_B] - \nabla_{[A,B]}
\ee
or in component form
\bea \la{D-riemann}
{{R_{AB}}^C}_D &=& \langle E^C, R(E_A, E_B) E_D \rangle \xx &=& E_A
{{\omega_B}^C}_D - E_B {{\omega_A}^C}_D + {{\omega_A}^C}_E
{{\omega_B}^E}_D - {{\omega_B}^C}_E {{\omega_A}^E}_D  + {f_{AB}}^E
{{\omega_E}^C}_D.
\eea

Imposing the condition that the metric \eq{D-metric} is covariantly
constant, i.e.,
\be \la{metric-condition}
\nabla_C \Bigl( \eta_{AB} E^A \otimes E^B \Bigr) = 0,
\ee
or, equivalently,
\be \la{spin-condition}
\omega_{CAB}= - \omega_{CBA},
\ee
the spin connections $\omega_{CAB}$ then have the same number of
components as $f_{ABC}$. Thus Eq.\eq{spin-structure} has a unique
solution and it is precisely given by Eq.\eq{spin-connection}. In
coordinate (holonomic) bases $\{\p_M, dX^M\}$, the curvature tensors
\eq{D-riemann} also coincide with Eq.\eq{cartan-curvature}. The
definition \eq{riemann-def} together with the metricity condition
\eq{spin-condition} immediately leads to the following symmetry
property
\be \la{curvature-anti-symm}
R_{ABCD} = - R_{ABDC} = -R_{BACD}.
\ee

As we remarked in Section 3.2, we want to represent the Riemann
curvature tensors in Eq.\eq{D-riemann} in terms of the gauge theory
basis $D_A$ in order to use the equations of motion
\eq{map-eom} and the Bianchi identity \eq{map-bianchi}. Using the
relation \eq{2-structure}, the spin connections in
Eq.\eq{spin-connection} are given by
\be \la{spin-conformal}
\lambda \omega_{ABC} = \half(\mathfrak{f}_{ABC} - \mathfrak{f}_{BCA}
+ \mathfrak{f}_{CAB}) - D_B \log \lambda \eta_{CA} + D_C \log
\lambda \eta_{AB}.
\ee
It is then straightforward to calculate each term in
Eq.\eq{D-riemann}. We list the results:
\bea \la{curvature1}
E_A \omega_{BCD} &=& - \frac{1}{2\lambda^2} D_A \log \lambda
(\mathfrak{f}_{BCD} - \mathfrak{f}_{CDB} + \mathfrak{f}_{DBC}) \xx
&& + \frac{1}{\lambda^2} \eta_{BD} D_A \log \lambda D_C \log
\lambda - \frac{1}{\lambda^2} \eta_{BC} D_A \log \lambda D_D \log
\lambda
\xx
&& + \frac{1}{2\lambda^2} D_A (\mathfrak{f}_{BCD} -
\mathfrak{f}_{CDB} +
\mathfrak{f}_{DBC}) \xx
&& + \frac{1}{\lambda^2} \Bigl( \eta_{BC} D_A D_D \log \lambda -
\eta_{BD} D_A D_C \log \lambda \Bigr),
\eea

\bea \la{curvature2}
\omega_{ACE} {{\omega_B}^E}_D & = & \; \; \frac{1}{4 \lambda^2} \eta^{EF}
(\mathfrak{f}_{ACE} - \mathfrak{f}_{CEA} + \mathfrak{f}_{EAC})
(\mathfrak{f}_{BFD} - \mathfrak{f}_{FDB} + \mathfrak{f}_{DBF}) \xx
&& + \frac{1}{2 \lambda^2} \eta^{EF}
\Bigl( \eta_{AC}(\mathfrak{f}_{BED} - \mathfrak{f}_{EDB} +
\mathfrak{f}_{DBE}) -\eta_{BD}(\mathfrak{f}_{ACE} - \mathfrak{f}_{CEA} +
\mathfrak{f}_{EAC}) \Bigr) D_F \log \lambda \xx
&& + \frac{1}{2 \lambda^2} \Bigl((\mathfrak{f}_{ACB} -
\mathfrak{f}_{CBA} + \mathfrak{f}_{BAC})  D_D \log \lambda
- (\mathfrak{f}_{BAD} - \mathfrak{f}_{ADB} + \mathfrak{f}_{DBA})
 D_C \log \lambda \Bigr) \xx
&& + \frac{1}{\lambda^2} \Bigl( \eta_{BD}D_A \log \lambda D_C
\log \lambda - \eta_{AB}D_C \log \lambda D_D \log \lambda
+ \eta_{AC} D_B \log \lambda D_D \log \lambda \Bigr) \xx && -
\frac{1}{\lambda^2} \eta_{AC} \eta_{BD} \eta^{EF} D_E \log \lambda
D_F \log \lambda,
\eea

\bea \la{curvature3}
{f_{AB}}^E \omega_{ECD} & = & \;\; \frac{1}{2
\lambda^2} {\mathfrak{f}_{AB}}^E  (\mathfrak{f}_{ECD} -
\mathfrak{f}_{CDE} + \mathfrak{f}_{DEC}) \xx
&& + \frac{1}{\lambda^2} (\mathfrak{f}_{ABC} D_D
\log \lambda- \mathfrak{f}_{ABD} D_C \log \lambda) \xx
&& + \frac{1}{2 \lambda^2} \Bigl((\mathfrak{f}_{BCD} -
\mathfrak{f}_{CDB} + \mathfrak{f}_{DBC})  D_A \log \lambda
- (\mathfrak{f}_{ACD} - \mathfrak{f}_{CDA} + \mathfrak{f}_{DAC})
 D_B \log \lambda \Bigr) \xx
&& + \frac{1}{\lambda^2} \Bigl( \eta_{BC}D_A \log \lambda D_D \log
\lambda - \eta_{BD} D_A \log \lambda D_C \log \lambda \Bigr) \xx
&& + \frac{1}{\lambda^2} \Bigl( \eta_{AD} D_B \log \lambda D_C
\log \lambda - \eta_{AC}D_B \log \lambda D_D \log \lambda \Bigr).
\eea

Substituting these expressions into Eq.\eq{D-riemann}, the curvature
tensors are given by
\bea \la{curvature-detail}
R_{ABCD} &=& \frac{1}{\lambda^2}  \Big[ \Big\{\half D_A
(\mathfrak{f}_{BCD} - \mathfrak{f}_{CDB} + \mathfrak{f}_{DBC})
\xx && + \eta_{BC} D_A D_D \log \lambda -
\eta_{BD} D_A D_C \log \lambda \xx
&& + \frac{1}{4} \eta^{EF} (\mathfrak{f}_{ACE} -
\mathfrak{f}_{CEA} + \mathfrak{f}_{EAC}) (\mathfrak{f}_{BFD} -
\mathfrak{f}_{FDB} + \mathfrak{f}_{DBF}) \xx && + \frac{1}{2} \eta^{EF}
\Bigl( \eta_{AC}(\mathfrak{f}_{BED} - \mathfrak{f}_{EDB} +
\mathfrak{f}_{DBE}) -\eta_{BD}(\mathfrak{f}_{ACE} - \mathfrak{f}_{CEA} +
\mathfrak{f}_{EAC}) \Bigr) D_F \log \lambda \xx
&& + \frac{1}{2} \Bigl((\mathfrak{f}_{ACB} -
\mathfrak{f}_{CBA} + \mathfrak{f}_{BAC})  D_D \log \lambda
- (\mathfrak{f}_{BAD} - \mathfrak{f}_{ADB} + \mathfrak{f}_{DBA})
 D_C \log \lambda \Bigr) \xx
&& + \eta_{BD}D_A \log
\lambda D_C
\log \lambda + \eta_{AC} D_B \log \lambda D_D \log \lambda \xx && -
\eta_{AC} \eta_{BD} \eta^{EF} D_E \log \lambda
D_F \log \lambda \Big\} - \{ A \leftrightarrow B \} \Big] \xx && +
\frac{1}{\lambda^2} \Big[ \half {\mathfrak{f}_{AB}}^E  (\mathfrak{f}_{ECD} -
\mathfrak{f}_{CDE} + \mathfrak{f}_{DEC}) +
(\mathfrak{f}_{ABC} D_D \log \lambda- \mathfrak{f}_{ABD} D_C
\log \lambda) \Big].
\eea

Using Eq.\eq{curvature-detail}, the Ricci tensors $R_{AC} \equiv
\eta^{BD} R_{ABCD}$ and the Ricci scalar $R \equiv \eta^{AC} R_{AC}$
are accordingly determined as
\bea \la{ricci-tensor}
R_{AC} &=& \frac{1}{\lambda^2}  \Big[ - \half (D-4) (D_A D_C + D_C
D_A) \log \lambda - \eta_{AC} \eta^{BD} D_B D_D \log \lambda \xx &&
+(D - 2) D_A \log \lambda D_C \log \lambda - (D-4) \eta_{AC}
\eta^{BD} D_B \log \lambda D_D \log \lambda \xx && + \half (D-4)
\eta^{BD} \big( \mathfrak{f}_{ABC} - \mathfrak{f}_{BCA} \big) D_D \log \lambda \xx
&& - \half \eta^{BD} D_B (\mathfrak{f}_{ACD} -
\mathfrak{f}_{CDA} + \mathfrak{f}_{DAC}) \xx
&& + \frac{1}{4} \eta^{BD} \eta^{EF}
\mathfrak{f}_{BEC} \mathfrak{f}_{DFA} +
\frac{1}{2} \eta^{BD} {\mathfrak{f}_{AB}}^E (\mathfrak{f}_{ECD} -
\mathfrak{f}_{CDE}) \Big],
\eea

\bea \la{ricci-scalar}
R &=& \frac{1}{\lambda^2}  \Big[ -2(D-3) \eta^{AC} D_A D_C \log
\lambda - (D-2)(D-5) \eta^{AC} D_A \log \lambda D_C \log \lambda \xx
&& + \frac{1}{4} \eta^{AC} \eta^{BD} {\mathfrak{f}_{AB}}^E (2
\mathfrak{f}_{ECD} - \mathfrak{f}_{CDE}) \Big],
\eea
where we have used the relation \eq{D-structure-f} and
\bea \la{identity-structure}
&& \frac{1}{4} \eta^{BD} \eta^{EF} (\mathfrak{f}_{BCE} -
\mathfrak{f}_{CEB} + \mathfrak{f}_{EBC}) (\mathfrak{f}_{AFD} -
\mathfrak{f}_{FDA} + \mathfrak{f}_{DAF}) \xx
= && \half \eta^{BD} {\mathfrak{f}_{AB}}^E \mathfrak{f}_{DEC} -
\frac{1}{4} \eta^{BD} \eta^{EF} \mathfrak{f}_{BEC} \mathfrak{f}_{DFA}.
\eea

Up to now we have not used Eqs.\eq{map-eom} and \eq{map-bianchi}. We
have simply calculated curvature tensors for an arbitrary metric
\eq{D-metric}. Now we will impose on the curvature tensors the
equations of motion Eq.\eq{map-eom} and the Bianchi identity
\eq{map-bianchi}. First note the following identity
\bea \la{1-Bianchi}
&& R(E_A, E_B) E_C + R(E_B, E_C) E_A + R(E_C, E_A) E_B \xx &=&
 [E_A, [E_B, E_C]] + [E_B, [E_C, E_A]] + [E_C, [E_A, E_B]]
\eea
which can be derived using the condition \eq{torsion-free}. The
Jacobi identity then implies $R_{[ABC]D} = 0$. Since $D_A = \lambda
E_A$, we have the relation $[D_{[A}, [D_B, D_{C]}]] = \lambda ^3
[E_{[A}, [E_B, E_{C]}]]$ where all the terms containing the
derivations of $\lambda$ cancel each other. Thus the first Bianchi
identity $R_{[ABC]D} = 0$ follows from the Jacobi identity $[D_{[A},
[D_B, D_{C]}]]=0$. Then Eq.\eq{map-bianchi} confirms that the guess
\eq{bianchi} is pleasingly true, i.e.,
\be \la{bianchi-true}
\widehat{D}_{[A} \widehat{F}_{BC]} = 0 \quad \Longleftrightarrow \quad
R_{[ABC]D} = 0.
\ee
One can also directly check Eq.\eq{bianchi-true} using the
expression \eq{curvature-detail}:
\be \la{1-bianchi-check}
R_{[ABC]D} = \frac{1}{\lambda^2} \Bigl( D_{[A} \mathfrak{f}_{BC]D} -
{\mathfrak{f}_{[BC}}^E
\mathfrak{f}_{A]ED} \Bigr) = 0.
\ee

Let us summarize the algebraic symmetry of curvature tensors
determined by the properties about the torsion and the tangent-space
group:
\bea \la{curvature-symm1}
&& R_{ABCD} = - R_{ABDC} = -R_{BACD}, \\
\la{curvature-symm2}
&& R_{[ABC]D} = 0, \\
\la{curvature-symm3}
&& R_{ABCD} = R_{CDAB}
\eea
where the last symmetry can be derived by using the others.
Therefore it is obvious that the vector fields $D_A \in TM$
satisfying Eq.\eq{map-bianchi} describe a usual (pseudo-)Riemannian
manifold.

Some useful properties can be further deduced. Contracting the
indices $C$ and $D$ in Eq.\eq{map-bianchi} leads to
\be \la{bianchi-cont}
D_A \rho_B - D_B \rho_A + {\mathfrak{f}_{AB}}^C \rho_C = D_C
{\mathfrak{f}_{AB}}^C
\ee
and the left-hand side identically vanishes due to
Eq.\eq{structure-d} with Eq.\eq{D-structure-f}. Thus we get
\be \la{identity-divergence}
D_C {\mathfrak{f}_{AB}}^C = 0.
\ee
Similarly, from Eq.\eq{map-eom}, we get
\be \la{identity-eom}
\eta^{AB} D_A D_B \log \lambda = \half D_A \rho^A = - \half \eta^{AB}
{\mathfrak{f}_{AC}}^D {\mathfrak{f}_{BD}}^C.
\ee
Eq.\eq{identity-divergence} now guarantees that the Ricci tensor
\eq{ricci-tensor} is symmetric, i.e., $R_{AC} = R_{CA}$.
(It should be the case since the symmetry property
\eq{curvature-symm3} shows that $R_{AC} = \eta^{BD} R_{ABCD} =
\eta^{DB} R_{CDAB} = R_{CA}$. Recall that the property \eq{curvature-symm3} results from
the Bianchi identity \eq{1-bianchi-check}.)

In order to check the conjecture \eq{einstein-energy}, we first
consider the Euclidean $D=4$ case since we already know the answer
for the self-dual case. For the Euclidean space we will not care
about raising and lowering indices. Using Eqs.\eq{D-structure-f},
\eq{map-eom} and
\eq{identity-eom}, the Ricci tensor \eq{ricci-tensor} can be rewritten as follows
\bea \la{ricci-eom}
R_{AC} &=& \frac{1}{2 \lambda^2}  \Big[ \delta_{AC}
\mathfrak{f}_{BDE} \mathfrak{f}_{BED}  + \mathfrak{f}_{BAB}
\mathfrak{f}_{DCD} - \mathfrak{f}_{BDA} \mathfrak{f}_{BCD} -
\mathfrak{f}_{BDC} \mathfrak{f}_{BAD} \xx
&& + \half \mathfrak{f}_{BDA} \mathfrak{f}_{BDC} +
\mathfrak{f}_{ABD} \mathfrak{f}_{DCB} - \mathfrak{f}_{ABD} \mathfrak{f}_{CBD}
\Big].
\eea
Now we decompose $\mathfrak{f}_{ABC}$ into self-dual and
anti-self-dual parts as in Eq.\eq{spin-sd-asd}
\be \la{f-sd-asd}
\mathfrak{f}_{ABC} = \mathfrak{f}_C^{(+)a} \eta^a_{AB}
+ \mathfrak{f}_C^{(-)a} \bar{\eta}^a_{AB}
\ee
where
\be \la{f-decomposition}
\mathfrak{f}_C^{(\pm)a} \eta^{(\pm)a}_{AB} = \half \Big(
\mathfrak{f}_{ABC} \pm \half {\varepsilon_{AB}}^{DE}
\mathfrak{f}_{DEC} \Big)
\ee
and introduce a completely antisymmetric tensor defined by
\be \la{3-form}
\Psi_{ABC} = \mathfrak{f}_{ABC} +
\mathfrak{f}_{BCA} + \mathfrak{f}_{CAB} \equiv \varepsilon_{ABCD} \Psi_D.
\ee
Using the decomposition \eq{f-sd-asd} and Eq.\eq{self-eta} one can
easily see that
\be \la{psi-decomp}
\Psi_A = - \frac{1}{3 !} \varepsilon_{ABCD} \Psi_{BCD} =
- (\mathfrak{f}_B^{(+)a} \eta^a_{AB} - \mathfrak{f}_B^{(-)a}
\bar{\eta}^a_{AB}),
\ee
while Eq.\eq{D-structure-f} leads to
\be \la{rho-decomp}
\rho_A = \mathfrak{f}_{BAB} =
\mathfrak{f}_B^{(+)a} \eta^a_{AB}
+ \mathfrak{f}_B^{(-)a} \bar{\eta}^a_{AB}.
\ee

The calculation of the Ricci tensor \eq{ricci-eom} can
straightforwardly be done using the decomposition \eq{f-sd-asd} and
the identities \eq{eta^2} and \eq{eta-ex} after rewriting the
following term
\bea \la{non-term}
\mathfrak{f}_{ABD} \mathfrak{f}_{DCB} &=&  \mathfrak{f}_{ABD}(
\Psi_{DCB} - \mathfrak{f}_{CBD} - \mathfrak{f}_{BDC}) \xx
&=& \varepsilon_{DCBE} (\mathfrak{f}_D^{(+)a} \eta^a_{AB} +
\mathfrak{f}_D^{(-)a} \bar{\eta}^a_{AB}) \Psi_E - \mathfrak{f}_{ABD}
\mathfrak{f}_{CBD} - \mathfrak{f}_{ABD} \mathfrak{f}_{BDC} \xx
&=& - \Psi_A \Psi_C - (\mathfrak{f}_A^{(+)a} \eta^a_{CD} -
\mathfrak{f}_A^{(-)a} \bar{\eta}^a_{CD})\Psi_D + \delta_{AC} \Psi_D
\Psi_D \xx
&& - \mathfrak{f}_{ABD} \mathfrak{f}_{CBD} - \mathfrak{f}_{ABD}
\mathfrak{f}_{BDC}
\eea
where Eq.\eq{self-eigen} was used at the last step. An interesting
thing is that Eq.\eq{non-term} cancels most terms in
Eq.\eq{ricci-eom} leaving a remarkably simple form
\bea \la{ricci-final}
R_{AC} &=& - \frac{1}{\lambda^2}  \Big[ \mathfrak{f}_D^{(+)a}
\eta^a_{AB} \mathfrak{f}_D^{(-)b} \bar{\eta}^b_{CB} + \mathfrak{f}_D^{(+)a}
\eta^a_{CB} \mathfrak{f}_D^{(-)b} \bar{\eta}^b_{AB} \xx
&& - \big(\mathfrak{f}_B^{(+)a}
\eta^a_{AB} \mathfrak{f}_D^{(-)b} \bar{\eta}^b_{CD} + \mathfrak{f}_B^{(+)a}
\eta^a_{CB} \mathfrak{f}_D^{(-)b} \bar{\eta}^b_{AD} \big) \Big].
\eea

Note that the right-hand side of Eq.\eq{ricci-final} is purely
interaction terms between the self-dual and anti-self-dual parts in
Eq.\eq{f-sd-asd}. (The same result was also obtained in
\ct{grant2}.) Therefore, if NC gauge fields satisfy the self-duality
equation \eq{sde-structure}, they describe a Ricci-flat manifold,
i.e., $R_{AC}=0$. Of course, this result is completely consistent
with that in Appendix A. Moreover we see the reason why self-dual NC
gauge fields satisfy the Einstein equation \eq{einstein-energy} with
vanishing energy-momentum tensor.

Finally we can calculate the Einstein tensor to find the form of the
energy-momentum tensor defined by Eq.\eq{einstein-energy}:
\bea \la{einstein-tensor}
E_{AB} &=& R_{AB} - \half \delta_{AB} R \xx &=& -
\frac{1}{\lambda^2}
\Big( \mathfrak{f}_D^{(+)a}
\eta^a_{AC} \mathfrak{f}_D^{(-)b} \bar{\eta}^b_{BC} + \mathfrak{f}_D^{(+)a}
\eta^a_{BC} \mathfrak{f}_D^{(-)b} \bar{\eta}^b_{AC} \Big) \xx
&& + \frac{1}{\lambda^2} \Big( \mathfrak{f}_C^{(+)a}
\eta^a_{AC} \mathfrak{f}_D^{(-)b} \bar{\eta}^b_{BD} + \mathfrak{f}_C^{(+)a}
\eta^a_{BC} \mathfrak{f}_D^{(-)b} \bar{\eta}^b_{AD} - \delta_{AB} \mathfrak{f}_D^{(+)a}
\eta^a_{CD} \mathfrak{f}_E^{(-)b} \bar{\eta}^b_{CE} \Big)
\eea
where the Ricci scalar $R$ is given by
\be \la{final-ricci-scalar}
R = \frac{2}{\lambda^2} \mathfrak{f}_B^{(+)a}
\eta^a_{AB} \mathfrak{f}_C^{(-)b} \bar{\eta}^b_{AC}.
\ee
We have adopted the conventional view that the gravitational field
is represented by the spacetime metric itself. The problem then
becomes one of finding field equations to relate the metric
\eq{D-metric} to the energy-momentum distribution. According to our
scheme, Eq.\eq{einstein-tensor} should correspond to such field
equations, i.e., the Einstein equations. In other words, if we are
clever enough, we should be able to find the NC gauge theory
described by Eqs.\eq{map-eom} and \eq{map-bianchi} starting from the
Einstein gravity described by Eqs.\eq{curvature-symm2} and
\eq{einstein-tensor} by properly reversing our above derivation as
we have explicitly demonstrated it for the self-dual case in
Appendix A.

As we explained in Section 3.2, we want to identify
Eq.\eq{einstein-tensor} with an energy-momentum tensor. First note
that the Ricci scalar $R$, \eq{final-ricci-scalar}, is nonvanishing
for a generic case. This means that there is an extra field
contribution to the energy-momentum tensor in addition to Maxwell
fields whose energy-momentum tensor is traceless. Since the extra
field energy-momentum tensor turns out to be basically a gradient
volume energy (see the latter part of Sec. 3.2), we call it the
``Liouville" energy-momentum tensor. A similar result was also
obtained in \ct{madore-poisson} where it was dubbed as the `Poisson'
energy. Since the first term in Eq.\eq{einstein-tensor} is traceless
due to Eq.\eq{eta-etabar}, it would be a candidate of the Maxwell
energy-momentum tensor while the second term would be the Liouville
energy-momentum tensor. So we tentatively make the following
identification for the Maxwell energy-momentum tensor $T_{AB}^{(M)}$
and the Liouville energy-momentum tensor $T_{AB}^{(L)}$
\bea \la{em-tensor-maxwell}
\frac{8 \pi G_4}{c^4} T_{AB}^{(M)} &=& - \frac{1}{\lambda^2}
\Big( \mathfrak{f}_D^{(+)a}
\eta^a_{AC} \mathfrak{f}_D^{(-)b} \bar{\eta}^b_{BC} + \mathfrak{f}_D^{(+)a}
\eta^a_{BC} \mathfrak{f}_D^{(-)b} \bar{\eta}^b_{AC} \Big), \xx
&=& - \frac{1}{\lambda^2}
\Big(\mathfrak{f}_{ACD} \mathfrak{f}_{BCD} - \frac{1}{4}
\delta_{AB} \mathfrak{f}_{CDE} \mathfrak{f}_{CDE} \Big), \\
\la{dark-energy}
\frac{8 \pi G_4}{c^4} T_{AB}^{(L)} &=& \frac{1}{\lambda^2}
\Big(\mathfrak{f}_C^{(+)a}
\eta^a_{AC} \mathfrak{f}_D^{(-)b} \bar{\eta}^b_{BD} + \mathfrak{f}_C^{(+)a}
\eta^a_{BC} \mathfrak{f}_D^{(-)b} \bar{\eta}^b_{AD} - \delta_{AB} \mathfrak{f}_D^{(+)a}
\eta^a_{CD} \mathfrak{f}_E^{(-)b} \bar{\eta}^b_{CE} \Big), \xx
&=& \frac{1}{2\lambda^2} \Big( \rho_A \rho_B - \Psi_A \Psi_B -
\frac{1}{2} \delta_{AB}(\rho_C^2 - \Psi_C^2) \Big)
\eea
where we have used the decomposition \eq{f-decomposition} and the
relation
$$ \mathfrak{f}_B^{(+)a} \eta^a_{AB} = \half(\rho_A - \Psi_A), \qquad
\mathfrak{f}_B^{(-)a} \bar{\eta}^a_{AB} = \half(\rho_A + \Psi_A).$$

We have anticipated that the energy-momentum tensor
\eq{em-tensor-maxwell} will be related to that of Maxwell fields
since both are definitely traceless in four dimensions. So our
problem is how to rewrite the energy-momentum tensor in terms of NC
fields in $\star$-algebra ${\cal A}_\theta$, denoted as
$\widehat{T}_{AB}({\cal A}_\theta)$, using the expression
\eq{em-tensor-maxwell} defined in $TM$, denoted as $T_{AB}(TM)$.
In other words, we want to translate $T_{AB}(TM)$ into an ${\cal
A}_\theta$-valued energy momentum tensor. This problem is quite
subtle.

Recall that NC fields are identified with vector fields in $TM$
through the map \eq{co-inner} at the leading order. For example, we
get the following identification from Eq.\eq{map-ncem}
\bea \la{nc-vec-identity}
-i[ \widehat{F}_{AB}, \widehat{f}]_\star &=& \{ F_{AB}, f
\}_{\theta} + \cdots =   [D_A, D_B][f] + \cdots \xx
&=& - {\mathfrak{f}_{AB}}^C D_C [f] + \cdots.
\eea
Note that Eq.\eq{nc-vec-identity} is nothing but the Lie algebra
homomorphism \eq{poisson-lie} for the Poisson algebra. But a NC
field regarded as an element of NC $\star$-algebra ${\cal A}_\theta$
in general lives in a Hilbert space ${\cal H}$, e.g., the Fock space
\eq{fock} while the vector fields $D_A$ in Eq.\eq{co-inner}
are defined in the real vector space $TM$. Furthermore we see from
Eqs.\eq{co-inner} and \eq{nc-vec-identity} that ``anti-Hermitian"
operators in NC algebra ${\cal A}_\theta$ such as the NC fields
$\widehat{D}_A$ and $-i \widehat{F}_{AB}$ are mapped to real vector
fields in $TM$. Thus we have the bizarre correspondence between
geometry defined in $TM$ and NC algebra ${\cal A}_\theta
$\footnote{It might be remarked that the transition from $TM$ to
${\cal A}_\theta$ is analogous to that from classical mechanics (an
$\mathbf{R}$-world) to quantum mechanics (a $\mathbf{C}$-world). See
Section 5.4 in \ct{mechanics} for the exposition of the similar
problem in the context of quantum mechanics.}
\be \la{map-htm}
{\rm Anti}-{\rm Hermitian \; operators \; on \; {\cal H}} \quad
\Leftrightarrow \quad {\rm Real \; vector \; fields \; on} \; TM.
\ee

In order to identify ${\cal A}_\theta$-valued quantities from
$TM$-valued ones, it is first necessary to analytically continue the
real vector space $TM$ to a complex vector space $TM_{{\bf C}}$. At
the same time, the real vector field $D_A$ is replaced by a
self-adjoint operator $\mathcal{D}_A$ in $TM_{{\bf C}}$ and the
structure equation
\eq{structure-d} instead has the form
\be \la{structure-complex}
[\mathcal{D}_A, \mathcal{D}_B] = i {\mathfrak{f}_{AB}}^C
\mathcal{D}_C.
\ee
Now we want to translate a quantity defined on $TM_{{\bf C}}$ such
as Eq.\eq{structure-complex} into a NC field defined on ${\cal H}$
as the Weyl-Wigner correspondence \ct{nc-review}. Since we have the
identification \eq{nc-vec-identity}, we need to relate the inner
product on the operator algebra ${\cal A}_\theta$, denoted as
$\langle \widehat{V}, \widehat{W}
\rangle_{{\cal A}_\theta}$ for $\widehat{V}, \widehat{W} \in {\cal
A}_\theta$ to the inner product $\langle V, W \rangle_{TM_{{\bf C}}}
\equiv
\overline{V} \cdot W$ on $TM_{{\bf C}}$ for $V, W \in TM_{{\bf C}}$,
both of which are defined to be positive definite. To do this, we
will take the natural prescription according to the correspondence
\eq{nc-vec-identity}
\be \la{product-map}
\langle \widehat{F}_{AB}, \widehat{F}_{CD} \rangle_{{\cal A}_\theta}
\; \Leftrightarrow \;
{\mathfrak{f}_{AB}}^E {\mathfrak{f}_{CD}}^F (\mathcal{D}_E
\cdot \mathcal{D}_F) + \cdots
\ee
where the ellipsis means that we need a general inner product for
multi-indexed vector fields, e.g., polyvector fields though the
leading term is enough for our purpose. Note that $\mathcal{D}_A =
\lambda \mathcal{E}_A$ carry the mass dimension, i.e., $[\mathcal{D}_A]
= [\mathcal{E}_A] = L^{-1}$ where $\lambda$ is chosen to be real
such that both $\mathcal{D}_A$ and $\mathcal{E}_A$ are self-adjoint
operators in $TM_{{\bf C}}$. Hence we will take into account the
physical dimension of the vector fields $\mathcal{D}_A$ in the
definition of the inner product
\eq{product-map}
\be \la{inner-product}
\mathcal{D}_A \cdot \mathcal{D}_B =
\lambda^2 (\mathcal{E}_A \cdot \mathcal{E}_B) =
\frac{\lambda^2}{|\rm{Pf} \theta|^{\frac{1}{n}}} \delta_{AB}.
\ee
Here the noncommutativity $|\theta|$ is the most natural
dimensionful parameter at our hands that can enter the definition
\eq{inner-product}.

Suppose that the analytic continuation was performed and we adopt
the prescription \eq{product-map}. Then the analytic continuation
from $TM$ to $TM_{{\bf C}}$ accompanies the $i$ factor in the
structure equation \eq{structure-complex} which will introduce a
sign flip in Eq.\eq{em-tensor-maxwell}.\footnote{To avoid any
confusion, we point out that it never means changing the sign of
Eq.\eq{em-tensor-maxwell} because Eq.\eq{em-tensor-maxwell} is
obviously defined on $TM$. It simply prescribes the analytic
continuation to get a correct definition of $\widehat{T}_{AB}({\cal
A}_\theta)$. Anyway we think that this perverse sign problem will
disappear (at the price of transparent geometrical picture) if we
work in the vector space $TM_{{\bf C}}$ from the outset using the
structure equation \eq{structure-complex}. It will also be useful to
clearly understand the structure of Hilbert space defining (quantum)
gravity, especially, in the context of emergent gravity. We hope to
address this approach in the near future.} And then $T_{AB}(TM_{{\bf
C}})$ will be identified using the prescription \eq{product-map}
with $\widehat{T}_{AB}({\cal A}_\theta)$. After taking the sign flip
into account, one can finally identify $\widehat{T}_{AB}({\cal
A}_\theta)$ from the Maxwell energy-momentum tensor
\eq{em-tensor-maxwell}
\be \la{em-maxwell}
\frac{8 \pi G_4}{c^4} \widehat{T}_{AB}^{(M)}({\cal A}_\theta) =
\frac{g^2_{YM} |\rm{Pf} \theta|^{\frac{1}{n}}}{\hbar^2
c^2 \lambda^4} \frac{\hbar^2 c^2}{g^2_{YM}}\Big( \widehat{F}_{AC}
\widehat{F}_{BC} -  \frac{1}{4} \delta_{AB} \widehat{F}_{CD}
\widehat{F}_{CD} \Big)
\ee
where we simply rewrote the global factor for later use. Recall that
we are taking the commutative limit $|\theta| \to 0$ (see the
paragraph in Eq.\eq{newton-constant}). Thus one can simply replace
the field strengths in Eq.\eq{em-maxwell} by commutative ones, i.e.,
$\widehat{F}_{AC}
\approx F_{AC} + {\cal O}(\theta)$, since the global factor
$|\rm{Pf} \theta|^{\frac{1}{n}}$ already contains ${\cal
O}(\theta)$. Therefore, in the commutative limit, the trace of NC
spacetime in Eq.\eq{em-maxwell} only remains in the global factor
which will be identified with the Newton constant. Thus we get the
usual Maxwell energy-momentum tensor at the leading order. It should
be pointed out that the energy momentum tensor
\eq{em-maxwell} is not quite the same as that derived from the
action \eq{matrix-action} since the background part $B_{MN}$ does
not appear in the result. We will see in Section 3.4 that this fact
bears an important consequence about the cosmological constant and
dark energy.

Note that the result \eq{em-maxwell} is independent of spacetime
dimensions including the front factor. By comparing the expression
\eq{em-maxwell} with Eq.\eq{einstein-energy}, we get the
identification of the Newton ``constant"
\be \la{Newton-constant}
G_D = \frac{c^2 g^2_{YM} |\rm{Pf} \theta|^{\frac{1}{n}}}{8
\pi \hbar^2 \lambda^4}.
\ee
Thereby we almost confirmed Eq.\eq{newton-constant} obtained by a
simple dimensional analysis except the dimensionless factor
$\lambda^4$. (Of course the dimensional analysis alone cannot fix
any dimensionless parameters.) Then Eq.\eq{Newton-constant} comes
with a surprise. It raises a question whether the Newton ``constant"
$G_D$ is a constant or not. If it is a constant, then it means that
$g_{YM}$ (or even $\hbar$ and $c$ ?) depends on $\lambda$ such that
$G_D$ is a constant. Or if $g_{YM}, \;c$ and $\hbar$ are really
constants, $G_D$ depends on the conformal factor (or the size of
spacetime) given by Eq.\eq{D-volume}. We prefer the former
interpretation since we know that $g_{YM}$ changes under a
renormalization group flow. Furthermore we note that $g_{YM}^2 $ in
NC gauge theory depends on an open string metric in $B$-field
background \ct{sw} and $\lambda^2$ is also related to the metric
$g_{MN}$ through the relation \eq{D-volume}. (In four dimensions
$\lambda^2 \sim
\sqrt{-g}$.) Nevertheless, we could not find any inconsistency for the
latter interpretation either, because it seems to be consistent with
current laboratory experiments since $\lambda = 1$ for any flat
spacetime.

In the course of our derivation, we have introduced a completely
antisymmetric tensor
\be \la{D-3-form}
\Psi_{ABC} = \mathfrak{f}_{ABC} +
\mathfrak{f}_{BCA} + \mathfrak{f}_{CAB}.
\ee
So one may identify it with a 3-form field
\be \la{3-form-h}
H \equiv \frac{1}{3!}
\Psi_{ABC} E^A \wedge E^B \wedge E^C = \frac{\lambda}{2}
f_{ABC} E^A \wedge E^B \wedge E^C
\ee
where we used Eq.\eq{2-structure}. But $H$ is not a closed 3-form in
general. Using the structure equation
\be \la{structure-equation}
dE^A = \half {f_{BC}}^A E^B \wedge E^C
\ee
one can show that instead it satisfies the following relation
\bea \la{3-form-eq}
dH &=& \frac{\lambda}{2} \big( E_A f_{BCD} - {f_{BC}}^E f_{AED}
\big) E^A \wedge E^B \wedge E^C \wedge E^D \xx
&& + \Big( \frac{1}{4\lambda} {\mathfrak{f}_{AD}}^E
\mathfrak{f}_{BCE} + \frac{3 \lambda}{2} E_A \log \lambda f_{BCD}
\Big) E^A \wedge E^B \wedge E^C \wedge E^D \xx
&=& \frac{|\rm{Pf} \theta|^{\frac{1}{n}}}{\lambda^3} F \wedge F + 3
d \log \lambda \wedge H
\eea
where we used the Jacobi identity $[E_{[A}, [E_B, E_{C]}]] = 0 $ to
show the vanishing of the first term and the map \eq{product-map}
for the second term. From Eq.\eq{3-form-eq} we see that
$\widetilde{H} \equiv \lambda^{-3} H = \frac{1}{3!}
\Psi_{ABC} D^A \wedge D^B \wedge D^C $ is closed, i.e., $d
\widetilde{H} = 0$, if and only if $F \wedge F = 0$. In this case
locally $\widetilde{H} = d \widetilde{B}$ by the Poincar\'e lemma.
Indeed the 3-form $\widetilde{H} = d \widetilde{B}$ is quite similar
to the Kalb-Ramond field in string theory while the conformal factor
$\lambda$ in Eq.\eq{D-structure-f} behaves like a dilaton field in
string theory. In its overall picture the emergent gravity is very
similar to string theory where a metric $g_{MN}$, an NS-NS 3-form
$H=dB$ and a dilaton $\Phi$ describe a gravitational theory in D
dimensions.

Now we go to the second energy-momentum tensor \eq{dark-energy}.
Note that $\rho_A$ is determined by the volume factor in
Eq.\eq{D-volume} evaluated in the gauge theory basis $\{ D_A \}$
while $\Psi_A$ is coming from the 3-form \eq{3-form-h}.
Eq.\eq{dark-energy} has an interesting property that they
identically vanish for flat spacetime and self-dual gauge fields
where $\rho_A = \pm \Psi_A$. This kind of energy has no counterpart
in commutative spacetime and would be a unique property appearing
only in NC spacetime. This exotic feature might be expected from the
beginning because the NC spacetime leads to a perplexing mixing
between short (UV) and large (IR) distance scales \ct{uv-ir}. To
illuminate the property of the energy-momentum tensor
\eq{dark-energy}, let us simply assume that its average (in a broad sense)
is $SO(4)$ invariant, i.e.,
\be \la{so4-invariant}
\langle \rho_A \rho_B \rangle = \frac{1}{4} \delta_{AB} \rho_C^2, \quad
\langle \Psi_A \Psi_B  \rangle = \frac{1}{4} \delta_{AB} \Psi_C^2.
\ee
Then the average of the energy-momentum tensor \eq{dark-energy} is
given by
\be \la{average-dark-energy}
\langle T_{AB}^{(L)} \rangle = -
\frac{c^4}{64 \pi G_4 \lambda^2} \delta_{AB}(\rho_C^2 - \Psi_C^2).
\ee
Note that the Ricci scalar \eq{final-ricci-scalar} is purely coming
from this source since Eq.\eq{em-maxwell} is traceless. For a
constant curvature space, e.g., de Sitter or anti-de Sitter space,
the Ricci scalar $R = \frac{1}{2 \lambda^2} (\rho_A^2 - \Psi_A^2)$
will be constant. In this case the energy-momentum tensor
\eq{average-dark-energy} precisely behaves like a cosmological
constant since $T_{AB}^{(L)} = - \frac{c^4}{32 \pi G_4} \delta_{AB}
R$. Of course this conclusion is meaningful only if
Eq.\eq{einstein-tensor} allows a constant curvature spacetime. But
the energy momentum tensor given by Eq.\eq{average-dark-energy} will
behave like a cosmological constant as ever for an almost constant
curvature space as shown in Eq.\eq{dark-energy-cal}.

Although we have taken the Euclidean signature for convenience, it
can be analytically continued to the Lorentzian
signature.\footnote{\la{wick-rotation}The Wick rotation will be
defined by $x^4 = ix^0$. Under this Wick rotation, $\delta_{AB} \to
\eta_{AB}=(-+++)$ and $\varepsilon^{1234} = 1 \to - \varepsilon^{0123} = -1$.
Then we get $\Psi_A^{(E)} = i \Psi_A^{(L)}$ according to the
definition \eq{3-form}.} For example, a crucial step in our approach
was the decomposition
\eq{f-sd-asd}. But that decomposition can also be done in the
Lorentzian signature by introducing an imaginary self-duality
$\eta^{(\pm)a}_{AB} = \pm \frac{i}{2} {\varepsilon_{AB}}^{CD}
\eta^{(\pm)a}_{CD}$ where $SU(2)_{L,R}$ is formally extended to
$SL(2, {\bf C})$. Indeed the proof in Appendix A can equally be done
using the imaginary self-duality as adopted in
\ct{mason-newman}. Or equivalently we can use the spinor
representation \ct{big-gravity} for an arbitrary anti-symmetric rank
2-tensor
\be \la{spinor}
F_{AB} = F_{a b \dot{a} \dot{b}} = \varepsilon_{\dot{a} \dot{b}}
\phi_{ab} + \varepsilon_{ab} \psi_{\dot{a} \dot{b}}
\ee
where $a, \dot{a}, \cdots$ are $SL(2, {\bf C})$ spinor indices. For
a real 2-form, $\psi = \bar{\phi}$. In this notation, the 2-form
dual to $F_{AB}$ is given by
\bea \la{dual-spinor}
{}^* F_{AB} &=& \half {\varepsilon_{AB}}^{CD} F_{CD} = {}^* F_{a b
\dot{a} \dot{b}}  \\
&=& -i \varepsilon_{\dot{a} \dot{b}} \phi_{ab} + i \varepsilon_{ab}
\psi_{\dot{a} \dot{b}},
\eea
that is,
\be \la{self-dual-spinor}
{}^* F_{a b \dot{a} \dot{b}} = i F_{a b \dot{b} \dot{a}} = - i F_{b
a \dot{a} \dot{b}}.
\ee

For the sake of completeness we will also consider $D=2$ and $D=3$
cases. For convenience we consider the Euclidean signature again for
both cases. (The $D=2$ case should be Euclidean in our context since
we don't want to consider time-space noncommutativity.) From now on
we set $\hbar = c = 1$.

In two dimensions, the analysis is simple. So we immediately list
the formulas:
\bea \la{2d-fomula-1}
&& \mathfrak{f}_{ABC} \equiv \varepsilon_{AB} \Psi_C, \\
\la{2d-fomula-2}
&& \rho_A = \mathfrak{f}_{BAB} = 2 D_A \log \lambda, \\
\la{2d-fomula-3}
&& \Psi_A = \varepsilon_{AB} \rho_B = 2 \varepsilon_{AB} D_B
\log \lambda, \\
\la{2d-fomula-4}
&& D_A \rho_A = - \rho_A \rho_A = - \Psi_A \Psi_A, \\
\la{2d-fomula-5}
&& D_A \Psi_A = 0, \\
\la{2d-fomula-6}
&& R_{ABCD} = \half \varepsilon_{AB} \varepsilon_{CD} R =
\half (\delta_{AC} \delta_{BD} - \delta_{AD} \delta_{BC}) R, \\
\la{2d-fomula-7}
&& R = \frac{2}{\lambda^2} (D_A D_A \log \lambda - 2 D_A \log
\lambda D_A \log \lambda).
\eea
Of course it is a bit lengthy to directly check Eq.\eq{2d-fomula-6}
from Eq.\eq{curvature-detail}.

Using the equation of motion \eq{2d-fomula-4}, the Ricci scalar
\eq{2d-fomula-7} can be rewritten as
\be \la{2d-ricci}
R = - \frac{2}{\lambda^2} \rho_A \rho_A = - \frac{2}{\lambda^2}
\Psi_A \Psi_A = - \frac{8}{\lambda^2} D_A
\log \lambda D_A \log \lambda.
\ee
The Einstein equation in two dimensions can be written as
\be \la{2d-einstein-eq}
R_{AB} = \half \delta_{AB} R = - \frac{1}{2\lambda^2}
\delta_{AB} \mathfrak{f}_{CDE} \mathfrak{f}_{CDE}.
\ee
An interesting thing in Eq.\eq{2d-ricci} is that the Ricci scalar is
always negative unlike as the 4-dimensional case where $R =
\frac{1}{2 \lambda^2} (\rho_A^2 - \Psi_A^2)$. Hence
Eq.\eq{2d-einstein-eq} describes only hyperbolic (negative
curvature) Riemann surfaces but most Riemann surfaces belong to this
class.

From Eq.\eq{2d-einstein-eq} one can see that the case with
$\widehat{F}_{AB} = 0$ corresponds to parabolic (curvature 0)
Riemann surfaces which include a plane ${\bf R}^2$ and a torus ${\bf
T}^2$. Then a natural question is where the different topology for
${\bf R}^2$ and ${\bf T}^2$ comes from. Note that there are still
background gauge fields given by Eq.\eq{vacuum-spacetime} although
the {\it fluctuations} are vanishing. (Two-dimensional gauge fields
do not have any physical degrees of freedom but encode only a
topological information. So the {\it fluctuations} here mean the
variation of a topological shape.) We observe that, though $B
\in H^2(M)$ in Eq.\eq{vacuum-spacetime} is constant,
it reveals its topology through the first cohomology group $H^1(M)$
which measures the obstruction for symplectic vector fields to be
globally Hamiltonian (see the footnote 3 in \ct{hsy2}). That is the
only source we can imagine for the origin of the topology of Riemann
surfaces. We believe that the topology of the fluctuation
$\widehat{F}_{AB}$ in Eq.\eq{2d-einstein-eq} similarly appears in
hyperbolic Riemann surfaces with a higher genus. Then a natural
question is about a rational (positive curvature) Riemann surface,
i.e., ${\bf S}^2$. It may be necessary to introduce a mass term as a
potential term. We leave it for a future work.\footnote{In this
respect, the work \ct{shimada} by H. Shimada should be interesting.
He showed that the topology of a membrane in matrix theory can be
captured by a Hamiltonian function defined on a Riemann surface. The
Hamiltonian function for a nontrivial Riemann surface is in general
given by a Morse function containing several nondegenerate critical
points, e.g., a height function, where the topology of a membrane is
realized as the Morse topology.}

Now we go over to $D=3$ case. In three dimensions
$\mathfrak{f}_{ABC}$ have totally 9 components. We will decompose
them into $9 = 1 + 3 + 5$ as follows
\be \la{3d-structure}
\mathfrak{f}_{ABC} = \varepsilon_{ABC} \Psi + \varepsilon_{ABD}
(\rho_{DC} + \varphi_{DC})
\ee
where the first term is totally anti-symmetric part like
Eq.\eq{D-3-form} and the second term is anti-symmetric, $\rho_{DC} =
- \rho_{CD}$, and the third term is symmetric, $\varphi_{DC} =
\varphi_{CD}$, and traceless, $\varphi_{CC} = 0$. Eq.\eq{D-structure-f}
then leads to the relation $\rho_{AB} = \half
\varepsilon_{ABC} \rho_C$. Therefore we get the following decomposition
\be \la{3d-structure-decomp}
\mathfrak{f}_{ABC} = \varepsilon_{ABC} \Psi +
\half (\delta_{AC} \rho_B - \delta_{BC} \rho_A) + \varepsilon_{ABD}
\varphi_{DC}.
\ee
In other words, the symmetric part can be deduced from
Eq.\eq{3d-structure-decomp} as follows
\be \la{3d-symm-part}
\varphi_{AB} = \half \varepsilon_{ACD} \mathfrak{f}_{CDB}
- \half \varepsilon_{ABC} \rho_C - \delta_{AB} \Psi.
\ee

Using the variables in Eq.\eq{3d-structure-decomp}, the equations of
motion \eq{map-eom} can be written as
\bea \la{3d-eom}
D_B \mathfrak{f}_{BCA} & = & -2 \delta_{AC} \Psi^2 - \Psi
\varphi_{AC} + \frac{1}{4}(\delta_{AC} \rho_B \rho_B - \rho_A
\rho_C) \\
&& + \frac{3}{2} \varepsilon_{ACB} \Psi \rho_B + \varepsilon_{CBD}
\rho_B \varphi_{DA} + \half \varepsilon_{ACB} \rho_D  \varphi_{BD} +
 \varphi_{AB} \varphi_{CB}.
\eea
Contracting the indices $A$ and $C$ in the above equation leads to
the relation
\be \la{3d-eom-trace}
D_A \rho_A = 6 \Psi^2 - \frac{1}{2} \rho_A \rho_A -  \varphi_{AB}
\varphi_{AB}.
\ee
Using the above results, it is straightforward though a bit lengthy
to calculate the Ricci tensor \eq{ricci-tensor}
\bea \la{3d-ricci}
R_{AC} &=& - \frac{1}{\lambda^2} \Big(\mathfrak{f}_{ABD}
\mathfrak{f}_{CBD} -
\frac{1}{4} \delta_{AC} \mathfrak{f}_{BED} \mathfrak{f}_{BED} \Big) \xx
&& + \frac{1}{4\lambda} (\nabla_A \rho_C + \nabla_C \rho_A) +
\frac{1}{2 \lambda^2} \rho_A \rho_C
\eea
and the Ricci scalar \eq{ricci-scalar}
\be \la{3d-ricci-scalar}
R = \frac{1}{\lambda} \nabla_A \rho_A +
\frac{1}{2 \lambda^2} \Big( \rho_A \rho_A - 9 \Psi^2 \Big).
\ee
Since the first term in Eq.\eq{ricci-tensor} is nonvanishing while
it was absent in four dimensions, we introduced the covariant
derivative of the ``Liouville" field $\rho_A$ defined by
\be \la{3d-covariant-der}
\nabla_A \rho_C = E_A \rho_C - {{\omega_A}^B}_C \rho_B
\ee
and then we used the following relation derived from
Eq.\eq{spin-conformal}
\be \la{3d-cov-id}
\nabla_A \rho_C + \nabla_C \rho_A = \frac{1}{\lambda} \Big( D_A
\rho_C + D_C \rho_A - (\mathfrak{f}_{ABC} + \mathfrak{f}_{CBA})
\rho_B + \delta_{AC} \rho_B \rho_B - \rho_A \rho_C \Big).
\ee
Also the expression \eq{3d-ricci-scalar} has been achieved after
using the relation
\be \la{3d-f2}
\mathfrak{f}_{ABC} \mathfrak{f}_{ABC}= 18 \Psi^2 - 2 \lambda  \nabla_A
\rho_A.
\ee

Finally we can get the 3-dimensional Einstein equation induced from
the NC $U(1)$ gauge fields
\bea \la{3d-einstein-eq}
E_{AB} &=& R_{AB} - \half \delta_{AB} R \xx &=& 8 \pi G_3 \big(
T^{(M)}_{AB} + T^{(L)}_{AB} \big)
\eea
where the Maxwell energy-momentum tensor and the Liouville
energy-momentum tensor are, respectively, given by
\bea \la{3d-maxwell-em}
T^{(M)}_{AB} &=& - \frac{1}{8 \pi G_3 \lambda^2}
\Big(\mathfrak{f}_{ACD} \mathfrak{f}_{BCD} -
\frac{1}{4} \delta_{AB} \mathfrak{f}_{CDE} \mathfrak{f}_{CDE} \Big)  \\
\la{3d-liouville-em}
T^{(L)}_{AB} &=& \frac{1}{16 \pi G_3 \lambda^2} \Big(
\half \big( \widetilde{\nabla}_A \rho_B + \widetilde{\nabla}_B \rho_A
+ \rho_A \rho_B \big) \xx && - \delta_{AB}
\big( \widetilde{\nabla}_C \rho_C  + \half( \rho_C \rho_C - 9 \Psi^2) \big) \Big)
\eea
where $\widetilde{\nabla}_A = \lambda \nabla_A$.

Following the exactly same strategy as the four dimensional case,
one can identify $\widehat{T}_{AB}^{(M)}({\cal A}_\theta)$ from
Eq.\eq{3d-maxwell-em} getting the same form as Eq.\eq{em-maxwell}.
Once again we get an exotic form of energy described by
Eq.\eq{3d-liouville-em} in addition to the usual Maxwell
energy-momentum tensor. This energy density is also related to the
gradient volume energy. (See Section 3.2.) But the explicit form is
different from the four dimensional one, Eq.\eq{dark-energy}. This
difference is due to the fact that the first term in
Eq.\eq{ricci-tensor}, which appears as the covariant derivative
terms in Eq.\eq{3d-liouville-em}, is absent in four dimensions. An
interesting thing in Eq.\eq{3d-liouville-em} is that $\rho_A$
behaves like a massive field whose mass is vanishing in flat
spacetime since $\lambda = 1$ in that case. We further discuss in
Section 3.4 about the physical implications of the Liouville
energy-momentum tensor.

In higher $D \geq 5$ dimensions, the calculation of the
energy-momentum tensor from Eq.\eq{ricci-tensor} becomes more
complicated. The 3-form field \eq{3-form-h} contributes nontrivially
to the energy-momentum tensor. We have not tried to find its
concrete form. We hope to attack this problem in the near future.

\newpage


\nc{\npb}[3]{Nucl. Phys. {\bf B#1} (#2) #3}

\nc{\plb}[3]{Phys. Lett. {\bf B#1}(#2) #3}

\nc{\prl}[3]{Phys. Rev. Lett. {\bf #1} (#2) #3}

\nc{\prd}[3]{Phys. Rev. {\bf D#1} (#2) #3}

\nc{\ap}[3]{Ann. Phys. {\bf #1} (#2) #3}

\nc{\prep}[3]{Phys. Rep. {\bf #1} (#2) #3}

\nc{\epj}[3]{Eur. Phys. J. {\bf #1} (#2) #3}

\nc{\ptp}[3]{Prog. Theor. Phys. {\bf #1} (#2) #3}

\nc{\rmp}[3]{Rev. Mod. Phys. {\bf #1} (#2) #3}

\nc{\cmp}[3]{Commun. Math. Phys. {\bf #1} (#2) #3}

\nc{\mpl}[3]{Mod. Phys. Lett. {\bf #1} (#2) #3}

\nc{\cqg}[3]{Class. Quant. Grav. {\bf #1} (#2) #3}

\nc{\jhep}[3]{J. High Energy Phys. {\bf #1} (#2) #3}

\nc{\atmp}[3]{Adv. Theor. Math. Phys. {\bf #1} (#2) #3}

\nc{\hepth}[1]{{\tt hep-th/{#1}}}


\end{document}